\documentclass[11pt,a4paper]{article}

\usepackage[T1]{fontenc}
\usepackage[utf8]{inputenc}
\usepackage{lmodern}
\usepackage[margin=2.5cm]{geometry}
\usepackage{amsmath,amssymb,amsfonts}
\usepackage{graphicx}
\usepackage[normalem]{ulem}
\usepackage[numbers,sort&compress]{natbib}
\usepackage[hidelinks]{hyperref}

\title{On Photon Propagation in Strong Magnetic Fields: Time Delays from Pulsars}
\author{%
	A. W. Romero Jorge$^{1,2,3,*}$, A. P\'erez Mart\'inez$^{4}$,\\
	E. Rodr\'iguez Querts$^{4}$, and H. P\'erez Rojas$^{4}$\\[1.0em]
	{\small $^{1}$ Frankfurt Institute for Advanced Studies, Ruth Moufang Str.\ 1, 60438 Frankfurt, Germany}\\
	{\small $^{2}$ Institut f\"ur Theoretische Physik, Johann Wolfgang Goethe University,}\\
	{\small Max-von-Laue-Str.\ 1, 60438 Frankfurt, Germany}\\
	{\small $^{3}$ Helmholtz Research Academy Hessen for FAIR (HFHF), GSI Helmholtz Center}\\
	{\small for Heavy Ion Physics, Campus Frankfurt, 60438 Frankfurt, Germany}\\
	{\small $^{4}$ Instituto de Cibern\'etica, Matem\'atica y F\'isica (ICIMAF),}\\
	{\small Calle E esq a 15, Vedado 10400, La Habana, Cuba}\\[0.4em]
	{\small $^{*}$ Corresponding author}
}
\begin{document}
	\maketitle
	
	\begin{abstract}
		We study photon dispersion in the presence of a strong external magnetic field for propagation both perpendicular and parallel to the field, considering (i) the magnetized QED vacuum and (ii) a magnetized electron-positron medium at finite density (nonzero chemical potential). The analysis is performed in one-loop QED at finite temperature and density, and we extract polarization-dependent dispersion relations and phase velocities from the eigenmodes of the photon polarization operator. Using these phase velocities, we estimate propagation-induced time delays for photons emitted from neutron stars in the strong-field regime, adopting a simplified magnetosphere model with a toroidal magnetic field geometry and a radially varying chemical potential profile in the degenerate limit. We compare the predicted delays with observational timing constraints and discuss how strong-field QED and plasma corrections can generate measurable departures from the light cone, with potential signatures relevant for magnetars and modern polarimetric and timing observations. It should be noted that the time-delay estimates derived here are based on the phase velocity and serve as indicative order-of-magnitude predictions.
	\end{abstract}
	
	%\noindent\textbf{Keywords:} photon dispersion; strong magnetic fields; magnetars; QED; electron-positron plasma; finite temperature and density; phase velocity; time delay
	
	\bigskip
	\section{Introduction}\label{sec1}
	
	Maxwell's theory predicts that electromagnetic waves propagate in vacuum at the universal speed $c$.  Quantum electrodynamics (QED), however, implies that the
	vacuum is \emph{polarizable}: virtual electron-positron fluctuations couple to
	external fields and modify the propagation of photons
	\cite{Euler35,Schwinger1951}.  In the presence of sufficiently intense
	background fields, these radiative corrections promote classical electrodynamics
	to an effective nonlinear theory, enabling phenomena such as vacuum
	birefringence, photon splitting, and field-induced pair creation
	\cite{Rojas79,Shabad72,DiPiazza:2021rum}.
	
	A constant magnetic field breaks isotropy and splits photon propagation into
	distinct polarization eigenmodes with different dispersion relations.  In
	vacuum, the classic one-loop calculations of the polarization operator
	$\Pi_{\mu\nu}$ in a magnetic field established the mode decomposition and the
	corresponding eigenvalues that control the deformation (or preservation) of the
	light cone condition \cite{Rojas79,Shabad72}.  In this context, a key qualitative
	feature is that strictly \emph{parallel} propagation along the symmetry axis
	defined by $\mathbf{B}$ remains tied to the light cone, whereas for
	\emph{perpendicular} propagation the transverse modes acquire nontrivial
	dispersive corrections, giving rise to Cotton-Mouton-type birefringence
	\cite{PhysRevD.12.1132,Shabad72, Rojas79,Heyl_1997}. 
	
	These strong-field QED effects are actively pursued experimentally.  Vacuum
	magnetic birefringence remains extremely small at laboratory field strengths,
	but dedicated setups (e.g.\ PVLAS and related efforts) have produced steadily
	improving bounds approaching the QED prediction \cite{Ejlli:2020yhk}.  In
	parallel, neutron stars provide a natural arena where magnetic fields can reach
	(or exceed) the QED critical scale $B_c\simeq 4.41\times 10^{13}\,\mathrm{G}$,
	particularly in magnetars \cite{PhysRevLett.91.071101}-\cite{Turolla2017}.  The first optical
	polarimetry-based evidence consistent with vacuum birefringence was reported for the isolated neutron star (NS) RX~J1856.5-3754 \cite{Turolla2017}.
	
	From an astrophysical standpoint, photon propagation near neutron stars is
	shaped not only by vacuum polarization but also by magnetospheric plasma.  More
	realistic magnetosphere models admit electron-positron populations  that can
	partially or fully fill the environment, introducing finite-density and
	finite-temperature scales (e.g.\ chemical potential $\mu$) that must be treated
	on the same footing as the external field \cite{Petri16,Turolla2017}.  In
	such a magnetized medium, the  photon self-energy
	$\Pi_{\mu\nu}(k;B,\mu,T)$ generally develops off-diagonal components, and the
	physical propagation modes follow from the eigenvalues of $\Pi_{\mu\nu}$ after
	its diagonalization \cite{Rojas79,Rojas87}.  
	
	Besides the direct calculation of the one-loop polarization operator, an alternative and widely used description of photon propagation in external electromagnetic fields is provided by the effective-action approach. In the low-energy regime, the nonlinear interaction of photons is encoded in the Euler-Heisenberg effective action, whose derivatives with respect to the electromagnetic invariants determine the coefficients of the quadratic expansion in the photon field and, consequently, the corresponding polarization operator. This formulation has been extensively developed for magnetized vacuum and subsequently generalized to include matter effects, finite temperature, and finite density, providing a complementary framework for the study of photon propagation and nonlinear electromagnetic phenomena in strong-field environments\cite{PhysRevD.42.2881,PhysRevLett.71.480,
		Persson_1995,PhysRevD.91.085027,PhysRevD.88.085033,Karbstein2013}
	
	A direct observable consequence of dispersive propagation is a frequency- and
	polarization-dependent \emph{time delay}\cite{Kim2023EPJC}. Standard pulsar timing already
	exhibits the familiar interstellar plasma dispersion delay
	$\Delta t\propto \omega^{-2}$, which mainly probes the line-of-sight electron
	column density \cite{Push10,Wang2004}.  Here we focus instead on a delay
	accumulated \emph{near the source}, where strong-field QED and magnetospheric
	plasma effects can induce additional phase velocity shifts and potentially
	observable delays for magnetar-scale parameters. 
	
	In this work we pursue two objectives.  First, we solve the one-loop dispersion
	equation for photons propagating parallel and perpendicular to $\mathbf{B}$ in
	(i) the magnetized QED vacuum and (ii) a cold electron-positron medium at
	finite chemical potential, emphasizing the eigenmode structure and the regimes
	where medium-induced mode mixing becomes essential.  Second, we translate the
	resulting phase velocities into time-delay estimates in a simplified neutron
	star magnetosphere model with a radially varying $\mu(r)$ and a large-scale
	magnetic field geometry.
	Building on our earlier study of time delays in a magnetized vacuum \cite{romero2020photon}, we extend the analysis to include a cold (degenerate) electron–positron medium and quantify the resulting dispersion for both parallel and perpendicular propagation.
	
	Our calculations for the magnetized medium present at least three  aspects to be highlighted. The first concerns an improved treatment of the one-loop corrections to the polarization operator, refining previous results for the dispersion relations and phase velocity of photons propagating in a magnetized medium in the degenerate limit (zero temperature and nonzero chemical potential). In particular, for perpendicular propagation, the eigenvalues have been computed for the first time. In contrast, for parallel propagation, they have been obtained without relying on some of the simplifying assumptions adopted in earlier work \cite{Lidice13}. The resulting eigenvalues are analyzed in two limiting regimes: the long-wavelength approximation and the infrared limit.
	The second contribution establishes the connection between the one-loop calculation and the effective-action formalism. Specifically, we extend to finite density the quadratic expansion of the Euler-Heisenberg effective action \cite{Shabad:2011hf,Jorge:2024tok} and derive the corresponding photon-self energy in the infrared (IR) limit. We then verify that this approach reproduces the one-loop dispersion relation in the same approximation, providing an independent consistency check of both formulations while recovering the standard vacuum result in the appropriate limit.

	Finally, we aim to illustrate the effect of quantum corrections on observable astrophysical phenomena\cite{PhysRevLett.91.071101,Lai_2003,10.1111/j.1365-2966.2009.14895.x, Lai:2022knd,Kim2023AstronRep,Kim:2025uhi,Kim2023EPJC}.  Specifically, the quantum corrections to the phase velocity derived within the perturbative approach have been used to explain the delay of photons emerging from the magnetosphere of neutron stars. Naturally, the description neglects many features of realistic neutron star magnetospheres while retaining the essential physics. We consider two representative scenarios: one assuming a magnetized vacuum and the other treating it as a magnetized cold electron medium. Despite these simplifications, our model successfully explains the delay, demonstrating that medium corrections are larger than those in vacuum and are consistent with the expected microsecond-scale delay.

	The paper is organized as follows. Section~\ref{sec2} introduces the  photon self-energy  in a strong magnetic field and its diagonalization into propagation eigenmodes. Section~\ref{vacuum0} presents analytic approximations and numerical solutions for the dispersion relations in vacuum. Section~\ref{sec3} presents the solution in a magnetized medium in the degenerate, finite-density limit. Two regimes are examined: the long-wavelength and infrared limits. Numerical results for the solutions of the dispersion equation are given in Section~\ref{Numerical}, where phase velocities are obtained for parallel and perpendicular photon propagation.
	
	Section~\ref{sec5} applies the resulting phase velocities to a simplified neutron star magnetosphere model and computes the associated photon time delays. We conclude in Sec.~\ref{sec7}. Natural units are used throughout, except where stated otherwise.
	
	The Appendix complements the details of the calculations. In particular, Appendix~B presents a solution of the dispersion equation in vacuum using a power-series expansion of the photon self-energy in terms of hypergeometric functions. Appendix~D presents the extension to a magnetized medium at finite density and zero temperature of the quadratic expansion of the Euler–Heisenberg effective action in the photon field, from which the infrarred regime photon self-energy  is obtained.
	
	\section{Study of photon propagation in a magnetized vacuum and medium}
	\label{sec2}
	
	In this section, we examine how photons propagate under a strong external magnetic field in both a magnetized vacuum, 
	and an electron-positron plasma with nonzero chemical potential. We begin by briefly reviewing the propagation of electromagnetic waves in vacuum according to Maxwell's classical theory, followed by an introduction to quantum corrections encoded in the one-loop photon self-energy~$\Pi_{\mu\nu}$. By diagonalizing $\Pi_{\mu\nu}$, we identify the distinct eigenmodes of photon propagation and derive explicit dispersion relations, which exhibit departures from the standard light cone condition in the form of modified phase velocities and refractive indices. These deviations provide a quantitative measure of vacuum polarization and medium effects in strong magnetic fields, laying the groundwork for subsequent discussions on finite temperature and density regimes (see Sec.~\ref{sec4}) and their application in astrophysical contexts.
	
	\subsection{Propagation of light in the  vacuum:  Classical Maxwell Theory}
	
	The propagation of an electromagnetic wave in a vacuum at the speed of light is described by classical electrodynamics using the Lagrangian \cite{Jackson1998,Peskin:1995ev},
	\begin{equation}\label{ddkedyd} \mathcal{L}= -\dfrac{1}{4}F_{\mu\nu}F^{\mu\nu},
	\end{equation}
	where $\rho, \nu, \mu, \nu = 0,1,2,3$, $F_{\nu\rho} = \partial_\nu a_\rho - \partial_\rho a_\nu$ is the electromagnetic field tensor, and $a_\nu$ is the four-potential vector of the wave field.
	From this Lagrangian, Maxwell's equations for the vacuum without external currents can be derived as
	$\partial_\nu F^{\nu \rho}=0 $, in momentum space,
	$  k^2 a^{\rho}(x)=0\label{rdfouriedr} $, where $a_{\nu}(x)$ are the electromagnetic propagation modes of small amplitude, defined as $\langle a_{\nu}(x) \rangle$ associated with the wave field. In this paper, the Lorentz gauge, $\partial_\mu a^{\mu}=0$, is used in Fourier space, $k_\mu a^\mu=0$.
	The last equation is a system of homogeneous equations, whose trivial solution is $a_{\nu}(x)=0$. For a non-trivial solution, it must satisfy $\det{(k^2g_{\rho \nu})}=0$, whose solution is the light cone $k^2=0$, corresponding to electromagnetic waves propagating at the speed of light in a vacuum,
	\begin{equation}
		k_\perp^2+k_3^2-\omega^2=0,
	\end{equation}
	where $k_\perp^2$ and $k_3^2$ are the perpendicular and parallel components of the wave momentum, respectively, and $\omega$ is the photon energy.
	
	Despite the success of classical electrodynamics in unifying special relativity and electromagnetism, it has limitations at high energies and intense magnetic fields. Quantum Electrodynamics (QED) addresses these limitations by accurately describing quantum effects at energy scales comparable to the electron mass and corresponding electromagnetic fields
	\cite{Schwinger1951,DittrichLibro}.
	\subsection{Propagation of light in the magnetized vacuum of QED}
	In the presence of an external magnetic field, the propagation of an electromagnetic wave in the vacuum of the QED is described by the  Lagrangian density \cite{Jackson1998,Peskin:1995ev}
	\begin{equation}\label{lagEDC}
		\mathcal{L}=\bar{\psi}[i \gamma^\mu(\partial_\mu-i e A_\mu)-m_e]\psi -\dfrac{1}{4}F_{\mu\nu}F^{\mu\nu}%-ie\bar{\psi}\gamma_\mu A_\mu \psi,
	\end{equation}
	where $ \psi(x) $, $ \bar{\psi}(x) $ are the four-spinors of the electron-positron field, $ e $ is the electron charge, $ \gamma^\mu $ are the gamma matrices and $ m_e$ the electron mass, $F_{\mu\nu}=\partial^{\mu}A^{\nu}(x)-\partial^{\nu}A^{\mu}(x)$ is the electromagnetic field tensor, $A^{\nu}(x)$ the four-potential vector and $\mu,\nu=0,1,2,3$.

	\subsection{Propagation of light in the magnetized vacuum and medium. Photon self-energy.}
	
	The propagation of a wave in the magnetized medium is described by the Maxwell equation  \cite{Jackson1998},
	
	\begin{equation}
		[\partial_{\mu}^2-\partial_{\mu}\partial_{\nu} + \Pi_{\mu \nu}(\textbf{x},B,\mu,T)]a_{\mu} (x)=0,
	\end{equation}
	where $\Pi\textbf{(x},B,\mu,T)$ is the photon self-energy  in a magnetized medium which is obtained using corrections, in particular the photon self-energy 
	in the presence of the magnetic field and depends also on the temperature and particle density.
	To do this, we must calculate with the
	Schwinger-Dyson (SD) equation for the photon, in particular the corrections of the self-energy of the photonic field or photon photon self-energy \cite{Jorgito2021,Shabad84,Rojas79,Rojas87}, when doing the Fourier transform for Eq. of SD for the photon, we obtain
	\begin{equation}\label{kesssy}
		[ k^2 g_{\mu \nu} + \Pi_{\mu \nu}]  a^\nu (k)=0,
	\end{equation}
	where 
	$g_{\mu \nu}$ the metric tensor (We use the Minkowski metric  with  sign convention (-,+,+,+)).
	These calculations are made using quantum field theory at finite temperature and density and the imaginary time formalism, more details can be seen in \cite{Jorgito2021}. 
	The self-energy of the photon has the form
	{\small
		\begin{equation}\label{kesssy2}
			\Pi_{\mu\nu}(x,y)=e^2 Tr \int \gamma_\mu G(x,z) \Gamma_\nu (z,y^{\prime},y) G(y^{\prime},x) d^4 z d^4 y^{\prime}, 
	\end{equation}}
	with $ \gamma_\nu $ are the gamma matrices. To calculate the components of $ \Pi_{\mu\nu} $, we take (\ref{kesssy}) in the one-loop approximation with  the vertex $\Gamma_\nu=\gamma_\nu \delta(z-y^{\prime }) \delta(z-y)$ and the finite temperature Green's functions $ G(x,y|A_\nu^{ext}) $, which are solutions to the Dirac equation in a constant magnetic field $ \mathbf{B} $, such that $ A_\mu^{ext}=B x_1 \delta_{\nu,2}$,
	\begin{equation}\label{ecDiracGreen} [\gamma_\nu(\partial_\nu-i e A_\nu^{ext})+m_e] G(x,y|A_\nu^{ext}) = \delta(x-y), \end{equation}  $ \psi $ and $ \bar{\psi} $ are the Dirac spinors describing the electron-positron field, $ e $ is the electron charge, and $ m_e$ is the electron mass.
	Here, $ \partial_4=\partial/\partial x_4-\mu $, where $ \mu $ is the chemical potential of the electron-positron gas. Equation (\ref{ecDiracGreen}) defines the temperature-dependent Green's function for fermions with $ x_4 $ in the interval from $ -\beta $ to $ \beta $, where $ \beta=1/T $. The energy spectrum for the electron-positron gas is $ \varepsilon_{p_3,n} =\sqrt{m_e^2+p_3^2+2en B}$, where $ p_3 $
	is the momentum in the $ x_3 $ direction, and 
	$n$ are the Landau levels.
	
	To obtain the dispersion equation, we first need to diagonalize the photon self-energy  $ \Pi_{\mu\nu} $.  The structure of the photon self-energy  was obtained and thoroughly discussed in \cite{Rojas79}, \cite{Shabad84}, \cite{Rojas87}, showing that it can be expanded in terms of nine independent perpendiculartensors $ \Psi^{(i)}_{\mu\nu} $, six of which are symmetric tensors and three antisymmetric.
	This diagonalization is only nontrivial in a magnetized \emph{medium}
	($\mu\neq0$ and/or $T\neq0$), where $\Pi_{\mu\nu}(k;B,\mu,T)$ generally contains
	off-diagonal terms that mix the perpendicularbasis vectors; hence one must
	explicitly diagonalize $\Pi_{\mu\nu}$ to obtain the physical eigenvectors
	$c_\mu^{\prime(j)}$ and eigenvalues $\kappa^{(j)}$ (see also
	Refs.~\cite{Rojas79,Shabad84,Rojas87}). 
	
	Thus, its diagonalization takes the form,
	\begin{equation}
		\Pi_{\mu\nu}=\sum_{j=1}^{4} \kappa^{(j)} \dfrac{c_\mu^{\prime (j)} c_\nu^{\prime (j)}}{c_\rho^{\prime (j)} c_\rho^{ \prime (j)}}, 
	\end{equation}
	where $\kappa^{(j)}$ and $c_{\mu}^{ \prime  (j)}$ are the eigenvalues and eigenvectors of the photon self-energy  and obey the secular equation $\Pi_{\mu\nu}c_\nu^{\prime (j)}=\kappa^{(j)}c_\mu^{ \prime(j)}$.
	
	The photon self-energy $\Pi_{\mu\nu}$ in the rest frame, considering finite density and temperature equal to zero  \cite{Rojas79,Shabad84,Rojas87,Shabad72,Shabad75}, is reduced to the following expression, depending only on  four independent scalars ,
	
	\begin{equation}
		\Pi_{\mu\nu}=\left[\begin{array}{ccc}
			p & 0 & r\\
			0 & s & 0\\
			-r & 0 &  t^- \end{array}\right],\label{polarization}
	\end{equation}
	in the vector basis
	\begin{equation} \nonumber
		{c_{\mu}}^{(1)}(k)=({F_{\mu\nu}}^{2}k^{\nu})k^{2}-k_{\mu}(kF^{2}k),
		\hspace{0.2cm} {c_{\mu}}^{(2)}(k)=F_{\mu\nu}^{*}k^{\nu}, \hspace{0.2cm}
		{c_{\mu}}^{(3)}(k)=F_{\mu\nu}k^{\nu}, \hspace{0.2cm}
		{c_{\mu}}^{(4)}(k)=k_{\mu}.
	\end{equation}
	Note that  the transversality of the  polarization operator,
	$\Pi_{\mu\nu}(k)k^{\nu}=0$,  implies that
	${a_{\mu}}^{(4)}=k_{\mu}$ and $\kappa_{4}=0$. This mode is purely longitudinal and has no physical significance; therefore, it will be omitted in what follows. Depending on the direction of photon propagation relative to the magnetic field, the corresponding eigenvectors and eigenvalues take different forms. For perpendicular propagation, the eigenvectors are
	\begin{equation}
		c^{\prime (1,3)}_{\mu}=r c_{\mu}^{(1)}+(p-\kappa^{(1,3)}_M)c_{\mu}^{(3)},
		\quad c^{\prime (2)}_{\mu}=c^{(2)}_{\mu}.
		\label{AMPO}
	\end{equation}
	The corresponding eigenvalues are
	\begin{eqnarray}
		\kappa^{(1,3)}_M=\dfrac{p+t^{-}}{2}\pm \dfrac{1}{2}\sqrt{(p-t^{-})^2-4r^2},
		\label{eqdispperp}
		\quad\quad\quad
		\kappa^{(2)}_M=s.
		\label{eigenvalues_general}
	\end{eqnarray}
	For parallel propagation, the scalars satisfy $p=t^{-}$, and the eigenvalues become
	\begin{eqnarray}
		\kappa^{(1,3)}_M=t^{-}\pm \sqrt{-r^2},
		\quad\quad
		\kappa^{(2)}_M=s.
		\label{eigenvalues_general2}
	\end{eqnarray}
	In this case, the eigenvectors in Eq.~(\ref{AMPO}) become
	\begin{equation}
		c^{\prime (1,3)}_{\mu}=(c_{\mu}^{(1)}\pm ic_{\mu}^{(3)}),
		\quad
		c^{\prime (2)}_{\mu}=c^{(2)}_{\mu}.
		\label{AMPO2}
	\end{equation}
	The modes $c^{\prime (1,3)}_{\mu}$ describe circularly polarized waves propagating along the magnetic field, with opposite senses of circular polarization. The mode $c^{\prime (2)}_{\mu}$ corresponds to a longitudinal wave propagating along the magnetic field, with eigenvalue $\kappa^{(2)}_M=s$.
	
	To study the dispersion equations separately in vacuum and in the medium, we decompose the total photon self-energy into its vacuum and medium contributions as
	
	\begin{equation}
		\Pi^T_{\mu \nu}(\omega,\textbf{k},B,\mu,T)=\Pi^{V}_{\mu \nu}(\omega,k,B)+\Pi^{M}_{\mu \nu}(\omega,\textbf{k},B,\mu,T),\label{PiT}
	\end{equation}
	where $k^\mu=(\omega,\textbf{k})$, $ \Pi^{V}_{\mu \nu}(B)$ and $\Pi^{M}_{\mu \nu}(\omega,\textbf{k},B,\mu,T)$ is the vacuum and medium contribution, respectively. 
	The corresponding dispersion equation for Eq. (\ref{PiT}) is an algebraic dispersion equation
	
	\begin{equation}
		\textbf{k}^2-\omega^2 =\kappa^{(i)}=\kappa_V^{(i)}(\omega,\textbf{k},B)+\kappa_{M}^{(i)}(\omega,\textbf{k},B,\mu,T),\label{disprel}
	\end{equation}
	where (i) denotes the polarization mode, $\kappa_V^{(i)}$ and $\kappa_M^{(i)}$ are the eigenvalues of the photon self-energy, corresponding to its vacuum and medium contributions, respectively.
	
	\section{Eigenvalues of photon self-energy $\kappa_V^{(i)}$ and phase velocity in a magnetized vacuum}
	\label{vacuum0}
	We first consider the vacuum limit, $\mu=T=0$, before turning to the cold medium ($T=0$, $\mu\neq 0$). In this limit, $r=0$, and the photon self-energy in Eq.~(\ref{polarization}) becomes diagonal in the basis $c_\mu^{(i)}$, $i=1,2,3$,
	\begin{equation}
		\Pi_{\mu\nu}=\left[\begin{array}{ccc}
			p_0 & 0 &0\\
			0 & s_0 & 0\\
			0 & 0 &  t^-_0
		\end{array}\right].
		\label{polarization_0}
	\end{equation}
	\footnote{For the vacuum case, we denote $p$, $s$, and $t^{-}$ with the subscript $0$ to distinguish them from the corresponding quantities in the medium.}
	The corresponding eigenvalues are
	\begin{align}
		\kappa^{(1)}_V=p_0,\quad
		\kappa^{(2)}_V=s_0,\quad
		\kappa^{(3)}_V=t^{-}_0.
		\label{eigenvalues_generalvac}
	\end{align}
	The explicit form of the eigenvalues of the photon self-energy  in the vacuum  \cite{Shabad75,Romero20} take the explicit form 
	\begin{equation}	
		\kappa^{(i)}_V (\omega,\textbf{k},B)=\dfrac{2 \alpha}{\pi}\int_{0}^{\infty}d\tau\int_{-1}^{1}d\eta  \,  e^{-\dfrac{\tau}{b}}   \left[    \rho_{i} e^{\zeta}+\dfrac{k^2\bar{\eta}^2}{2t}  \right] , \hspace{1cm} i=1,2,3,\label{vacuumpolarizationtensor2}
	\end{equation}
	where the functions $\rho_i$ depend on the propagation mode, $ \alpha $ is the fine structure constant,  $b$ is the normalized magnetic field $b=\frac{B}{B_{c}}$ and 
	\begin{equation}
		\small
		\rho_1=-\dfrac{ k^2}{2}\varrho,
		\quad
		\rho_2=\dfrac{\omega^2-k^2_{3}}{2}\bar\eta^2 \coth{\tau}-\dfrac{k^2_{\perp}}{2}\varrho,  \quad \rho_3=\dfrac{\omega^2-k^2_{3}}{2}\varrho-\dfrac{k^2_{\perp}}{2}\dfrac{\varsigma}{\sinh^2{\tau}},
	\end{equation}
	\begin{equation}
		\small
		\zeta=\dfrac{(\omega^2-k^2_{3})\bar{\eta}^2\tau-k^2_{\perp}\xi}{bm^2},
	\end{equation}
	\begin{equation}
		\small
		\xi=\dfrac{\sinh{(\eta_+\tau)}\sinh{(\eta_-\tau)} }{\sinh{\tau}},
		\quad \eta_\pm=\frac{1\pm\eta}{2},\label{xivarrho}
	\end{equation}
	\begin{equation}
		\small      \varrho=\dfrac{\sinh{(\eta_+\tau)}\cosh{(\eta_+\tau)} }{\sinh^2{\tau}}\eta_-,
		\quad 
		\bar{\eta}=\sqrt{\eta_+ \eta_-},\label{etaeta}
		\noindent
	\end{equation}
	
	The dispersion equation for photon propagation in the magnetized vacuum is
	\begin{equation}
		\mathbf{k}^2-\omega^2=\kappa_V^{(i)}(\omega,\mathbf{k},B).
		\label{dispersionequationV}
	\end{equation} 
	For parallel propagation all the three eigenvalues become  proportional to $k^2$, that is $\rho_{i}(\omega^2,k^2_{3}, k^2_{\perp},\eta, \tau)\vert_{k_{\perp}=0} =k^2 \theta_i (\eta, \tau)$,  leading to the light cone solutions of the dispersion equations
	\begin{equation}	
		k^2-\kappa^{(i)}_V= k^2\left(1- \dfrac{2 \alpha}{\pi}\int_{0}^{\infty}d\tau\int_{-1}^{1}d\eta  \,  e^{-\dfrac{\tau}{b}}   \left[    \theta_{i} e^{\zeta}+\dfrac{\bar{\eta}^2}{2t}  \right] \right)=0, \hspace{1cm} \label{vacuumDE}
	\end{equation}
	Thus, for photon propagation parallel to the magnetic field, the phase velocity remains equal to $c$.
	For perpendicular photon propagation, $\mathbf{k}\perp\mathbf{B}$, the physical modes are $c_\nu^{(2)}$ and $c_\nu^{(3)}$, while $c_\nu^{(1)}$ corresponds to the longitudinal mode. The corresponding dispersion relations are obtained from Eq.~(\ref{dispersionequationV}).
	The electric and the magnetic fields,
	$\mathbf{e}^{i}=\left(\frac{\partial \mathbf{c}^{(i)}}{\partial x_{0}}\right)-\left(\frac{\partial c_{0}^{(i)}}{\partial \mathbf{x}}\right)$ and 
	$\mathbf{h}^{(i)}=\nabla\times\mathbf{c}^{(i)}$, respectively, associated with the different photon modes are:
	
	\begin{equation} \nonumber
		\mathbf{e}^{(1)}=-\mathbf{k}_{\perp}k^{2}\omega,\hspace{0.4cm}
		\mathbf{h}^{(1)}=[\mathbf{k}_{\perp}\times\mathbf{k}_{3}]k^{2},
	\end{equation}
	
	\begin{equation}
		\mathbf{e}_{\perp}^{(2)}=\mathbf{k}_{\perp}k_{3}^{2},\hspace{0.4cm}
		\mathbf{e}_{3}^{(2)}=\mathbf{k}_{3}(k_{3}^{2}-\omega^{2}),\hspace{0.4cm}
		\mathbf{h}^{(2)}=[\mathbf{k}_{\perp}\times\mathbf{k}_{3}]\omega,
	\end{equation}
	
	\begin{equation}\nonumber
		\mathbf{e}^{(3)}=[\mathbf{k}_{\perp}
		\times\mathbf{k}_{3}]
		\omega,\hspace{0.4cm}
		\mathbf{h}_{\perp}^{(3)}=-\mathbf{k}_{\perp}k_{3}^{2},\hspace{0.4cm}
		\mathbf{h}_{3}^{(3)}=\mathbf{k}_{3}k_{\perp}^{2},
	\end{equation}
	where $k_{3}$ and $k_{\perp}$ are the components of the photon propagation vector parallel and perpendicular to the magnetic field, respectively.
	
	With respect to the direction of radiation propagation, two fundamental cases can be distinguished. The first corresponds to propagation perpendicular to the external magnetic field ($\mathbf{k}_{3}=0$), for which modes (2) and (3) are transversely polarized ($\mathbf{e}^{(2,3)}\perp\mathbf{k}$) in two mutually orthogonal planes ($\mathbf{e}^{(2)}\perp\mathbf{e}^{(3)}$), while the first mode is longitudinally polarized and its magnetic field vanishes ($\mathbf{h}^{(1)}=0$), and therefore it does not correspond to a physical photon. In the second case, the propagation direction is parallel to $\mathbf{B}$ ($\mathbf{k}_{\perp}=0$); modes (1) and (3) are perpendicular($\mathbf{e}^{(1,3)}\perp\mathbf{B}$) and their electric fields are mutually orthogonal ($\mathbf{e}^{(1)}\perp\mathbf{e}^{(3)}$), whereas the second mode is longitudinal and does not represent a real photon. When $k_{\perp}=0$, the only preferred direction in space is that of $\mathbf{B}$, and due to the symmetry of the system, the two eigenvalues corresponding to the two mutually orthogonal polarizations coincide ($\kappa_{1}=\kappa_{3}$).
	\subsection{Phase velocity for photons of low energy: Infrared approximation}
	The dispersion equation (\ref{dispersionequationV}) is difficult to solve analytically and one usually resorts to approximation schemes; consequently, the solution of the dispersion equation depends on the approximation adopted.
	We first present the infrared approximation to get the phase velocity. For small  deviations from the light cone relation
	\[
	k^\mu k_\mu = k^2 = \mathbf{k}^2 - \omega^2 = 0 ,
	\]
	and for photon of low energy in the low-energy limit ($\omega<<2 m$)  ($\omega \sim kc$) we expand $e^{\zeta}\approx 1+\zeta$  leading 
	\begin{eqnarray}
		\kappa^{(2,3)}_V  =\dfrac{2 \alpha}{\pi}\int_{0}^{\infty}d\tau\int_{-1}^{1}d\eta  \,  e^{-\dfrac{\tau}{b}}   \left[    \rho_{2,3} (1+\zeta)+\dfrac{k^2\bar{\eta}^2}{2t}  \right] ,\label{appr2223}
	\end{eqnarray}
	with
	\begin{equation}
		%\rho_1=
		\rho_2=-\dfrac{ k_\perp^2}{2}\varrho, \quad  
		\rho_3=-\dfrac{k^2_{\perp}}{2}\dfrac{\varsigma}{\sinh^2{\tau}},
		\quad
		\zeta=-\dfrac{k^2_{\perp}\xi}{bm^2},
	\end{equation}
	The functions $\xi,\varrho,\eta_\pm$, and $\bar{\eta}$ remain consistent with Eq. (\ref{xivarrho}) and (\ref{etaeta}). By integrating Eq. (\ref{appr2223}) over $\eta$ and disregarding the second-order terms in $k_{\perp}^2$, we can derive the phase velocity $v_{(i)}=\omega/k_\perp$ from the dispersion equation (see for details appendix \ref{appen1}).
	
	The phase velocity can be obtained  for arbitrary values of the magnetic field yielding the explicit form \cite{Aurora2023}
	
	\begin{eqnarray}
		v^{(2)}_V(B)  = \left( 1- \dfrac{1}{2}
		s_0 B^2
		\right),\label{wf_velexact2} \\
		v^{(3)}_V(B)  = \left( 1- 
		\dfrac{1}{2} t_0  B^2
		\right),\label{wf_velexact3}
	\end{eqnarray}
	
	where $s_0$  and $t_0^{-}$ are 
	\begin{align}
		s_0&=\frac{\alpha}{2\pi B^2}\left (-\frac{1}{3}-\frac{2}{3}\left( \psi(1+h) -2h^2 +(3h)^{-1}\right ) +8\zeta^{\prime}(-1,h)\right .
		\\
		&\hspace{3cm}-\left .4h {\rm ln} \Gamma(h) + 2h {\rm ln} 2\pi +2h {\rm ln} h \right )\nonumber\\
		t^{-}_0&=\frac{\alpha}{2\pi B^2}\left (\frac{2}{3} +
		4 h^2 \psi(1+h)-2h-4h^2-4h {\rm ln} \Gamma(h) +2h {\rm ln} 2\pi -2h{\rm ln} h\right )
	\end{align}
	\noindent where  $h=\frac{B_c}{2B}$, and the quantum corrections are  proportional to the fine structure constant,  $\alpha$. $\psi$ denotes the polygamma o digamma function (first derivative of ${\rm ln}\, \Gamma$) and $\zeta^{\prime}$ is the first derivative of the Hurwitz zeta function with respect to the first argument. 
	
	The expressions (\ref{wf_velexact2})-(\ref{wf_velexact3}) considering weak field limit ($eB<<m^2$) reduce to   (see Appendix \ref{appen1})
	\begin{eqnarray}\label{wf_vel22}
		v_V^{WF,(1)} =v^{WF,(2)}_V &\approx& \left( 1- \dfrac{ \alpha}{90\pi}
		\dfrac{B^2}{B_c^2}
		\right),\\ \label{wf_vel2}
		v_V^{WF,(3)}  &\approx& \left( 1- \dfrac{ \alpha}{30\pi}
		\dfrac{B^2}{B_c^2}
		\right).
	\end{eqnarray}
	
	For the strong magnetic field limit ($m^2<<eB$),
	\begin{align}\label{sf_vel2}
		v_V^{SF,(2)}(B)
		&\approx \ \left (1-\frac{\alpha}{3\pi}\frac{B}{B_{c}}\right ),\\ 
		v_V^{SF,\,(3)}(B) &\approx
		\left (1-\frac{\alpha}{3\pi}\right). \label{sf_vel22}
	\end{align}
	Notice that, to leading order, $v_V^{SF,\,(3)}(B)$ is independent of $B$, reflecting the fact that the dynamics is dominated by the lowest Landau level. Eqs. (\ref{wf_vel22}-\ref{sf_vel22}) are consistent with the phase velocities computed in \cite{Elizabeth} and refraction indexes obtained earlier by Adler \cite{Adler71} and Dittrich \cite{Dittrich98}  from weak field limit of non-linear Euler-Heisenberg Lagrangian.  Moreover, these results are also obtained from velocities obtained for arbitrary magnetic field obtained in the previous paper \cite{Aurora2023}.

	\begin{figure*}[t]
		\centering
		\includegraphics[width=0.6\linewidth]{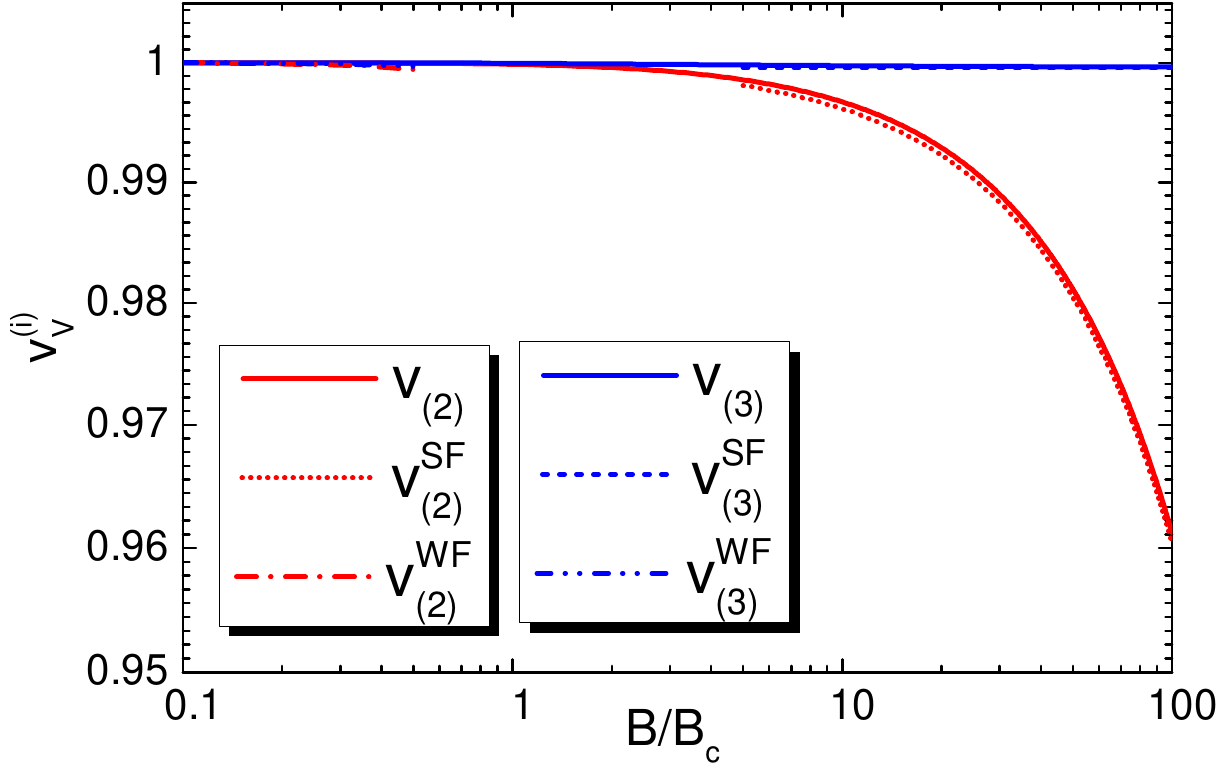}
		\caption{Phase velocity $v_V^{(i)}$ as a function of the magnetic field for the propagation modes $i=2$ (red) and $i=3$ (blue), obtained from Eq.~(\ref{vacvelo}) for perpendicular propagation ($\vec{k} \perp \vec{B}$) in a magnetized vacuum. The figure shows the weak-field (WF) and strong-field (SF) limits, illustrating the smooth transition between the two regimes described by Eqs.~(\ref{wf_vel2}) and~(\ref{sf_vel}).} 
		\label{velocitymodes}
	\end{figure*}
	
	Figure~\ref{velocitymodes} illustrates the photon phase velocity in vacuum, $v_V^{(i)}$, as a function of the magnetic field strength in the infrared limit. The left panel shows the two perpendicularpolarization modes, $i=2,3$, for perpendicular propagation ($\vec{k} \perp \vec{B}$), together with their weak-field (WF) and strong-field (SF) limits. Both modes exhibit a smooth transition between the two regimes. Mode $i=2$ (red curves) shows a stronger reduction in phase velocity with increasing magnetic field, reflecting a stronger vacuum-polarization effect, whereas mode $i=3$ (blue curves) remains closer to the speed of light.
	
	\subsection{Numerical solution of dispersion equation and phase velocity magnetized vacuum}
	
	Let us now solve the dispersion equation  under the condition on small deviation on light cone but considering the eigenvalues $\kappa_V^{(2,3)}$ as a power series in $k^2$.
	
	\begin{equation}
		\kappa^{(2,3)}_V(\omega,\mathbf{k},B)
		=
		\sum_{l=0}^{\infty} \chi_{il}(\omega,B)\,(k^2)^l ,
	\end{equation}
	where the coefficients $\chi_{il}(\omega,B)$ encode the dependence on the frequency and the external magnetic field and has the form 
	\begin{eqnarray} 
		\chi_{il}(\omega,B)=\dfrac{2 \alpha}{\pi}\int_{0}^{\infty}d\tau\int_{-1}^{1}d\eta \,  e^{-\dfrac{\tau}{b}} [ e^{\zeta_0} \left (\dfrac{\xi}{b m_{e}^2}\right )^{l} \nonumber \\  \dfrac{(-1)^{l}}{l!} \left( \rho_{0i}-l\left (\dfrac{\xi}{b m_{e}^2}\right )^{-1} \theta_{i} \right)+\delta_{1l}\dfrac{\bar{\eta}^2}{2\tau}   
		],
	\end{eqnarray}
	$ \rho_{0i} $ and $ \theta_{i} $ are the new scalar functions for every mode of propagation bound to the previous ones by  $ \rho_{i}=\rho_{0i} + k^2\theta_{i} $
	\begin{equation}
		\zeta=\zeta_0-\dfrac{k^2\xi}{b m_{e}^2},
		\quad
		\zeta_{0}=\dfrac{(\omega^2-k^2_{3})(\bar{\eta}^2 \tau-\xi)}{b m_{e}^2},
	\end{equation}
	\begin{equation}
		\rho_{01}=0,
		\quad
		\rho_{02}=\dfrac{\omega^2-k^2_3}{2}(\bar{\eta}^2 \coth{\tau}-\varrho ),
	\end{equation}
	\begin{equation}
		\rho_{03}=\dfrac{\omega^2-k^2_{3}}{2}(\varrho-\dfrac{\xi}{\sinh^2{\tau}} ),
	\end{equation}
	\begin{equation}
		\theta_{1}=\theta_{2}=-\dfrac{\varrho}{2},
		\quad \quad
		\theta_{3}=-\dfrac{\xi}{2 \sinh^2{\tau}}.
	\end{equation}
	
	To solve the dispersion equation, we truncate the series at $l=3$, which leads to a cubic equation in $k^2$. Rather than solving this cubic equation directly, which can be algebraically involved, we employ a method based on hypergeometric functions \cite{Sturmfels20}, which provides a simpler and more accurate solution. A detailed description of this procedure is presented in Appendix~\ref{VacuumPowerExpansion}.
	As we are considering perpendicular propagation  ($ k_{3}=0 $) we can compute the phase velocity of the photons $v^{(2,3)}_{V}(\omega,B)=\omega /k^{\perp, 2,3)}_V (\omega,B)$ as
	\begin{equation}
		\small
		v^{(2,3)}_{V}=\left (1-\dfrac{1}{\omega^2}\sum_{j_2,j_3=0}^{\infty}\dfrac{(-1)^{j_1}j_1!}{(j_0+1)!j_2!j_3!}\dfrac{\chi_{i0}^{j_0+1} \chi_{i2}^{j_2} \chi_{i3}^{j_3}  }{(\chi_{i1}-1)^{j_1+1}}   \right )^{-1/2},\label{vacvelo}
	\end{equation}
	where $   j_0=j_2+2j_3 $ and $  j_1=2j_2+3j_3 $.
	
	In view of Eq. (\ref{vacuumDE}), 
	the photon phase velocity for  parallel propagation %\parallel \mathbf{B}$, 
	is equals $c$.
	
\begin{figure*}[t]
		\centering
		\includegraphics[width=0.4\linewidth]{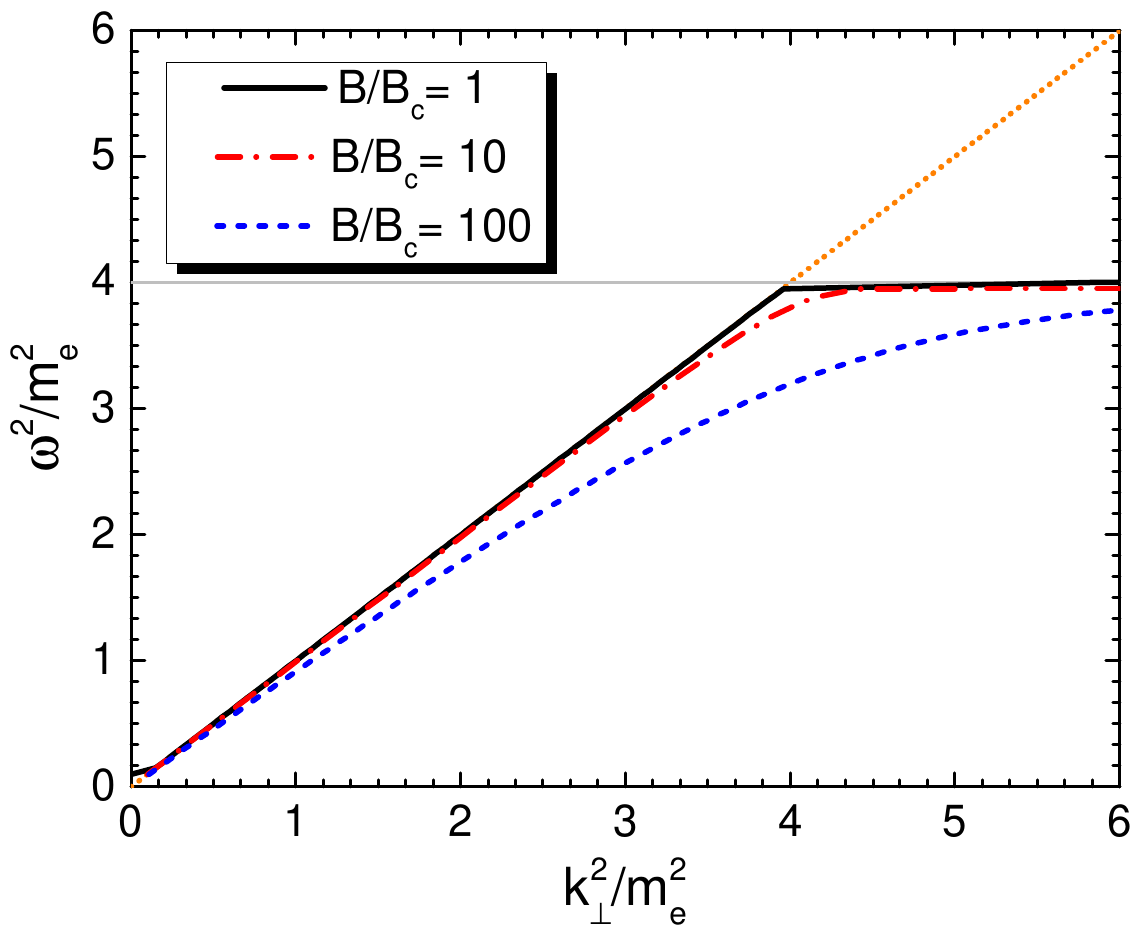}
		\includegraphics[width=0.49\linewidth]{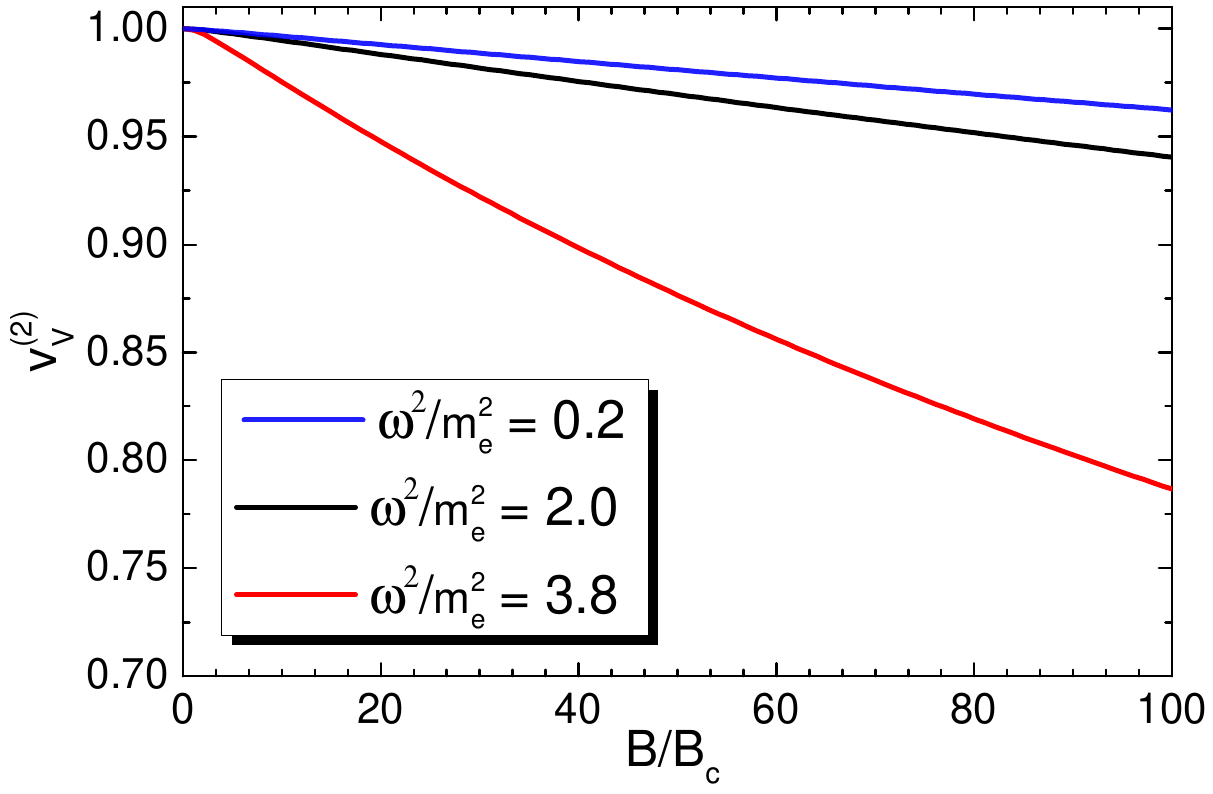}
		\caption{\textbf{Left panel:} Light cone for the second propagation mode ($i=2$)   in vacuum showing the squared photon energy, $\omega^2/m_e^2$, as a function of the squared perpendicular momentum, $k_\perp^2/m_e^2$, for different values of the normalized magnetic field strength, $B/B_c$. The solid black line corresponds to $B/B_c = 1$, the red dashed-dotted line to $B/B_c = 10$, and the blue dashed line to $B/B_c = 100$. The orange dashed line represents the classical light cone, given by $\omega^2 = k_\perp^2$. 
			\textbf{Right panel:} Photon phase velocity, $v_V^{(i)} = \omega/k_\perp$, for the mode $i=2$ as a function of the normalized magnetic field, $B/B_c$, at three different photon energies: $\omega^2/m_e^2 = 0.2$ (blue line), $\omega^2/m_e^2 = 2.0$ (black line), and $\omega^2/m_e^2 = 3.8$ (red line).}
		\label{lightconephasevelocityvac1}
	\end{figure*}

	Using Eq.~(\ref{vacvelo}), we plot the dispersion relation of the second polarization mode ($i=2$) for different values of the external magnetic field, as shown in Figure~\ref{lightconephasevelocityvac1}.
	The left panel shows $\omega^2/{m_e}^2$ as a function of $k_{\bot}^2/{m_e}^2$ for $B/B_c=1,10,100$. For $B/B_c=1$, the dispersion relation closely follows the classical light cone, $\omega^2=k_{\perp}^2$, whereas increasing $B$ produces increasingly significant deviations due to vacuum polarization effects. These corrections increase the photon energy at fixed momentum, producing a
	clear deviation from the classical light cone relation.

	The right panel of Figure~\ref{lightconephasevelocityvac1} shows the photon phase velocity, $v_p=\omega/k_\perp$, as a function of $B/B_c$ for $\omega^2/m_e^2=0.2,2.0,$ and $3.8$. At lower photon energies, $v_p$ remains close to unity, with only a slight decrease as $B$ increases, whereas higher-energy photons exhibit a more pronounced reduction in phase velocity. This behavior reflects the increasing refractive effects of vacuum polarization in strong magnetic fields.

	\section{Eigenvalues of photon self-energy $\kappa_M^{(i)}$  and phase velocity in a magnetized  medium } \label{sec3}
	We now extend our analysis to photon propagation in a magnetized medium composed of a dense electron-positron plasma at finite temperature and chemical potential.  
	To quantify these effects, we compute the photon self-energy eigenvalues, $\kappa_M^{(2,3)}(k,\omega,\mu,T,B)$.
	The full dispersion relation would contain both vacuum and medium contributions, as in Eq.~(\ref{disprel}), but in  this section, we isolate the medium contribution and solve the dispersion equation using only $\kappa_M^{(i)}$,
	\begin{equation}
		\textbf{k}^2-\omega^2=\kappa^{(2,3)}_M(\textbf{k},\omega,\mu,T,B). \label{dispersionequationM} 
	\end{equation}
	The explicit form for the eigenvalues   in a medium for finite temperature and density $\mu \ne 0$ can be written as  
	\begin{align}\label{escalares}
		&p(k\mid
		A;\mu,\beta^{-})=\left(1-\frac{k_{\perp}^2}{\omega^2}\right)t^+,\\
		&t^{\pm}(k\mid
		A;\mu,\beta^{-})=-\frac{e^3B}{2\pi^2\beta}\sum_{p_4}\hspace{-0.1cm}\sum_{n,n^{\prime}}
		\int\limits_{-\infty}^{\infty}\hspace{-0.2cm}\frac{dp_3C_{11,\,22}}{(p_4^{\prime\,2}+\varepsilon_p^2)((p_4^{\prime}+k_4)^2+
			\varepsilon_p^2)}\\
		&s(k\mid
		A;\mu,\beta^{-})=-\frac{e^3B}{2\pi^2\beta}\sum_{p_4}\sum_{n,n^{\prime}}
		\int\limits_{-\infty}^{\infty}\frac{dp_3C_{33}
		}{(p_4^{\prime\,2}+\varepsilon_p^2)((p_4^{\prime}+k_4)^2+\varepsilon_p^2)},\\
		&r(k\mid
		A;\mu,\beta^{-})=-\frac{e^3B}{2\pi^2\beta}\sum_{p_4}\sum_{n,n^{\prime}}
		\int\limits_{-\infty}^{\infty}\frac{dp_3C_{12,\,21}
		}{(p_4^{\prime\,2}+\varepsilon_p^2)((p_4^{\prime}+k_4)^2+\varepsilon_p^2)},
	\end{align}
	with 
	\begin{align}    
		&C_{11,\,22}=p_4^{\prime}[p_4(p_4^{\prime}+k_4)
		+p_3(p_3^{\prime}+k_3)+m^2]F^{(2)}_{n,n^{\prime}}
		\pm 2(k_1^2-k_2^2)N^{(1)}_{n,n^{\prime}},\nonumber\\
		&C_{12,\,21}=\pm i[p_4(p_4^{\prime}+k_4)
		+p_3(p_3^{\prime}+k_3)+m^2]F^{(3)}_{n,n^{\prime}},\nonumber\\
		&C_{33}=\mp p_4^{\prime}[p_4(p_4^{\prime}+k_4)
		\pm p_3(p_3^{\prime}+k_3)+m^2]F^{(1)}_{n,n^{\prime}}
		+2eB(nn^\prime)^{1/2}G^{(1)}_{nn^{\prime}},
	\end{align}
	and
	%....................vartheta...........
	%....................F1...............................................
	\begin{equation}
		\label{FyN}
		F^{(1)}_{n,n^{\prime}}(\vartheta)=
		\begin{cases}
			2\left(\delta_{n,n^{\prime}}+\delta_{n,0}\right),
			& \vartheta\to 0,
			\\[2mm]
			\displaystyle
			\left[
			\left(L_{n^{\prime}-1}^{\,n-n^{\prime}}(\vartheta)\right)^2
			+\frac{n^{\prime}}{n}
			\left(L_{n^{\prime}}^{\,n-n^{\prime}}(\vartheta)\right)^2
			\right]
			\frac{(n^{\prime}-1)!}{(n-1)!}\,
			\vartheta^{\,n-n^{\prime}}e^{-\vartheta},
			& \vartheta\neq 0.
		\end{cases}
	\end{equation}
	
	%....................F2,3.............................................
	\begin{equation}
		\label{FyN2}
		F^{(2,3)}_{n,n^{\prime}}(\vartheta)=
		\begin{cases}
			\delta_{n,n^{\prime}-1}\pm\delta_{n-1,n^{\prime}},
			& \vartheta\to 0,
			\\[2mm]
			\displaystyle
			\left[
			\frac{\vartheta}{n}
			\left(L_{n^{\prime}-1}^{\,n-n^{\prime}+1}(\vartheta)\right)^2
			\pm
			\frac{n^{\prime}}{\vartheta}
			\left(L_{n^{\prime}}^{\,n-n^{\prime}-1}(\vartheta)\right)^2
			\right]
			\frac{(n^{\prime}-1)!}{(n-1)!}\,
			\vartheta^{\,n-n^{\prime}}e^{-\vartheta},
			& \vartheta\neq 0.
		\end{cases}
	\end{equation}
	
	%....................G1...............................................
	\begin{equation}
		\label{FyN3}
		G^{(1)}_{n,n^{\prime}}(\vartheta)=
		\begin{cases}
			2\left(\delta_{n,n^{\prime}}-\delta_{n,0}\right),
			& \vartheta\to 0,
			\\[2mm]
			\displaystyle
			-2\left(\frac{n^{\prime}}{n}\right)^{1/2}
			\frac{(n^{\prime}-1)!}{(n-1)!}\,
			\vartheta^{\,n-n^{\prime}}
			L_{n^{\prime}-1}^{\,n-n^{\prime}}(\vartheta)
			L_{n^{\prime}}^{\,n-n^{\prime}}(\vartheta)
			e^{-\vartheta},
			& \vartheta\neq 0.
		\end{cases}
	\end{equation}
	
	%....................N1...............................................
	\begin{equation}
		\label{FyN4}
		N^{(1)}_{n,n^{\prime}}(\vartheta)=
		\begin{cases}
			0,
			& \vartheta\to 0,
			\\[2mm]
			\displaystyle
			-\frac{n^{\prime}!}{(n-1)!}\,
			\vartheta^{\,n-n^{\prime}-1}
			L_{n^{\prime}-1}^{\,n-n^{\prime}+1}(\vartheta)
			L_{n^{\prime}}^{\,n-n^{\prime}-1}(\vartheta)
			e^{-\vartheta},
			& \vartheta\neq 0.
		\end{cases}
	\end{equation}

	where $\vartheta=k_{\perp}^2/2eB$ and $L^{m^{\prime}}_{m}(x)$ are the generalized  Laguerre polynomials,
	After performing the sum over $p_4$, we obtain integral expressions for the eigenvalues, with the remaining integrations over $p_3$. The explicit general expressions are given in Appendix~\ref{appEigenfinite}, Eqs.~(\ref{scalar_q})-(\ref{scalar_s}). In the next subsection, we solve the dispersion equation for the magnetized degenerate case.

	\subsection{Photon propagating in a degenerate magnetized quantum medium}\label{sec4}
	
	To solve the dispersion equation in a media we have to compute the eigenvalues $\kappa^{(i)}_M$, it means to calculate the integrals of Eqs. 
	(\ref{scalar_q}-\ref{scalar_s}) over $p_3$ and later summing over the Landau levels. 
	As our study is focused on astrophysics scenarios, a degenerate gas or low-temperature limit is a suitable approximation to solve the integrals. For compact stars the plasma turns out to be strongly degenerate, for example, for neutron stars where temperatures are
	of the order of $T/m_e \sim 10^{-4}-10^{-3}$, the densities are $\mu/m_e \sim 1-10^{2}$, fulfilling that $T<< \mu$ \cite{Lidice13,Petri16,Zhu19}.
	
	For the degenerate gas limit ($T \ll \mu$) the integral over $p_3$ can be solved analytically. The distribution function for electrons  becomes   a Heaviside function $ n_{e} \rightarrow \theta(\mu-\varepsilon_{p_3,n}) $ and for positrons goes to zero $n_{p} \rightarrow 0$. 
	The integral over $p_3$  limits  the interval
	$ -p_{F}\leq p_3\leq p_{F} $, where $ p_{F} $ is the     Fermi momentum given by the condition $ \mu-\varepsilon_{p_3,n}=0 $, as $p_{F}=\sqrt{\mu^2-m_e^2-2neB}$ and the sum  over  Landau numbers, is now limited up to
	\begin{equation}\label{nivelesmax}
		n_{max}=I\left[\dfrac{\mu^2-m_e^2}{2eB}\right],
	\end{equation}
	given by the condition $p_{F}>0$ and $I$ is the integer part.     
	For degenerate limit the scalar functions  $p$,$r$,$t^{\pm}$,$s$ becomes in 
	%********
	
	\begin{align}
		& r(k_\perp,k_3,\omega,\mu,B)
		=
		-iC_B
		\left(1-\frac{k_{\perp}^2}{\omega^2}\right)^{1/2} 
		\sum_{n=0}^{n_{\max}}\sum_{n'=0}^{n_{\max}}
		F_{n,n'}^{(3)}
		J^{+}_{n,n'} I^{(1)},
		\label{scalar_r_sol}
		\\
		& t^{\pm}(k_\perp,k_3,\omega,\mu,B)
		=
		-C_B
		\sum_{n=0}^{n_{\max}}\sum_{n'=0}^{n_{\max}}
		\left[
		4 F_{n,n'}^{(2)} I^{(3)}
		+
		F_{n,n'}^{(2)} J^{+}_{n,n'} J^{-}_{n,n'} I^{(2)}
		\right.
		\nonumber\\
		&\qquad\qquad\left.
		\mp 4 k_\perp^2
		N_{n,n'}^{(1)} 
		J^{-}_{n,n'} I^{(2)}
		\right],
		\label{scalar_t_sol}
		\\
		& s(k_\perp,k_3,\omega,\mu,B)
		=
		C_B 
		\sum_{n=0}^{n_{\max}}\sum_{n'=0}^{n_{\max}}
		\Bigg[
		\frac{eB(n'-n)}{\omega^2}
		F_{n,n'}^{(1)} I^{(3)}
		\nonumber\\
		&
		+
		\left(
		k_\parallel^2
		\left[
		-\frac{e^2B^2(n'-n)^2}{\omega^2}
		+
		2m_e^2+eB(n+n')
		\right]
		F_{n,n'}^{(1)}
		+
		2eB\sqrt{nn'}\,G_{n,n'}^{(1)}
		\right)
		\frac{J^{-}_{n,n'}}{4\omega^2}
		I^{(2)}
		\Bigg],
		\label{scalar_s_sol}
	\end{align}
	
	with $C_B=\dfrac{e^3 B}{2\pi^2}$, $k^2_{\parallel}= k^2_3-\omega^2$ and 
	\begin{eqnarray}
		&J^{+,-}_{n,n^{\prime}}=2eB(n^{\prime}\pm  n)+k^2_{\parallel},
	\end{eqnarray}	
	and the  independent integrals $I^{(1)}, I^{(2)}$ and $I^{(3)}$ have the explicit form  
	\begin{eqnarray}
		I^{(1)}&=&\int
		\frac{dp_3}{Q_{n,n^{\prime}}
			(p_3) }\theta(\mu-\varepsilon_{p_3,n}) , \label{Int1}\\  
		I^{(2)}&=&\int
		\frac{dp_3}{\varepsilon_{p_3,n}}
		\frac{1}{Q_{n,n^{\prime}}(p_3)}\theta(\mu-\varepsilon_{p_3,n}),
		\\ 	
		I^{(3)}&=&\int
		\frac{dp_3}{\varepsilon_{p_3,n}}
		\theta(\mu-\varepsilon_{p_3,n}),\label{Int3}
	\end{eqnarray}
	with
	$Q_{n,n^{\prime}}(p_3)=[2p_{3}k_{3}+k_{\parallel}^2+2eB(n^\prime-n)]^{2}-4\omega^{2}\varepsilon_{p,n}^{2}.$
	
	Let us solve first the eigenvalues using long wavelength approximation and infrared regime. 
	\subsection{Long wavelength approximation}
	
	Firstly, let us consider some approximations over the scalars. We first consider the approximations relevant to the scalar functions. we consider low-energy photons,
	$\omega\ll 2m_e$, with small deviations from the light cone,
	$\omega\sim k_\perp$. In this regime, one finds for perpendicular propagation, $k_3=0$, 
	$q,r\ll t^-$. The eigenvalues of Eq.~(\ref{eqdispperp}) then reduce to
	the simpler form \cite{Rojas08}
	\begin{align}
		\kappa^{(1,3)}_M(k_{\bot},\omega,B,\mu,0)
		&=&
		\frac{t^-}{2}\mp\frac{t^-}{2},
		\nonumber\
		\kappa^{(1)}_M
		&=&
		0,
		\qquad
		\kappa^{(3,\bot)}_M=t^-,
		\nonumber\
		\kappa^{(2,\perp)}_M(k_{\bot},\omega,B,\mu,0)
		&=&
		s.
	\end{align}
	and for parallel propagation
	\begin{equation}
		p=t^-.
	\end{equation}
	\noindent $r$  is imaginary we redefine $r^{\prime}=Im[r]$ and  
	\begin{align}
		\kappa^{(3,\parallel)}_M= t\pm r'.
	\end{align}
	%\subsection  
	For parallel propagation, $t$ combines with $r$, giving rise to circularly polarized modes. 
	As $k_\perp^2$ is zero, the functions
	$F_{nn'}^{(2,3)}(0)=\delta_{n,n'-1}\pm\delta_{n-1,n'}
	$ and
	$N_{nn'}^{(1)}(0)=0$. The long wavelength approximation means to take $k_3=0$ in \ref{scalar_t_sol}-\ref{scalar_r_sol} and neglect in the integrals the term $2p_3k_3$. These approximation reduce the expressions  to the following form:
	\begin{align}
		t(0,\omega,B,\mu,0)
		&=
		\frac{e^{3}B}{2\pi^2}
		\left(
		\sum_{n=0}^{n_{\max}}
		\alpha_n
		\int_0^{p_F}
		\frac{
			(\omega^2+2eB)(-\omega^2+4eBn)\,dp_3
		}{
			\varepsilon_n
			\left[
			(\omega^2+2eB)^2
			-
			4\omega^2\varepsilon_n^2
			\right]
		}
		\right)
		\nonumber\\
		&=
		\frac{e^{3}B}{2\pi^2}
		\Bigg[
		\sum_{n=0}^{n_{\max}}
		\alpha_n
		\frac{
			(\omega^2-2eB(2n+1))(\omega^2-2eB)
		}{
			4m_n
			\sqrt{
				\left(\frac{2eB-\omega^2}{2m_n}\right)^2-1
			}
		}
		\ln\left|
		\frac{
			\dfrac{p_F}{m_n}
			-
			\sqrt{
				\mu^2
				\left(\frac{2eB-\omega^2}{2m_n}\right)^2-1
			}
		}{
			\dfrac{p_F}{m_n}
			+
			\sqrt{
				\mu^2
				\left(\frac{2eB-\omega^2}{2m_n}\right)^2-1
			}
		}
		\right|
		\nonumber\\
		&\quad+
		\sum_{n=0}^{n_{\max}}
		\alpha_n
		\frac{
			(\omega^2-2eB(2n-1))(2eB-\omega^2)
		}{
			4m_n
			\sqrt{
				\left(\frac{2eB+\omega^2}{2m_n}\right)^2-1
			}
		}
		\ln\left|
		\frac{
			\dfrac{p_F}{m_n}
			-
			\sqrt{
				\mu^2
				\left(\frac{2eB+\omega^2}{2m_n}\right)^2-1
			}
		}{
			\dfrac{p_F}{m_n}
			+
			\sqrt{
				\mu^2
				\left(\frac{2eB+\omega^2}{2m_n}\right)^2-1
			}
		}
		\right|
		\Bigg].
		\label{eq:kappa3}
	\end{align}

	\begin{align}
		&r^{\prime} (\omega,B,\mu,0) =
		\frac{e^{3}B}{2 \pi^2}\left (\underset{n=0}{\overset{n_{max}}{\sum}}\alpha_n\frac{-\omega^2+2eB(2n+1)}{4m_n \omega^2\sqrt{(\frac{2eB-\omega^2}{2\omega m_n})^2-1}}\ln{\mid\frac{p_F/m_n-\sqrt{(\frac{2eB-\omega^2}{2\omega m_n})^2-1}}{p_F/m_n+\sqrt{(\frac{2eB-\omega^2}{2\omega m_n})^2-1}}\mid}\right.\nonumber\\
		& +\left.\underset{n=1}{\overset{n_{max}}{\sum}}\alpha_n\frac{\omega^2+2eB(2n-1)}{4m_n \omega^2\sqrt{(\frac{2eB+\omega^2}{2\omega m_n})^2-1}}\ln{\mid\frac{p_F/m_n-\sqrt{(\frac{2eB+\omega^2}{2\omega m_n})^2-1}}{p_F/m_n+\sqrt{(\frac{2eB+\omega^2}{2\omega m_n})^2-1}}\mid}\right).\label{eq:3Dhallintegradaparan1}
	\end{align}
	where $\alpha_n=2-\delta_{n0}$.
	The integral expressions $r$ and $t$ (\ref{scalar_r_sol}-\ref{scalar_t_sol}) respectively, are in agreement with the results presented in \cite{DOlivo:2002omk} (Eq. 4.31 and 5.4 for $\pi_T\equiv t$ and $\pi_L\equiv s$ when the limit $T=0$ $\mu\ne 0$ is performed).

	For propagation perpendicular  the long-wavelength limit means to take $k_{\perp}\to0$. In this limit, Eq.(\ref{scalar_s_sol}) simplifies thanks to $
	F^{(1)}_{nn'}(0)=
	2\delta_{n,n'}-\delta_{n,0}\delta_{n',0},
	\,\,\,
	G^{(1)}_{nn'}(0)=
	2(\delta_{n,n'}-\delta_{n,0})$. The explicit form of s is
	
	\begin{align}
		s(0,\omega,B,\mu,0)
		=\frac{e^{3}B}{2 \pi^2}\left(
		\sum_{n=0}^{n_{\max}} \alpha_n\int_0^{p_{F}}
		\frac{m_n^2}{\varepsilon_n(\omega^2+4\varepsilon_n^2\omega^2)}\right)\nonumber\\
		= \frac{e^{3}B}{2 \pi^2}
		\left(
		\sum_{n=0}^{n_{\max}} \alpha_n
		\left(
		1 + \frac{\omega^2}{2 m_n^2}
		\right)
		\ln \left|
		\frac{
			p_F - \mu \left(1 + \frac{\omega^2}{2 m_n^2}\right)
		}{
			p_F + \mu \left(1 + \frac{\omega^2}{2 m_n^2}\right)
		}
		\right|
		\right).
		\label{eq:kappa2}
	\end{align}
	\noindent while the scalar function $t$ takes the same explicit form as in the parallell case.\\
	
	\subsection{Infrared regime: perpendicular and parallel propagation}

	Following the prescription adopted in Refs.~\cite{Zeitlin:1995ex,PhysRevD.52.2533}, we first take the long-wavelength limit and subsequently consider the static limit, $\omega\to0$. With this ordering, the eigenvalues $\kappa^{(i)}$ consistently reproduce those obtained from the effective NLED description, as discussed above. 
	Consequently,in the infrared regime $k_{\perp}=\omega=0$,  the eigenvalues  for transverse propagation reduce  to
	\begin{align}
		\kappa^{(2,\perp)}_M(\mu,B)=
		s=
		\dfrac{2\alpha eB}{\pi^2}
		\sum_{n=0}^{n_{\rm max}}
		\alpha_n
		\dfrac{p_F}{\mu}.
		\label{eqdispaprox2}
	\end{align}
	\noindent and 
	\begin{equation}
		\kappa^{(3),\perp}_M(\mu,B)=\sum_{n=0}^{n_{\rm max}}
		\alpha_n
		\dfrac{2\alpha eB}{\pi^2}
		\left[
		\log\left(\frac{p_F+\mu}{m_n}\right)
		\right]. \label{eigeninfra3}
	\end{equation}
	The Eq(  \ref{eqdispaprox2})-(\ref{eigeninfra3}) are obtained in appendix using non linear electrodynamics Eq.(\ref{effecIR})-(\ref{Lagmu})
	
	\noindent For parallel propagation  has the form 
	\begin{equation}
		\kappa^{(1,3),\parallel}_M(\mu,B)=
		\dfrac{2\alpha eB}{\pi^2}\sum_{n=0}^{n_{\rm max}}
		\alpha_n
		\left[
		\log\left(\frac{p_F+\mu}{m_n}\right)
		\pm \omega p_F/(eB)
		\right].
	\end{equation}
	
	Their nonlinear counterparts are obtained from Eq. (\ref{Lagmu}) by taking the mixed derivative ($\partial^2\mathcal{L}^{\mu,0}/\partial B\partial\mu) $\cite{Felipe1990}.
	
	As shown in Appendix \ref{NLEDIR}, Eq. (\ref{NLEDT}) extends the vacuum result to finite density while preserving the symmetric structure of the photon self-energy for perpendicular propagation. By contrast, in the case of parallel propagation, the leading contribution also involves the antisymmetric scalar $r$ of the photon self-energy. Incorporating this contribution requires additional terms from Eq. (\ref{NLEDT}), which are associated with the mixed derivative $\partial^2\mathcal{L}^{\mu,0}/(\partial B \partial\mu)$ appearing in Eq. (\ref{Lagmu}). A detailed analysis of these contributions will be presented in a forthcoming paper. 
	
	\subsection{Medium phase velocity infrared regime} 
	The phase velocity of low energy photons (infrared regime) for perpendicular propagation of the modes $i=3,2$    are 
	\begin{equation}
		v_p^{\bot,(3)}=\omega/k_{\perp} \sim \left (1-\frac{Re[t]}{2\omega^2}\right ),\label{vper1}
	\end{equation}
	and
	\begin{equation}
		v_p^{\bot,(2)}\sim \left (1-\frac{Re[s]}{2\omega^2}\right ),\label{vper2}
	\end{equation}
	\noindent while for parallel propagation photon exhibits a phase velocity 
	\begin{equation}
		v_p^{\parallel, (1,3)}\sim \left (1-\frac{Re[t\pm r]}{2\omega^2}\right )\label{vpar12},
	\end{equation}
	for perpendicular modes $i=1,3$. 
	\textit{Parallel} photon propagation induces Faraday rotation. In that scenario, a plane-polarized wave splits into two counter-rotating circularly polarized modes traveling at different speeds, causing a rotation of the overall polarization vector. The magnitude of this effect depends on the propagation length, magnetic field intensity, and magneto-optical susceptibility of the medium. Conversely, \textit{perpendicular} photon propagation does not produce Faraday rotation \citep{Lidice13,Lidice2015}.

	\section{Numerical results for phase velocity of photons propagating parallel and perpendicular to a degenerate magnetized medium}
	\label{Numerical}
	This section is devoted to the numerical solution of the transcendental dispersion equation, Eq.~(\ref{dispersionequationM}).
	
	We begin by evaluating the integrals over $p_3$, which define the scalar functions for photon propagation parallel and perpendicular to the magnetic field, Eqs.~(\ref{scalar_t_sol})–(\ref{scalar_s_sol}).
	
	For parallel propagation, the eigenvalues are proportional to the scalars $r(k_3,\omega,B,\mu)$ and $t(k_3,\omega,B,\mu)$. In this case, the integral over $p_3$ is simplified by neglecting terms of the form $2 p_3 k_3$. Moreover, since $k_{\bot}=0$, the functions $F^{(2,3)}_{n,n'}(0)$ reduce to Kronecker deltas, which allows the sum over Landau levels to be carried out analytically. As a result, the scalar functions $r(k_3,\omega,B,\mu)$ and $t(k_3,\omega,B,\mu)$ are obtained directly and subsequently used to solve the dispersion equation numerically.
	
	In contrast, for perpendicularpropagation, $k_3 =0$ the integrals do not depend on terms of the form $2 p_3 k_3$, and therefore the integration over $p_3$ can be performed exactly. However, the dependence on $k_{\perp}$ of the scalar functions $t(k_{\perp},\omega,B,\mu)$ and $s(k_{\perp},\omega,B,\mu)$ enters through the functions $F^{(1,2)}_{n,n'}$, $N^{(1)}_{n,n'}$ $G^{(1)}_{n,n'}$ and must therefore be computed numerically. This requires fixing a set of parameters $(\omega,B,\mu)$, performing the sum over Landau levels from $n=0$ to $n{\text{max}}$, and obtaining the scalar functions $t^{\pm}(k_{\perp},\omega,B,\mu)$ and $s(k_{\perp},\omega,B,\mu)$. These are then used to solve the dispersion equation~(\ref{dispersionequationM}).%%(\ref{eqdispperp})-
	
	Finally, we note that the integrals $I^{(1,2)}$, Eqs.~(\ref{Int1}) and (\ref{Int3}), which involve the scalars $t^{\pm}(k_{\parallel},\omega,B,\mu)$, $r(k_{\parallel},\omega,B,\mu)$, $t^{\pm}(k_{\perp},\omega,B,\mu)$, and $s(k_{\perp},\omega,B,\mu)$, are in general complex. Their nature depends on the denominator $Q_{nn'}$, which may be real or complex depending on whether the system lies inside or outside a region of \textit{transparency}, respectively.
	
	The phase velocities are computed from the principal values of these integrals. For completeness, the form of the imaginary part is presented in the Appendix. As will be shown below, Figure~6 displays, for perpendicular propagation, both the phase velocity and the corresponding imaginary part for mode 3.
	
	In \cite{Fukushima:2024ete} the numerical study about the perpendicularpropagation of photon at finite temperature and density have been done.

	\begin{figure*}[h!]
		\centering
		\includegraphics[width=0.45\linewidth]{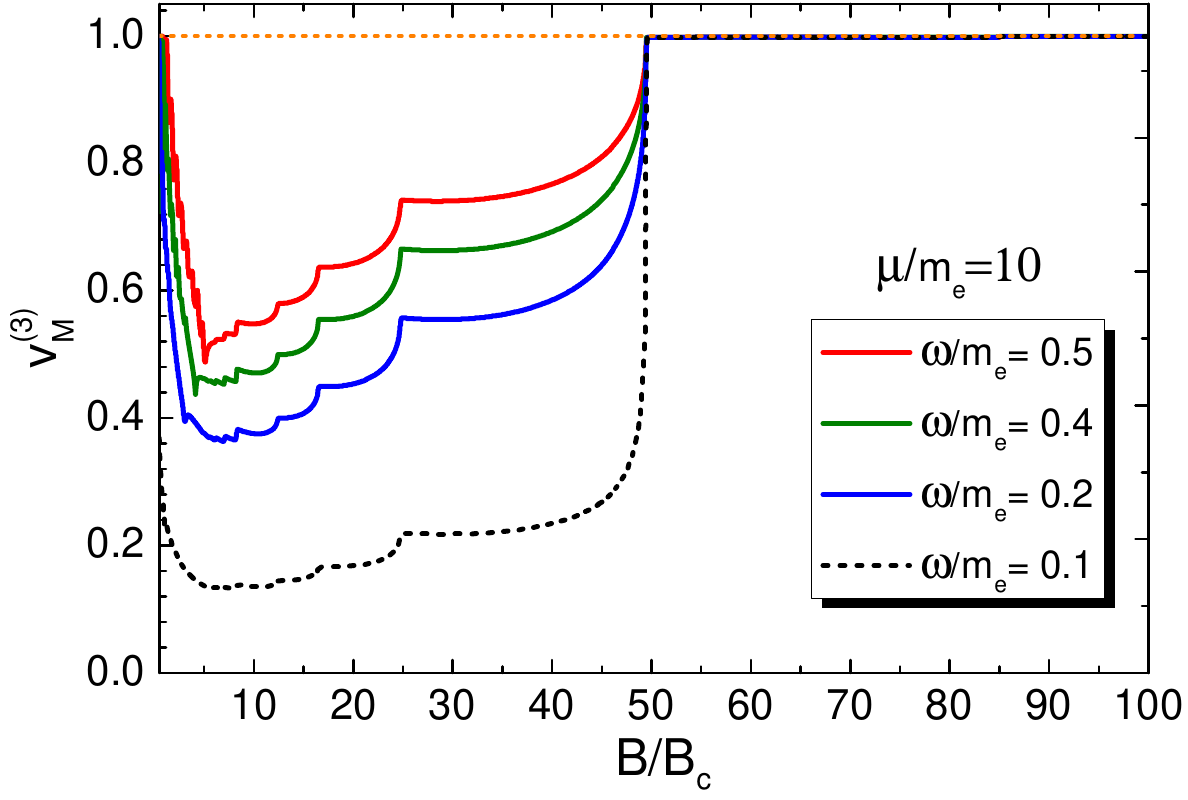}
		\includegraphics[width=0.45\linewidth]{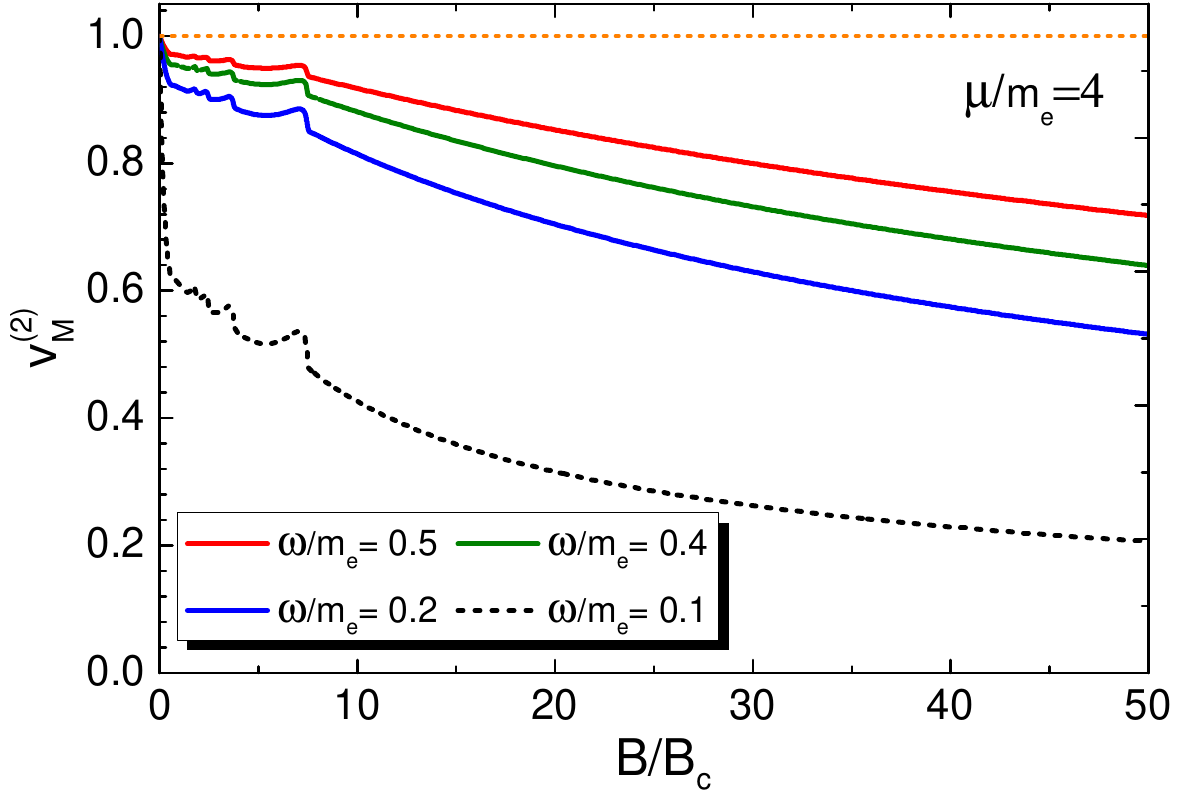}
		\includegraphics[width=0.45\linewidth]{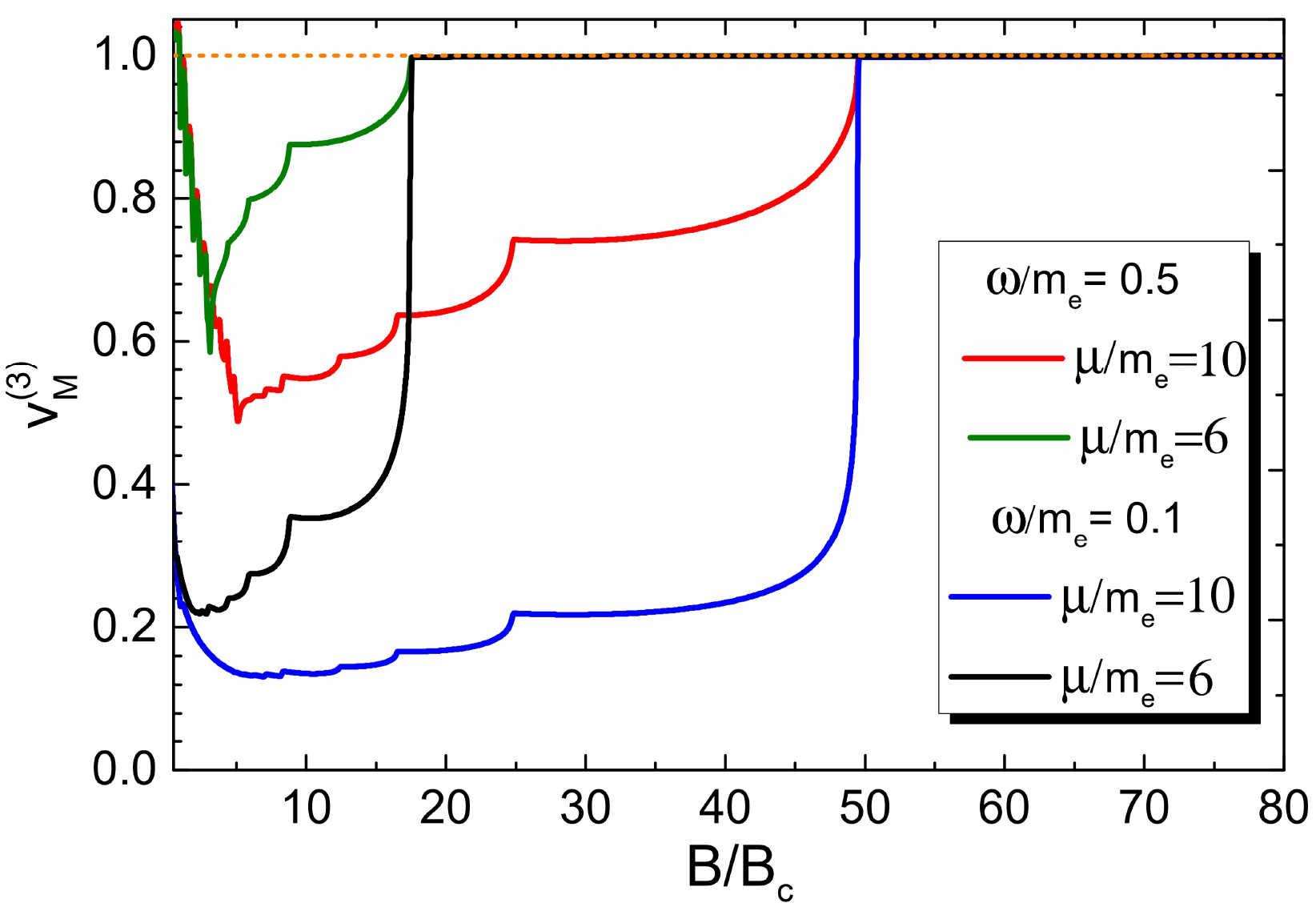}
		\includegraphics[width=0.45\linewidth]{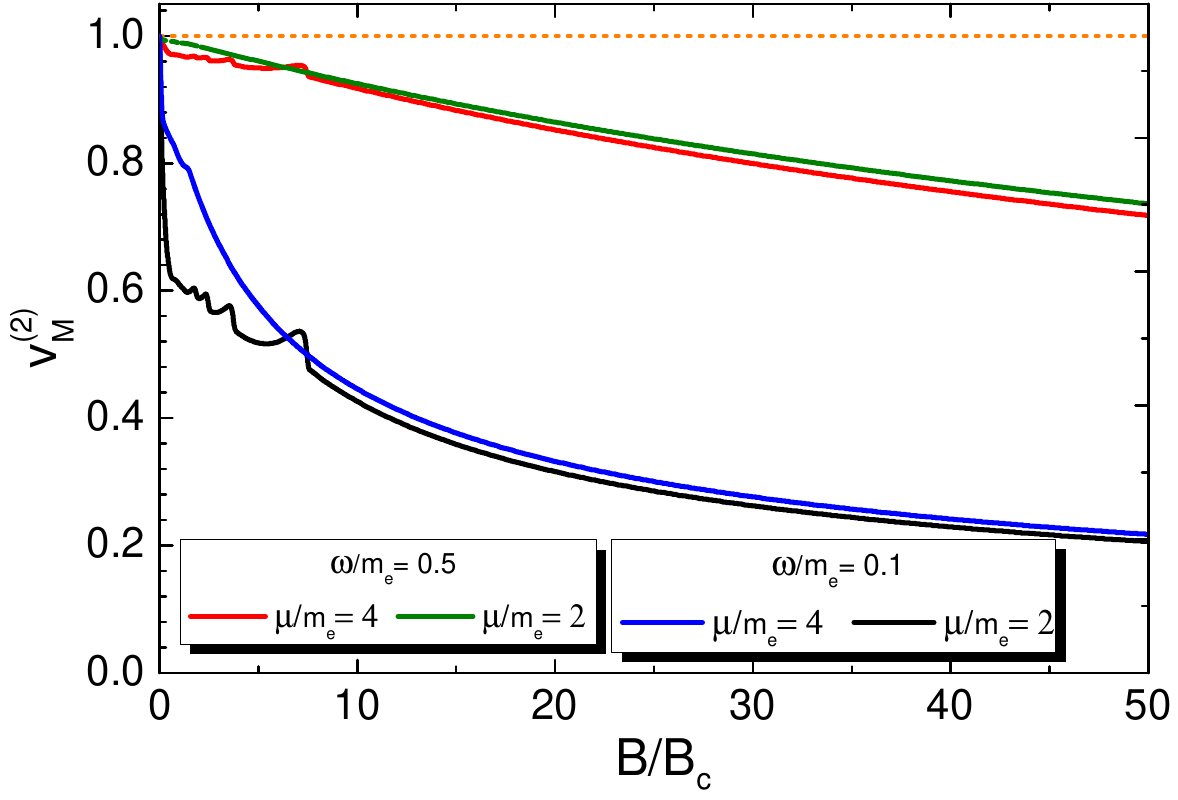}
		\caption{Perpendicular propagation photon phase velocity, $v_M^{(i)}$, as a function of the magnetic field strength $B/B_c$ for polarization modes $i=3$ (left panels) and $i=2$ (right panels). The top panels show curves for photon energies $\omega/m_e = 0.1, 0.2, 0.4, 0.5$ with two values of the chemical potential, $\mu/m_e = 10$ (left) and $\mu/m_e = 4$ (right). The bottom panels correspond to fixed photon energies, $\omega/m_e = 0.5, 0.1$, and various chemical potentials ($\mu/m_e = 6,10$ left; $\mu/m_e = 4,2$ right). For comparison, the orange dashed line represents the light cone photon phase velocity ($\omega = k_\perp$).
		}
		\label{velocidadcampo1}
	\end{figure*}
	To illustrate the behavior of the photon phase velocity eigenvalues in a degenerate magnetized medium, we analyze their dependence on the magnetic field strength, photon energy, and chemical potential. Figure~\ref{velocidadcampo1} shows for perpendicularpropagation the phase velocity as a function of $B/B_c$ (left panels) and chemical potential $\mu/m_e$ (right panels) for fixed photon energies.

	\begin{figure*}[h!]
		\centering
		\includegraphics[width=0.49\linewidth]{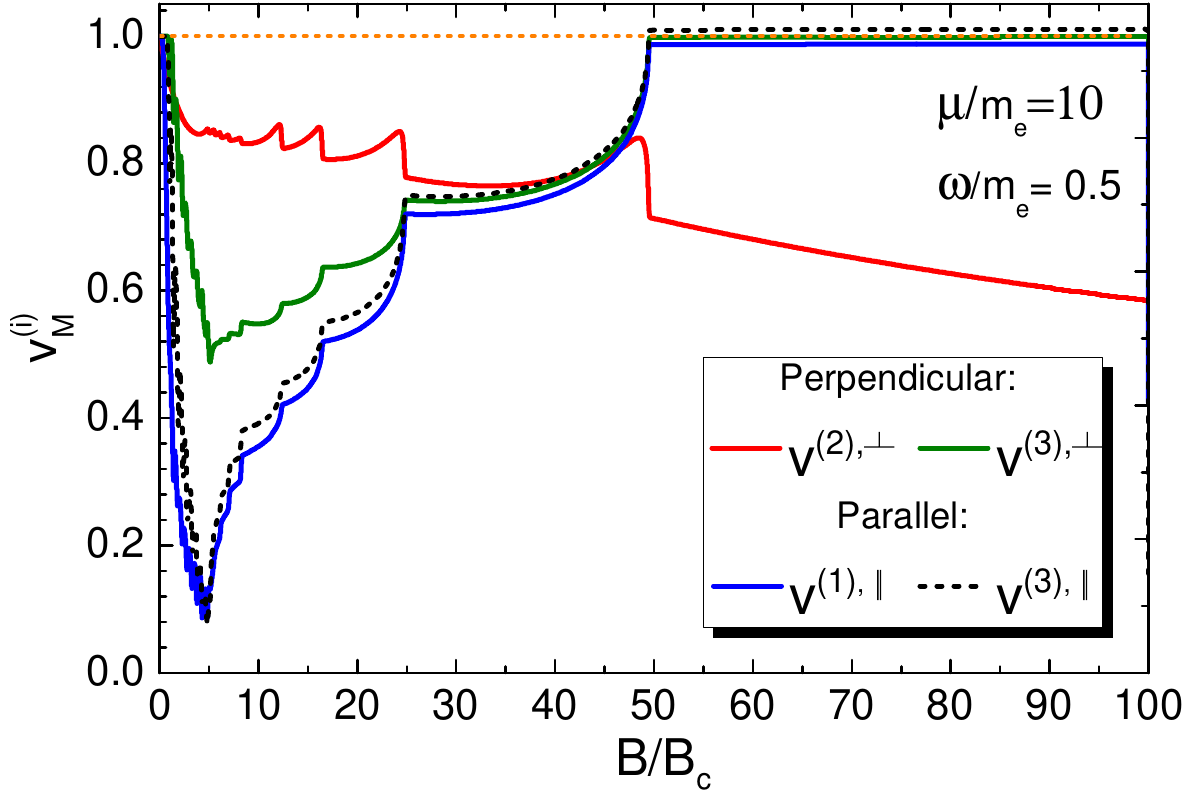}
		\hfill
		\includegraphics[width=0.49\linewidth]{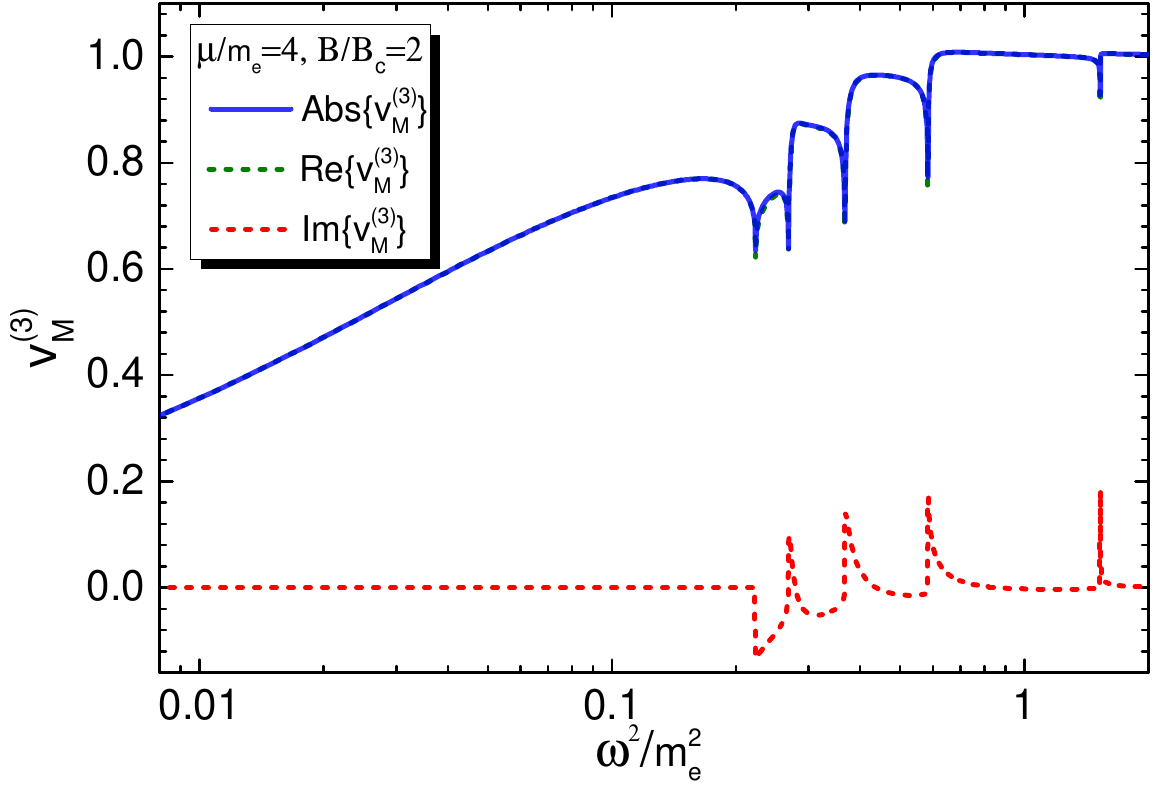}
		\caption{
			\textbf{Left panel:} Numerical photon phase velocity $v_M^{(i)}$ for
			propagation parallel and perpendicular to the magnetic field as a function
			of the magnetic-field strength, $B/B_c$. Polarization modes $i=2,3$ are
			shown for perpendicular propagation ($\vec{k}\perp\vec{B}$), and modes
			$i=1,3$ for parallel propagation ($\vec{k}\parallel\vec{B}$). The photon
			energy and chemical potential are fixed at $\omega/m_e=0.5$ and
			$\mu/m_e=10$, respectively. The orange dashed line denotes propagation at
			the speed of light (light cone).
			\textbf{Right panel:} Photon phase velocity $v_M^{(3)}$ as a function of
			the squared photon energy, $\omega^2/m_e^2$, for the third polarization
			mode ($i=3$) in perpendicular propagation ($\vec{k}\perp\vec{B}$).
			The solid blue line shows the absolute value of the phase velocity,
			$|v_M^{(3)}|$, while the dashed green and red lines depict its real and
			imaginary parts, respectively. The calculation is performed for
			$B/B_c=2$ and $\mu/m_e=4$.
		}
		\label{velocidadcampo2}
	\end{figure*}

	For polarization mode $i=3$, the phase velocity $v_M^{(3)}$ decreases rapidly at low magnetic fields and then saturates as $B/B_c$ increases. This behavior is a direct consequence of Landau-level quantization: in the strong-field regime, the maximum occupied Landau level approaches zero ($n_{\max} \to 0$), leaving only the lowest Landau level populated. As a result, photon-medium interactions are strongly suppressed, and the phase velocity approaches a constant value close to $c=1$.
	
	In contrast, mode $i=2$ exhibits a monotonic decrease of the phase velocity with increasing magnetic field, with an approximately linear dependence on $B/B_c$. This reflects the different coupling of this mode to plasma excitations, for which dispersive effects remain relevant even in the strong-field limit.
	Polarization mode 2 dominates over mode 3, leading to more pronounced variations in the phase velocity.
	
	The oscillatory features observed in both modes arise from transitions between discrete Landau levels and are characteristic of magnetized quantum systems. These oscillations depend sensitively on the photon energy and chemical potential and provide a clear signature of the underlying quantum structure of the medium.

	\begin{figure}[h!]
		\centering
		\includegraphics[width=0.6\linewidth]{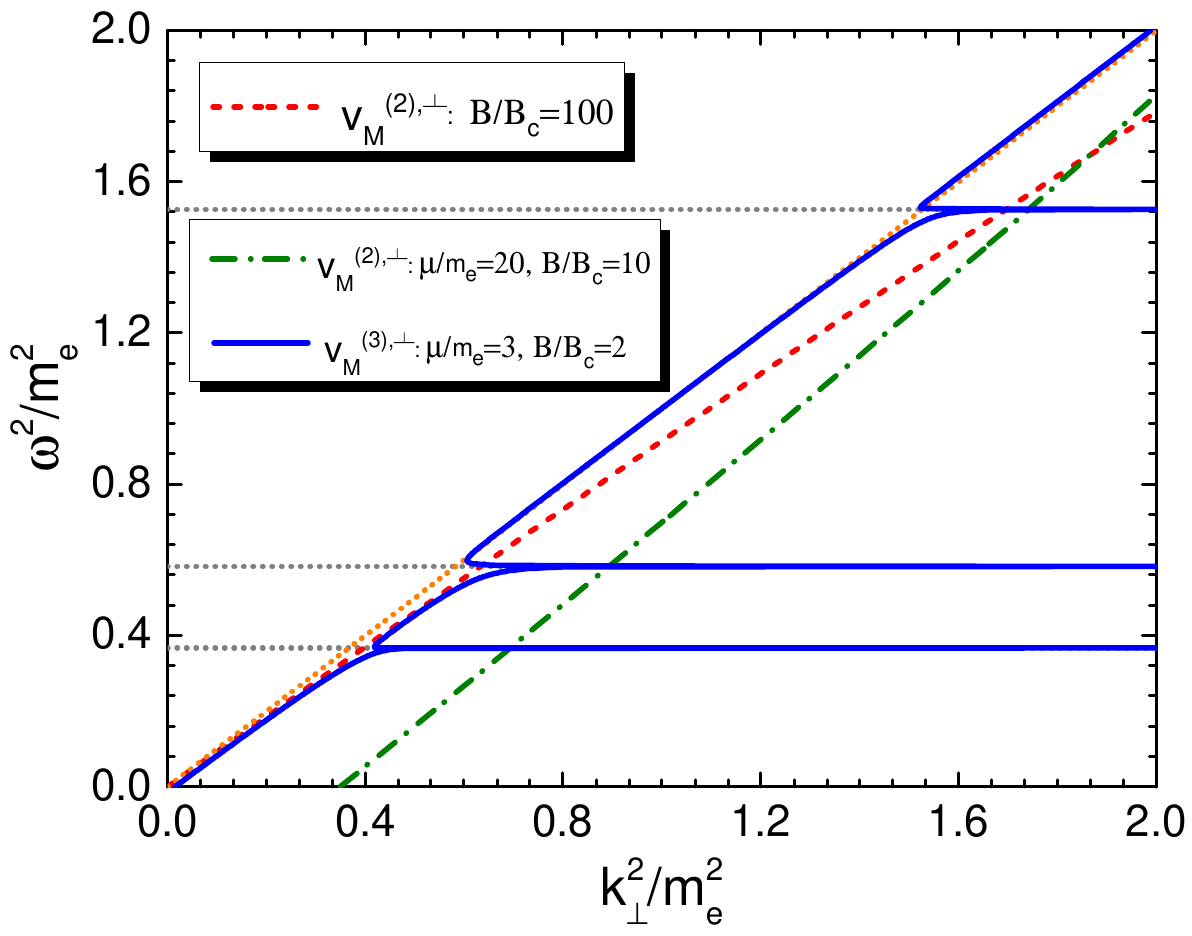}
		\caption{ Dispersion relation (light cone) for photon propagation perpendicular to the magnetic field  in a medium, displaying the squared photon energy $(\omega^2/m_e^2)$ as a function of squared perpendicular momentum ($k_\perp^2/m_e^2$). Curves correspond to polarization modes $i=2$ (red dashed and green dash-dotted lines) for magnetic field strengths $B/B_c=100$ and $10$, respectively, and mode $i=3$ (blue solid line) for $B/B_c=3$. Calculations are performed for fixed chemical potentials $\mu/m_e=20,3$, respectively. The orange dashed line represents the classical vacuum dispersion relation ($\omega^2 = k_\perp^2$), illustrating propagation at the speed of light.
		}
		\label{lightconephasevelocityvac2}
	\end{figure}
	\begin{figure*}[h!]
		\centering
		\includegraphics[width=0.49\linewidth]{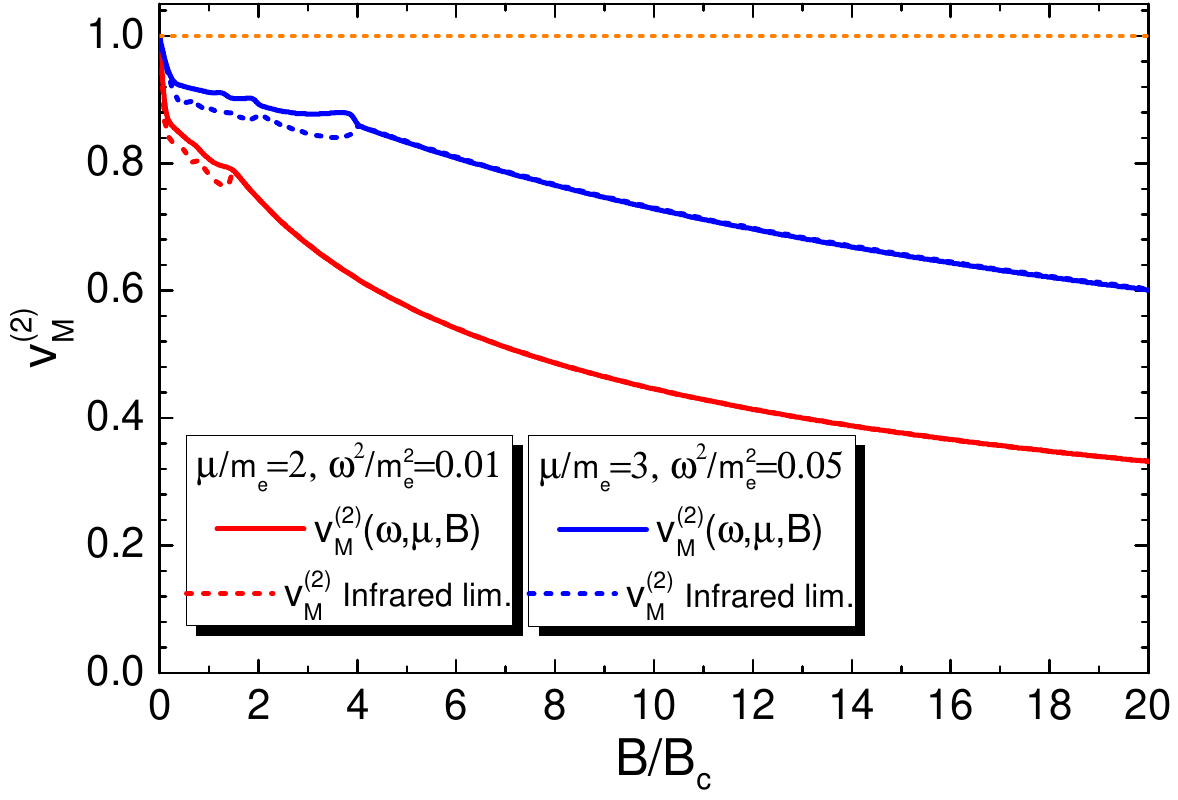}
		\includegraphics[width=0.49\linewidth]{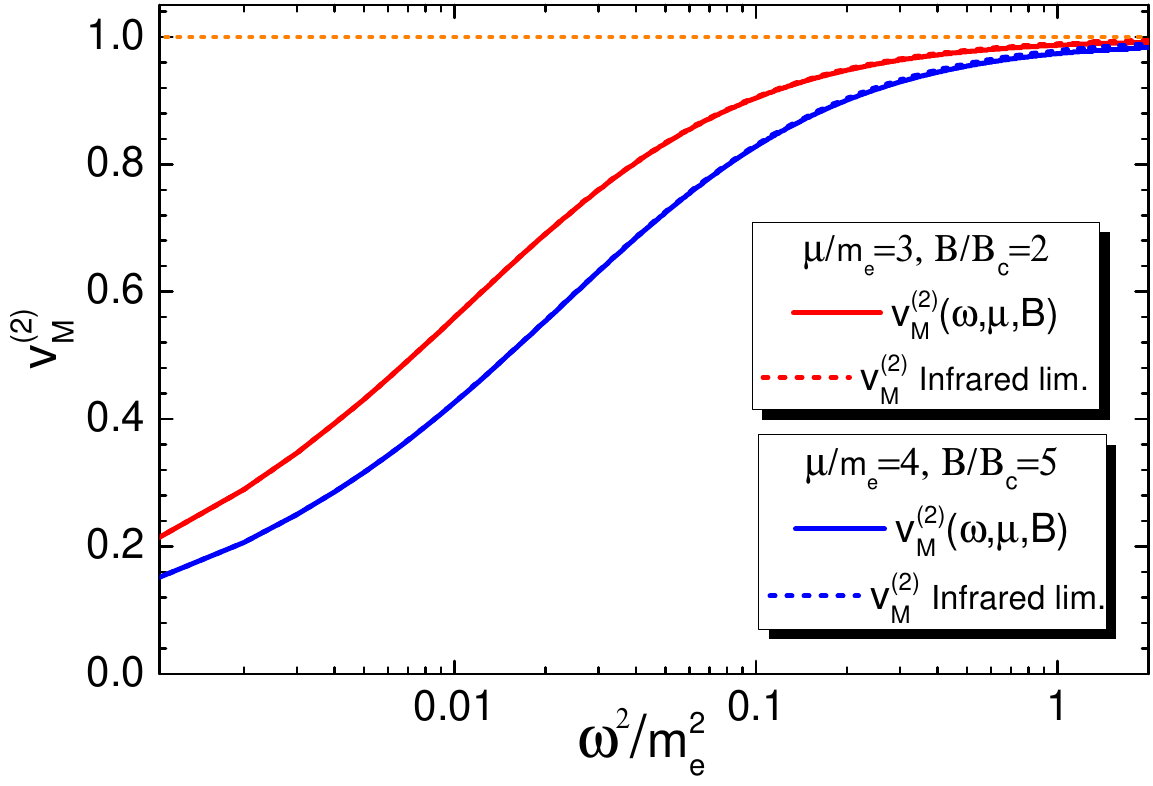}
		\includegraphics[width=0.49\linewidth]{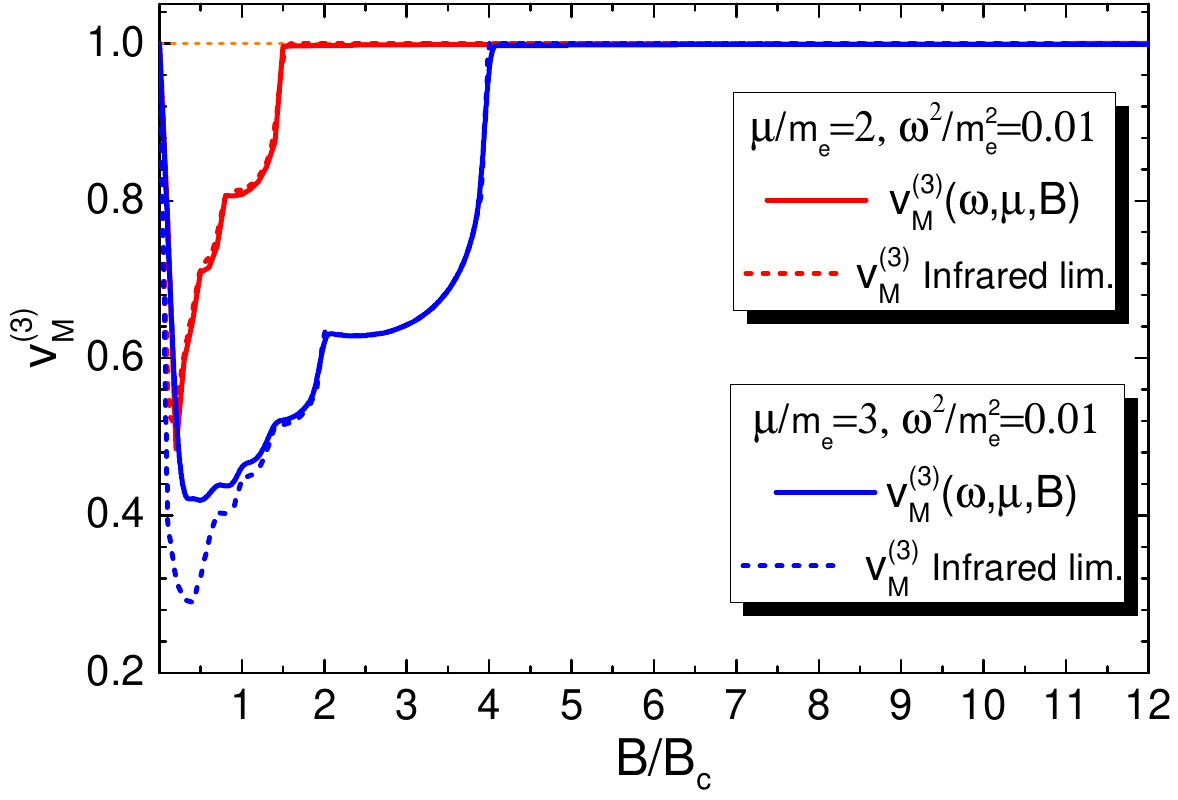}
		\includegraphics[width=0.49\linewidth]{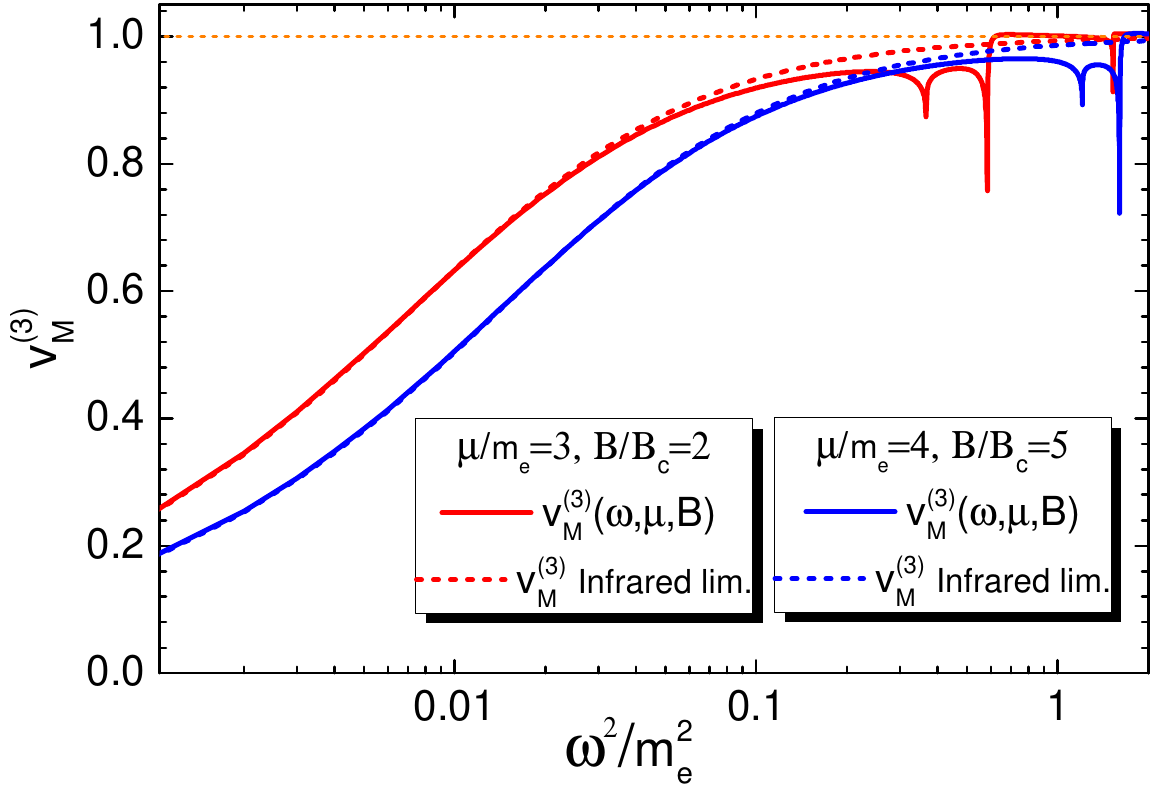}
		\caption{
			Phase velocity for perpendicular propagation  ($\vec{k} \perp \vec{B}$) in a magnetized medium.
			Top panels show $v_M^{(2)} = s$ for polarization mode $i=2$ as a function of the magnetic field (left) and photon energy (right).
			Bottom panels show $v_M^{(3)} = t$.
			Left panels correspond to $\mu/m_e = 2$, $\omega^2/m_e^2 = 0.01$ (red) and $\mu/m_e = 3$, $\omega^2/m_e^2 = 0.05$ (blue),
			while right panels correspond to $\mu/m_e = 3$, $B/B_c = 2$ (red) and $\mu/m_e = 4$, $B/B_c = 5$ (blue).
			Dashed lines indicate the infrared-limit approximation; the orange dashed line corresponds to photon propagation at the speed of light ($v_p = 1$).}
		\label{a6}
	\end{figure*}
	
	Figure \ref{velocidadcampo2} (left panel) compares photon propagation parallel and perpendicular to the magnetic field. For parallel propagation, modes 
	$i=1$ and 
	$i=3$ exhibit stronger deviations from the speed of light at high magnetic field strengths. In contrast, for perpendicular propagation, modes 
	$i=2$ and $i=3$ display milder, yet still significant, polarization-dependent refractive effects. This difference underscores the intrinsic anisotropy of photon propagation in strongly magnetized media.
	The figure shows that the phase velocities of modes for parallel propagation and mode 3 for perpendicular propagation  exhibit similar behavior, whereas mode 2 plays a distinct role in determining the phase velocity.

	Figure~\ref{velocidadcampo2} (right panel) shows the photon phase velocity for polarization mode $i=3$ propagating perpendicular to the magnetic field as a function of the photon energy $\omega^2/m_e^2$. The solid blue line denotes the magnitude of the phase velocity, which increases with energy and approaches the speed of light at high energies. The dashed green and red lines represent the real and imaginary parts, respectively. The imaginary component indicates photon absorption and damping, with peaks corresponding to resonant Landau-level transitions where photon energies match the level separations in the plasma. That is, this imaginary component stems directly from the imaginary part of the one-loop QED polarization operator, which via the optical theorem accounts for dissipation and absorption processes in the background magnetized medium. Specifically, the depicted sharp resonances correspond to the physical energy thresholds for cyclotron absorption in the magnetic field, where the photon energy satisfies the kinematic condition to excite these transitions (e.g., $\omega^2 \approx 2eB(2n+1)$, as can be seen in (\ref{eq:3Dhallintegradaparan1}).

	Figure ~\ref{lightconephasevelocityvac2} shows the dispersion relation for
	photon propagation perpendicular to the magnetic field in a magnetized medium.
	The curves display $\omega^2/m_e^2$ as a function of $k_\perp^2/m_e^2$ for
	fixed background parameters $(B/B_c,\mu/m_e)$ and different polarization modes.
	Thus, this figure represents full dispersion curves for selected values of the
	magnetic field and chemical potential, rather than phase velocities evaluated at
	a fixed photon energy. The orange dashed line denotes the classical light cone
	relation, $\omega^2=k_\perp^2$. 
	Mode $i=3$ exhibits plateau-like features in the dispersion relation, corresponding to resonant Landau-level transitions with enhanced photon absorption and damping (due to singularities and imaginary part). In contrast, mode $i=2$ shows smoother deviations, reflecting continuous dispersive effects. No singularities or imaginary parts are present.
	
	Figure~\ref{a6} compares the analytical infrared (IR) approximation with exact numerical solutions. For mode $i=2$ (top panels), the IR approximation reproduces the phase velocity at low magnetic fields but deviates at higher fields or energies. For mode $i=3$ (bottom panels), oscillations due to Landau-level transitions are evident, and at high fields, the phase velocity approaches the speed of light as only the lowest Landau level remains occupied. The IR approximation is valid at very low energies but fails at higher energies due to dispersion and photon absorption.

	\section{Photon propagation in the magnetosphere of neutron stars}\label{sec5}
	
	The aim of the section is to apply the previous results to the photon propagation in neutron stars magnetosphere and compute the photon time delay. 
	Therefore, we will use a simple model to describe the magnetosphere of the pulsar. Note we are not interested in  the internal composition of the star, just in  the photon propagation from the surface of the pulsar to the outside.  
	
	We consider a toroidal magnetic field 
	based on the description of the pulsar's magnetosphere (PM) using the “striped wind” model in
	\cite{Coroniti90,Michel71,Kirk09}, their geometrical structure follows the solution provided by \cite{Bogovalov99},
	\begin{equation}
		\textbf{B}(r,t)=B_0\dfrac{r_0}{r}\eta_\varphi\hat{\boldsymbol{\varphi}} 
		,\label{toroidal}
	\end{equation}
	where $ B_0 $ is the surface magnetic field at the surface $\textbf{B} (r=r_0)$, $ r_0 $ the radius of the star, 
	$ \textbf{r}  =(r,\theta,\varphi) $ is the 
	position in spherical  coordinates, $ \eta_\varphi $ is related to the Heaviside unit step function and takes the values, $ \pm 1 $ depending on the direction of the magnetic field, but this causes a discontinuity. Because of this, to make a smoother transition the function $ \eta_\varphi $ 
	is defined as 
	\begin{equation}
		\eta_\varphi=tanh(
		\psi), \quad  \psi=\cos{\theta}\cos{\chi}+\sin{\theta}\sin{\chi}\cos\left (
		\varphi-\Omega t
		\right ),
	\end{equation}
	where $ \chi $ is the angle between the axis of rotation of the neutron star and its magnetic moment, 
	$\Omega$ is the pulsar angular velocity. 
	
	The magnetic-field profile in Eq.~(\ref{toroidal}) should be understood as a
	simplified toy model rather than as a complete magnetospheric solution. In the
	inner magnetosphere, the field is often approximated by a dipolar geometry, for
	which $B(r)\propto r^{-3}$. Such a steeper radial dependence would confine the
	strongest refractive corrections to a narrower region close to the stellar
	surface and would therefore reduce the contribution accumulated at large radii.
	In contrast, the toroidal profile $B(r)\propto r^{-1}$, motivated by
	striped-wind-type geometries, allows us to explore how the propagation-induced
	delay behaves when strong magnetic-field effects persist over a larger radial
	range. The use of a toroidal component is also connected with recent studies of
	magnetar spin-down and magnetic-inclination evolution, where toroidally
	dominated magnetic configurations have been discussed as a possible ingredient
	in explaining low braking indices in highly magnetized neutron stars
	\cite{Li:2026ypc}. Therefore, the profile adopted here
	should be regarded as an idealized field configuration used to estimate the
	relative magnitude of vacuum and medium-induced propagation delays.

	Additionally, we assume the PM is 
	composed of a cold magnetized dense charged electron 
	plasma ( Heavy ions, protons, and neutrons are neglected, which are only a small part of the neutron stars  plasma \cite{Petri18}). 
	The  chemical potential of the plasma depends on the distance to the star $ \mu(\textbf{r},t) $ and has radial symmetry \cite{ Petri16, Petri11, Petri18}.
	The chemical potential in the PM is considered as a mixture of electrons
	and positrons $e^{+}-e^- $ 
	described using the chemical potential 
	given by \cite{ Petri11, Petri16, Petri18, Rojas04}:
	\begin{equation}\label{mudistance}
		\mu(r,t) = \dfrac{(\mu_{max}-\mu_{min}) sech^2(
			\psi )+\mu_{min}}{tanh(
			\psi)}\dfrac{r_0}{r}+\mu_0,
	\end{equation}
	where  $ \mu _{max} $ and $ \mu _{min} $ are the maximum and minimum chemical potential inside the layers of the magnetosphere respectively. We have added the last term in Eq. (\ref{mudistance}), $ \mu_0 $  should be understood as an average of the background chemical potential of the gas within the pulsar nebula, measured at a sufficient distance where the magnetic field effects produced by the star become negligible.
	Finally,
	NS is assumed as a very slowly rotated stars compared to the speed of radiation. Therefore, it can be considered a stationary model.  
	Furthermore, we take into account the  curvature produced by the star using the Schwarzschild metric assuming NS is spherical and static,
	{\small 
		\begin{eqnarray}
			ds^2=-\left (1-\dfrac{2M}{r}\right )dt^2+\left (1-\dfrac{2M}{r}\right )^{-1}dr^2+ 
			r^2 d\Omega^2,
	\end{eqnarray}}
	\noindent where $d\Omega^2=d^2\theta+\sin^2\theta d^2\phi$, $ M $ is the mass of the star. We used the sign convention ($-,+,+,+ $).
	\begin{figure*}[t]
		\centering
		% Chemical-potential figures removed from the current version
		% \includegraphics[width=0.49\linewidth]{Figures/muplot.pdf}
		\includegraphics[width=0.44\linewidth]{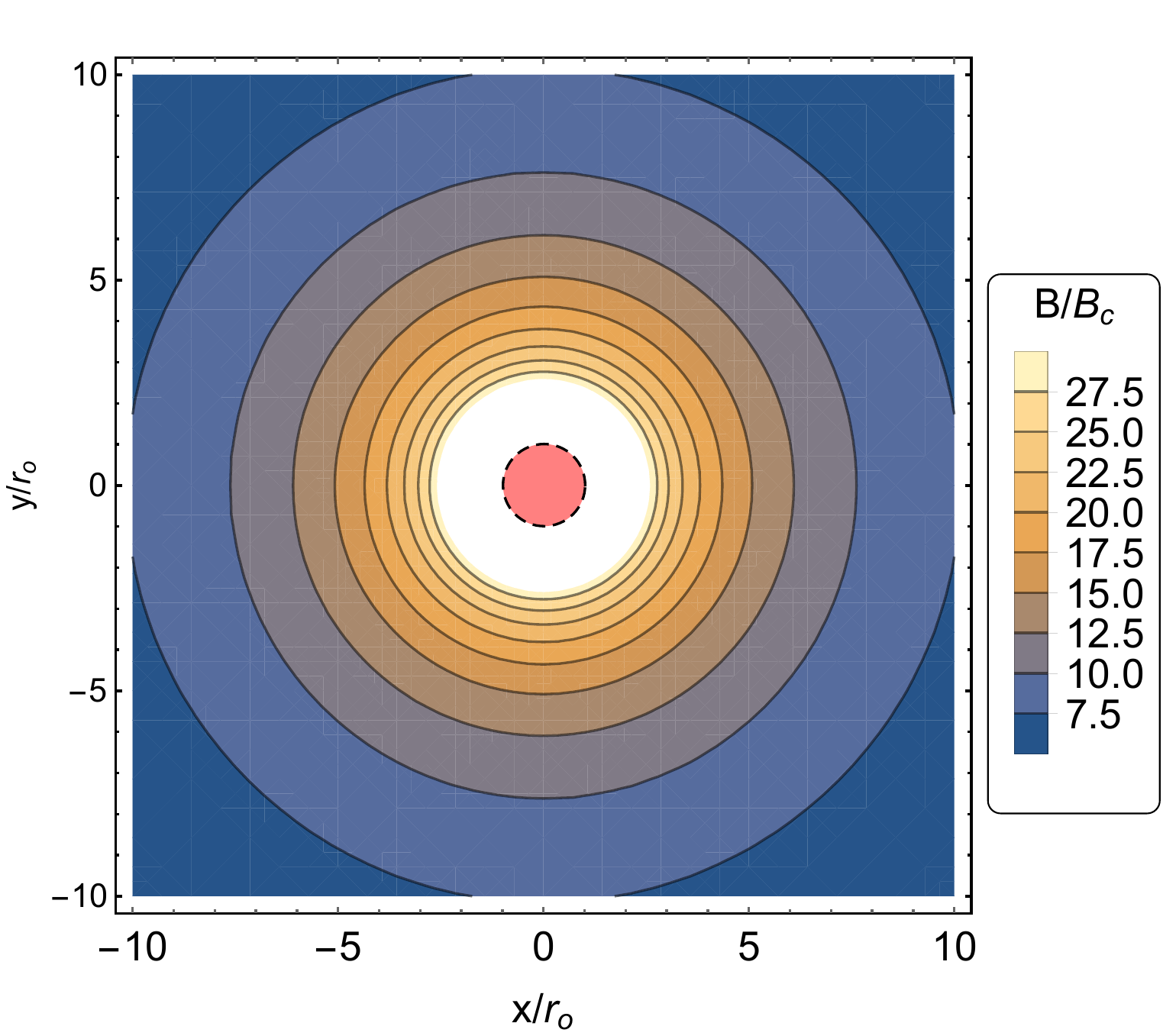}
		\includegraphics[width=0.52\linewidth]{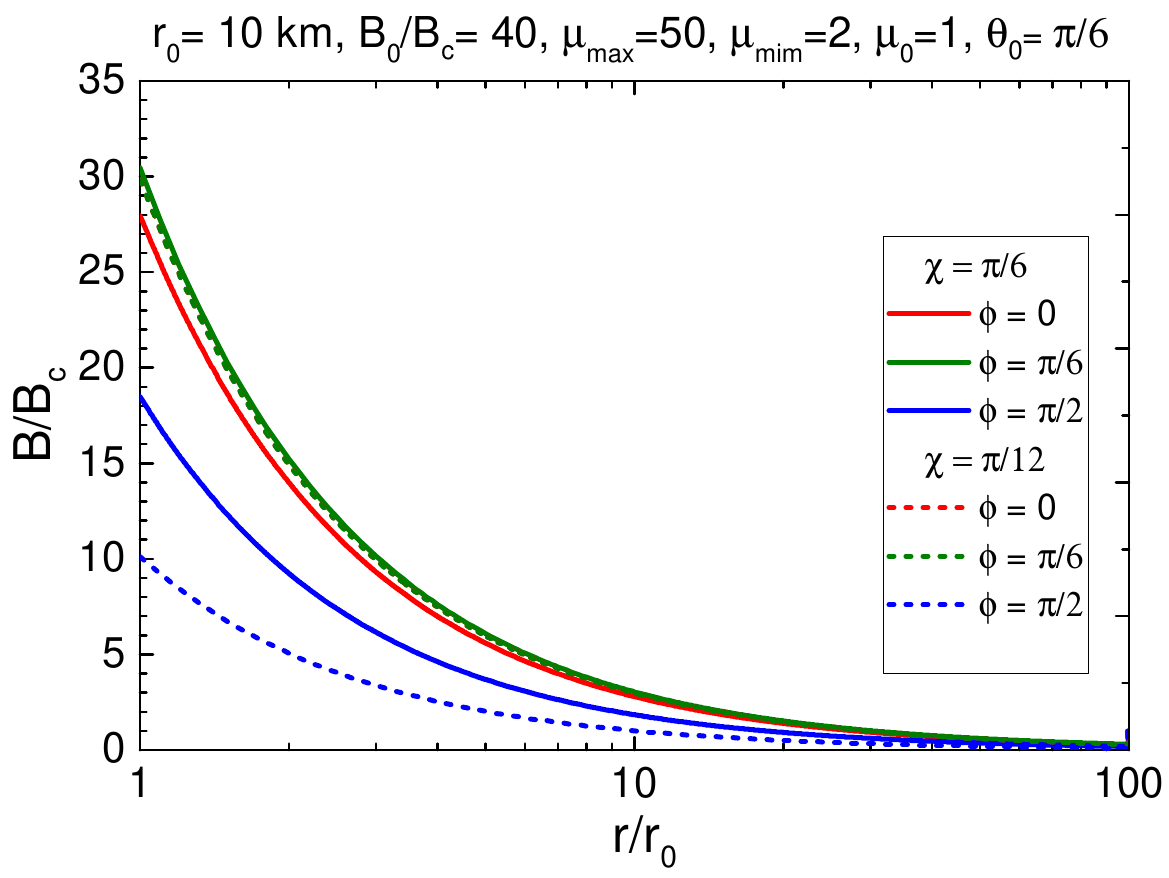}
		\caption{
			\textbf{Left panel:} Density plot of the magnetic-field strength $B/B_c$
			in the transverse plane ($x/r_0$, $y/r_0$). The dashed line indicates the
			stellar surface with radius $r_0=10$ km. Higher magnetic-field strengths
			are represented by warmer colors.
			\textbf{Right panel:} Radial dependence of the magnetic-field strength
			$B/B_c$ as a function of the normalized radial distance $r/r_0$ for
			different azimuthal angles $\theta$. The radial distance is measured from
			the stellar center, with $r_0=10$ km representing the stellar radius.
			The parameters used are $B_0/B_c=100$, $\mu_0/m_e=2$,
			$\mu_{\min}/m_e=2$, $\mu_{\max}/m_e=50$, $\varphi=0$, and
			$\chi=\pi/6$.
		}
		
		\label{muBdistance}
	\end{figure*}

	Figure~\ref{muBdistance} illustrates the spatial distribution and radial
	evolution of the magnetic-field strength $B/B_c$. The left panel shows its
	two-dimensional distribution in the transverse plane, illustrating the
	strong-field region close to the star and the decrease of the magnetic-field
	strength with increasing distance.

	The right panel shows the radial dependence of the magnetic field,
	highlighting its rapid decrease from strong fields near the stellar surface
	to weaker fields at larger distances. This decrease in magnetic-field
	strength influences the photon propagation characteristics and Landau
	quantization.

	\subsection{Photon time delay}
	Once we have a description of the pulsar magnetosphere using a simple model, the phase velocity and time delay can be computed. 
	The contribution  to  total photon time delay comes from: $ \Delta \tau _{V}$, $ \Delta \tau _{M}$, and $ \Delta \tau_{S}$, where $ \Delta \tau _{V} $ is the delay due to  vacuum effects (vacuum birefringence
	), $\Delta\tau_{S}$ is due to space-time curvature effect, $\Delta\tau_{M}$ is the contribution due to the medium effects (which includes absorption-emission processes). 
	We obtain the equations of motion of the photons $ x^\mu(\lambda) $ under Schwarzschild metric, the geodesic equation  $\ddot{x}^{\rho}+\Gamma^{\rho}_{\mu\nu} \dot{x}^{\mu}\dot{x}^{\nu}=0$ is used \cite{Pechenick}, where $ \Gamma^{\rho}_{\mu\nu} $ are Christoffel's symbols.
	In the case of photons, their equations of motion are given by $u^\mu =dx^\mu/d\lambda $ \cite{Falkner12}.
	In our calculation, we include gravitational redshift/time-dilation effects through a Schwarzschild correction factor, while neglecting light bending and treating the photon trajectory as radial.
	Assuming a photon with an energy $\omega$  emitted from a pulsar's surface with a  radius $ r_0$ and mass $ M_0 $, then the time delay between this photon passing through a dispersive medium  and another photon that does not disperse it is
	\begin{equation}
		\Delta\tau =\int_{r_0}^{r_0+L} dr\left( \dfrac{1}{v_{p}(\omega,\mu, B)} -1\right) \left (1-\dfrac{2M}{r}\right )^{-1}, \label{TDcontodo}
	\end{equation} 
	where $ v_{p} $ is the phase velocity of the wave. In the present work, we estimate the propagation delay using the phase velocity as an effective measure of the refractive correction. A more complete treatment of pulse arrival times would require the group velocity, $v_g=d\omega/dk$, together with the absorptive part of the photon self-energy . Therefore, the delay estimates presented below should be interpreted as order-of-magnitude propagation-induced delays within the transparency regime. 
	The photon time delay between two photons of different energies that pass through the same medium is given by the difference in time using (\ref{TDcontodo}), $\Delta\tau_2(\omega_2)-\Delta\tau_1(\omega_1)$. Note that we are working in natural units, and the time to travel a distance $ L $ at speed $ c $ is $ t_0=L/c $, but here $ c=1 $.
	We consider a NS with a mass of $M_0 = 2M_s$ and a radius of $r_0 = 10$ km.

	For vacuum we start from Eq. of dispersion (\ref{leyesdispersionvacio}) while for medium  dispersion Eq. (\ref{eqdispperp}) is considered  taking  the values of $\mu_0$, $\mu_{min}$, $\mu_{max}$, $\chi$, $\theta$ for fixed $\varphi$.

	\begin{figure}[h]
		\centering
		\includegraphics[width=0.6\linewidth]{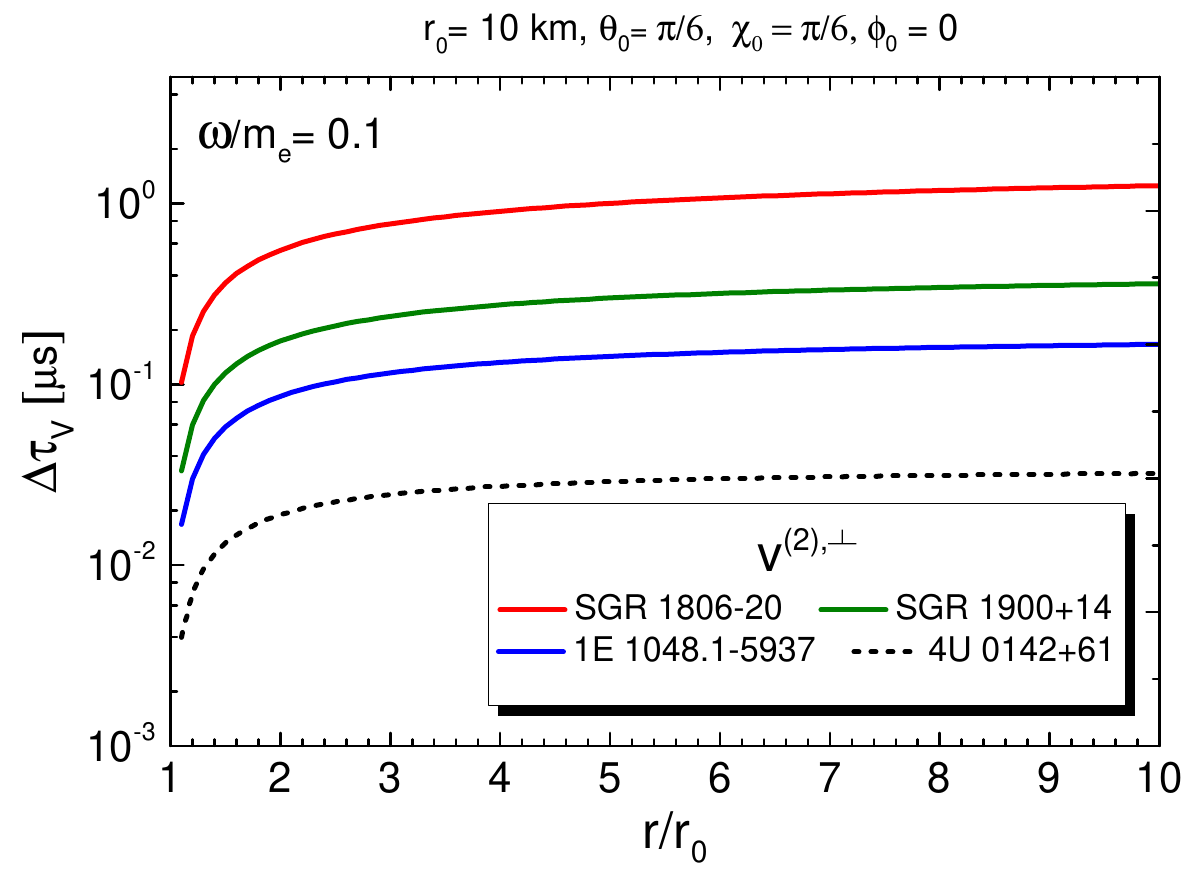}
		\caption{Photon delay time $\Delta \tau_V^{\perp}$ for perpendicular propagation  ($\vec{k}\perp\vec{B}$) as a function of radial distance $r/r_0$ from the stellar surface for different magnetars taken from the catalog (2014) \cite{Olausen2014},   SGR 1806-20 ($B_0/B_c$=45.35),
			SGR 1900+14 ($B_0/B_c$=15.47),
			1E 1048.1-5937 ($B_0/B_c$=8.84), and
			4U 0142+61 ($B_0/B_c$=2.94). Calculations assume a toroidal magnetic field described by Eq.~(\ref{toroidal}), with parameters $M=2M_s$, $r_0=10$~km, photon energy $\omega/m_e=0.1$, and angular configuration given by $\theta=\pi/6$, $\varphi=0$, and $\chi=\pi/6$.} \label{TD_magnetars_Vacuum_om01} 
	\end{figure}
	
	The phase velocity of photons propagating in a vacuum depends only on the magnetic field. This is valid for energies far from the first pair creation threshold $ \omega^2<<4m_e^2 $.
	At this point, we would like to emphasize that, because we have taken the low-frequency approximation for the nonlinear EH theory, we do not obtain a dependence of the phase velocity on the photon energy \cite{RomeroLicenciatura}.

	\begin{figure*}[t]
		\centering
		\includegraphics[width=0.49\linewidth]{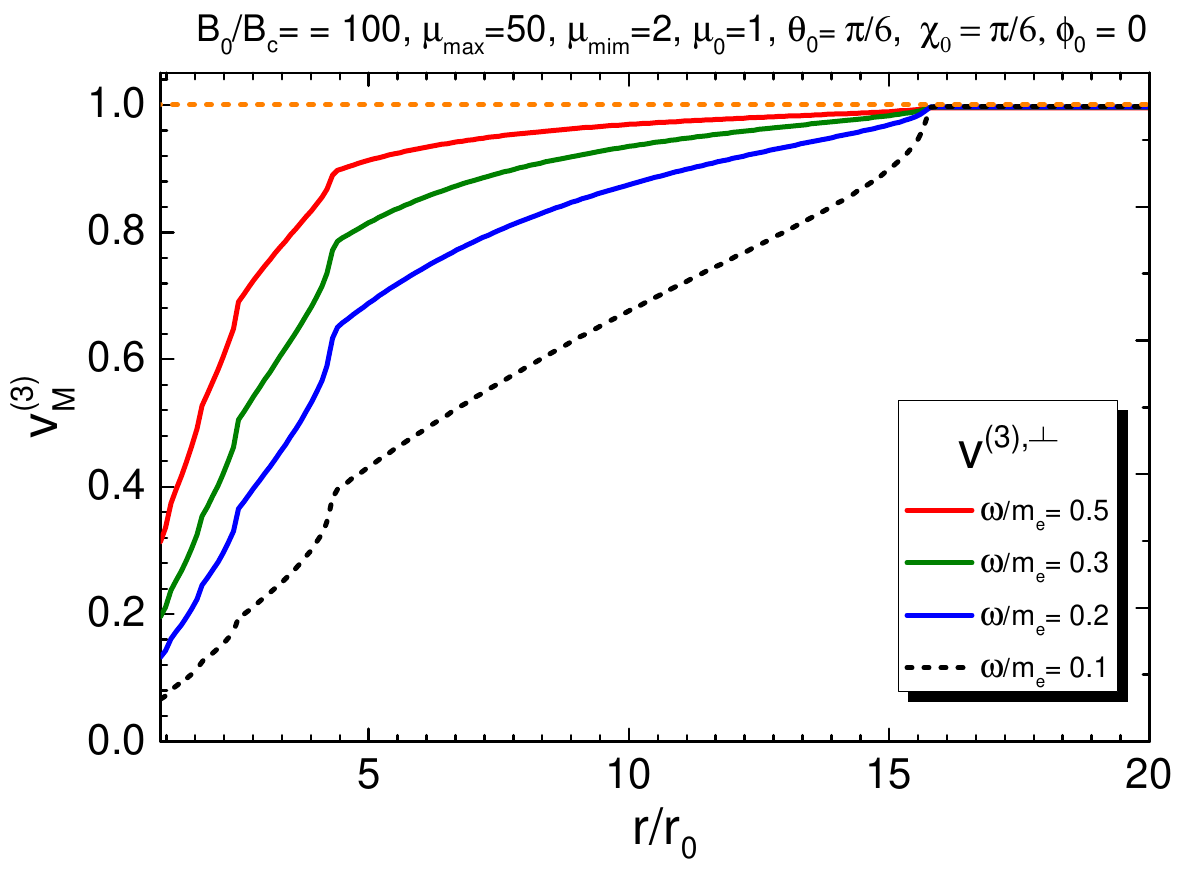}
		\includegraphics[width=0.49\linewidth]{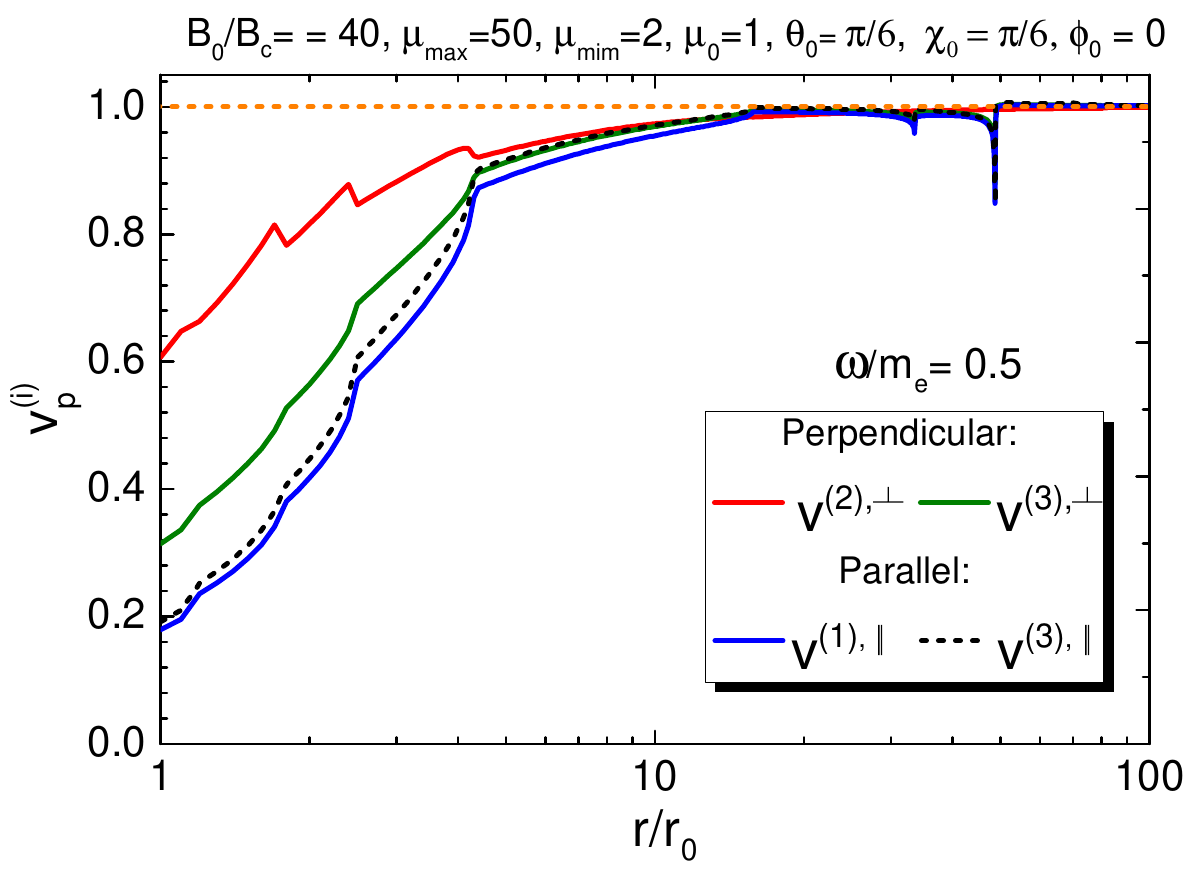}
		\caption{\textbf{Left panel:} Photon phase velocity as a function of the radial
			distance traveled from the stellar surface for perpendicular propagation
			($\vec{k}\perp\vec{B}$) and polarization mode $i=3$, evaluated at different
			photon energies: $\omega/m_e=0.5,\,0.3,\,0.2,\,0.1$. The calculation uses
			$B_0/B_c=100$, $\mu_{\max}/m_e=50$, $\mu_{\min}/m_e=2$,
			$\mu_0/m_e=1$, $\theta=\pi/6$, $\chi_0=\pi/6$, and $\phi_0=0$.
			\textbf{Right panel:} Comparison of photon phase velocities for perpendicular
			propagation, modes $i=2,3$, and parallel propagation, modes $i=1,3$, as a
			function of radial distance, for fixed photon energy $\omega/m_e=0.5$. The
			calculation uses $B_0/B_c=40$, $\mu_{\max}/m_e=50$,
			$\mu_{\min}/m_e=2$, $\mu_0/m_e=1$, $\theta_0=\pi/6$,
			$\chi_0=\pi/6$, and $\phi_0=0$. Dashed lines correspond to mode $i=2$,
			and solid lines correspond to mode $i=3$. Both panels assume a toroidal
			magnetic field profile, Eq.~(\ref{toroidal}), and the radially dependent
			chemical-potential profile, Eq.~(\ref{mudistance}).}
		\label{velocity_distance}
	\end{figure*}

	To assess the sensitivity of the accumulated time delay to the assumed radial
	magnetic-field profile, we consider the generalized form
	\begin{equation}
		B_n(r)=B_0\left(\frac{r_0}{r}\right)^n,
		\qquad n=1,3.
		\label{Bprofile_sensitivity}
	\end{equation}
	Here, $n=1$ corresponds to the toroidal profile adopted in the main
	calculation, while $n=3$ represents a steeper, dipole-like radial decrease.
	This comparison is not intended as a self-consistent model of a dipolar
	magnetosphere, but rather as a simple sensitivity estimate of the time-delay
	integral to the radial decay of the magnetic field.
	
	In the weak-refractive-correction regime, if the local refractive correction
	scales as $\delta n\propto B^q$, the accumulated delay scales as
	\begin{equation}
		\Delta\tau_n\propto
		\int_{r_0}^{R}dr
		\left(\frac{r_0}{r}\right)^{nq}
		=
		\frac{r_0}{nq-1}
		\left[
		1-\left(\frac{r_0}{R}\right)^{nq-1}
		\right].
		\label{delay_scaling_profile}
	\end{equation}
	For the weak-field vacuum contribution, for which $q=2$, and for
	$R\gg r_0$, this gives
	\begin{equation}
		\frac{\Delta\tau_{n=3}}{\Delta\tau_{n=1}}
		\simeq
		\frac{1}{5}=0.2.
	\end{equation}
	Thus, changing from the toroidal profile to a dipole-like radial decay reduces
	the accumulated weak-field vacuum delay by approximately a factor of five.
	This demonstrates that the absolute magnitude of the predicted delay is
	sensitive to the assumed radial extent of the strong magnetic field.
	
	For the finite-density contribution, the dependence on $B$ is more involved
	because of the Landau-level structure of the polarization operator and cannot,
	in general, be represented by a single power $B^q$. Therefore, the above
	estimate should not be interpreted as a quantitative prediction for the medium
	contribution. Nevertheless, it indicates that the millisecond-scale delays
	obtained with the toroidal profile should be regarded as model-dependent
	estimates associated with an extended strong-field configuration, rather than
	as geometry-independent predictions.

	\section{Phase velocity and photon time delay for different magnetars.}\label{sec6}

	We apply the preceding results to representative magnetars, using their surface
	magnetic fields from the McGill Magnetar Catalog \cite{Olausen2014}. Unless
	otherwise stated, we assume $M=2M_\odot$ and $r_0=10~\mathrm{km}$.

	In Figure~\ref{TD_magnetars_Vacuum_om01}, we present the photon delay time $\Delta \tau_V^{\perp}$ as a function of the radial distance $r/r_0$ traveled by photons propagating perpendicular to the magnetic field lines ($\vec{k}\perp\vec{B}$) for 4 different  magnetars. The delay arises due to quantum vacuum polarization and medium-induced refractive effects, which slow the photon propagation below the vacuum speed of light.

	The predicted delays increase with the surface magnetic field and can reach
	the microsecond scale within the present model. However, their direct
	observability cannot be inferred from these estimates alone, since the
	calculation employs an idealized magnetospheric geometry and phase velocity
	based propagation times, while a realistic arrival-time analysis would require
	the group velocity, emission geometry, plasma structure, and ray propagation
	to be treated consistently.
	
	Figure~\ref{velocity_distance} shows the radial evolution of the photon phase
	velocity for different energies and propagation modes. The largest deviations
	from $c$ occur close to the stellar surface, where both the magnetic field and
	the chemical potential are largest. As $r/r_0$ increases, the phase velocities
	approach $c$. The comparison also illustrates the polarization dependence of
	the medium correction and the differences between parallel and perpendicular
	propagation.

	\begin{figure}[h] \centering \includegraphics[width=0.6\linewidth]{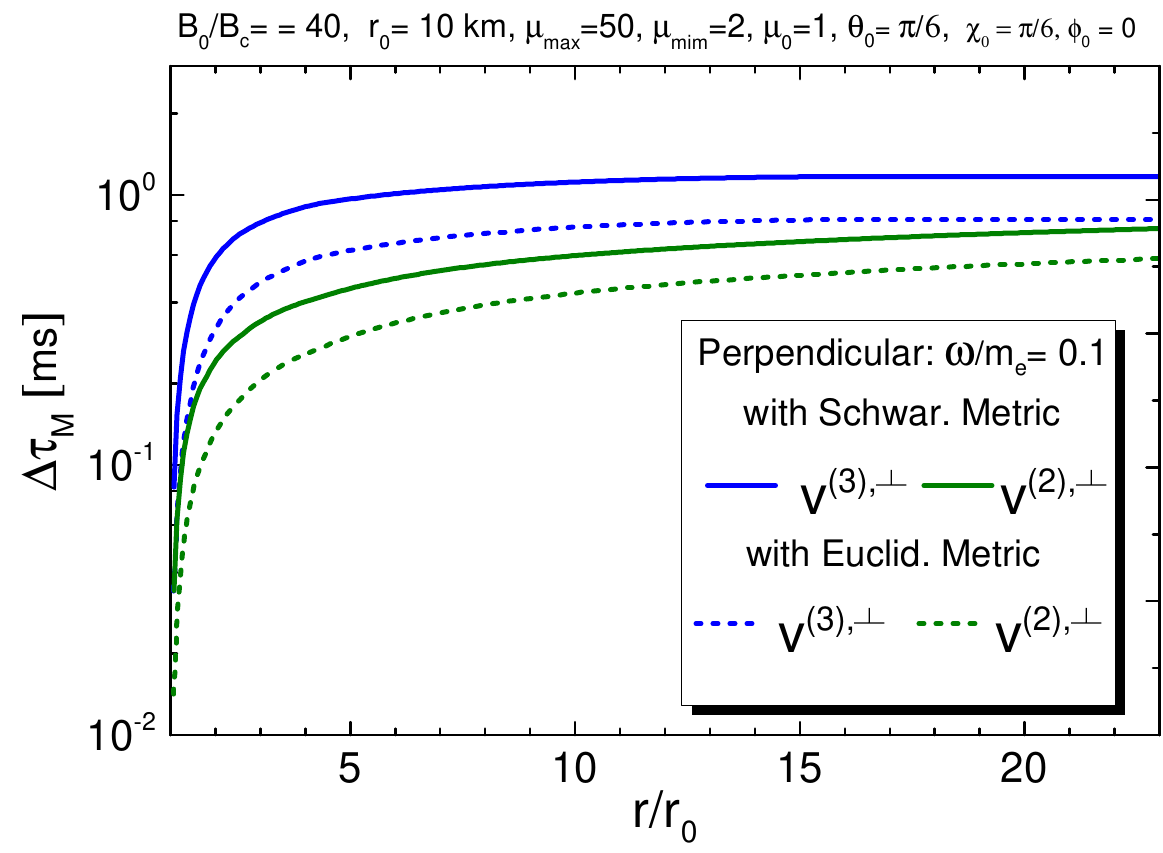} 
		\caption{Medium-induced photon time delay $\Delta\tau_M$ as a function of
			$r/r_0$ for perpendicular propagation and polarization modes $i=2,3$.
			Solid (dashed) curves correspond to Schwarzschild (flat) spacetime.
			Parameters: $B_0/B_c=40$, $\mu_0/m_e=1$,
			$\mu_{\min}/m_e=2$, $\mu_{\max}/m_e=50$,
			$\theta=\pi/6$, $\varphi=0$, and $\chi=\pi/6$.}
		\label{TD_M_C}
	\end{figure}

	Figure~\ref{TD_M_C} shows the effect of spacetime curvature on the photon
	delay time $\Delta\tau_M$ for the two polarization modes $i=2,3$ in
	perpendicular propagation ($\vec{k}\perp\vec{B}$). The comparison between
	the Schwarzschild and flat-spacetime calculations shows that spacetime
	curvature increases the delay for both polarization modes. Mode $i=2$
	exhibits a larger delay than mode $i=3$. The difference between the curved-
	and flat-spacetime results is largest near the stellar surface
	($r/r_0<10$), where gravitational effects are strongest, and decreases as
	the photon propagates outward. At larger distances ($r/r_0>10$), the delay
	times approach the corresponding flat-spacetime results.
	
	\begin{figure*}[t]
		\centering
		\includegraphics[width=0.49\linewidth]{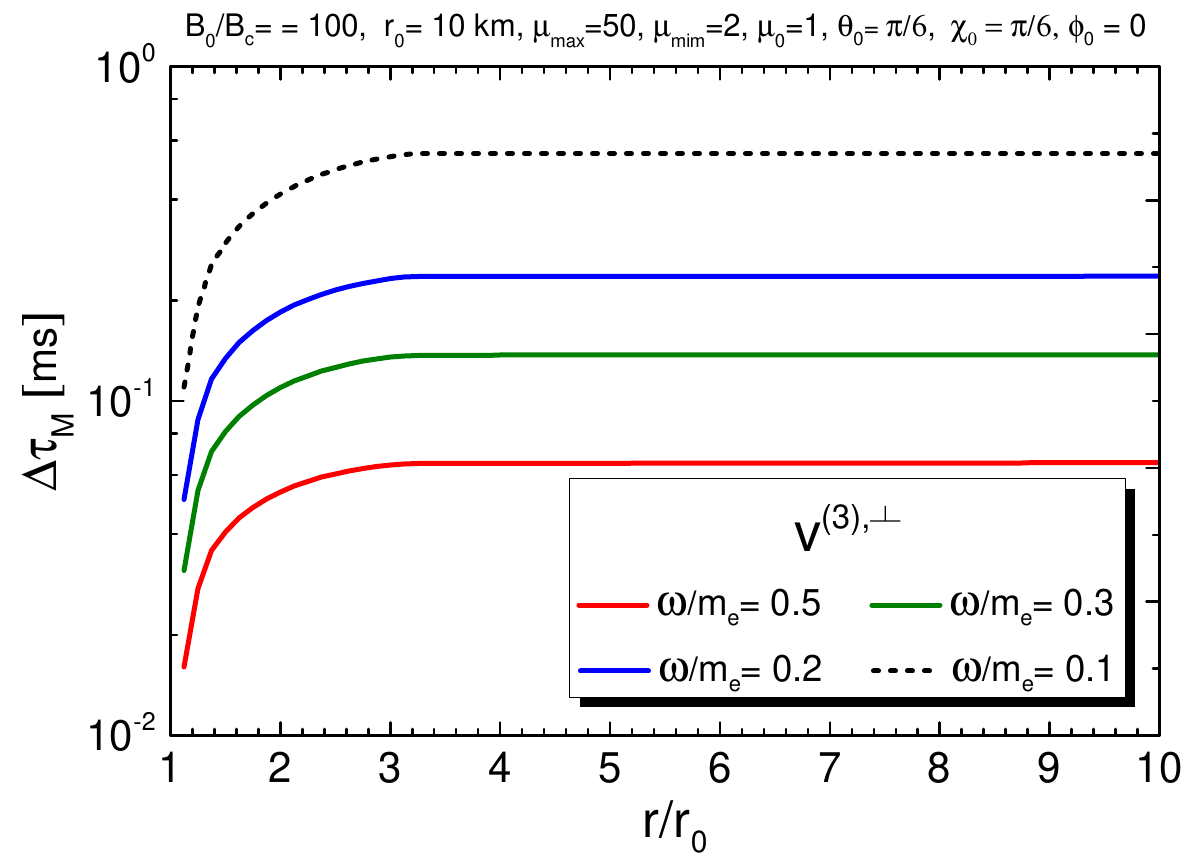}
		\hfill
		\includegraphics[width=0.49\linewidth]{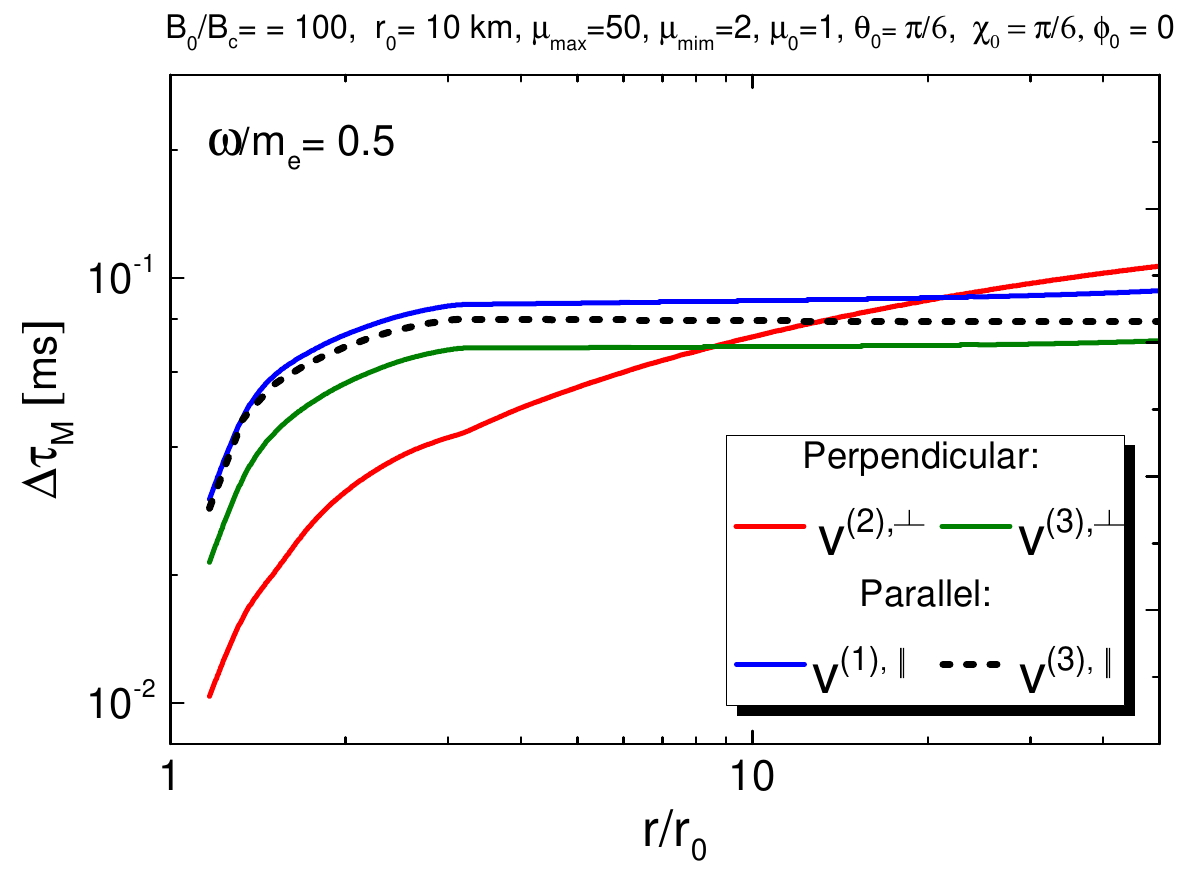}
		
		\caption{
			Photon delay time $\Delta\tau_M$ as a function of the normalized radial
			distance $r/r_0$ from the neutron star surface.
			\textbf{Left panel:} Delay time for photons with different energies.
			Lower-energy photons, in particular $\omega/m_e=0.1$ (black dashed curve),
			experience larger delays than higher-energy photons.
			\textbf{Right panel:} Delay time for different polarization modes and
			propagation directions. Modes $i=2,3$ correspond to perpendicular propagation
			($\vec{k}\perp\vec{B}$), while modes $i=1,3$ correspond to parallel
			propagation ($\vec{k}\parallel\vec{B}$). The photon energy is fixed at
			$\omega/m_e=0.5$ in the right panel.
			Both panels assume a toroidal magnetic-field profile
			[Eq.~(\ref{toroidal})] and the radially dependent chemical potential
			[Eq.~(\ref{mudistance})], with $B_0/B_c=100$, $\mu_0/m_e=1$,
			$\mu_{\min}/m_e=2$, $\mu_{\max}/m_e=50$, $\theta=\pi/6$,
			$\varphi=0$, and $\chi=\pi/6$.
		}
		\label{TRF1}
	\end{figure*}

	In Figure~\ref{TRF1} (left panel), the delay time $\Delta \tau_M$ induced by photon propagation through a magnetized plasma surrounding a neutron star is plotted as a function of the normalized radial distance $r/r_0$ from the star’s surface. The results show that lower-energy photons ($\omega/m_e=0.1$, dashed black curve) experience greater delays compared to higher-energy photons. This phenomenon arises due to stronger dispersive effects at low photon energies, as the refractive index of the magnetized plasma exhibits more significant quantum corrections at lower frequencies. 
	The larger delay obtained at lower photon energies reflects the stronger
	dispersive corrections predicted by the model in this energy range.
	
	Figure~\ref{TRF1} (right panel)  presents the photon delay time as a function of radial distance for different polarization modes and propagation directions relative to the magnetic field. Here we note distinct differences: perpendicular propagation  mode $i=2$ (red line) exhibits the smallest delay, reaching speeds closest to that of vacuum propagation for large distances. In contrast, perpendicular mode $i=3$ and parallel propagation modes ($i=1,3$) experience more significant delays. Such variations arise because the photon phase velocity depends explicitly on both the photon polarization state and its propagation direction relative to the magnetic field. This anisotropic dispersion is due to polarization-dependent interactions between photons and magnetized plasma, causing certain polarization states and propagation directions to experience stronger quantum electrodynamic corrections and therefore greater delays.

	\begin{figure}[t]
		\centering
		\includegraphics[width=0.6\linewidth]{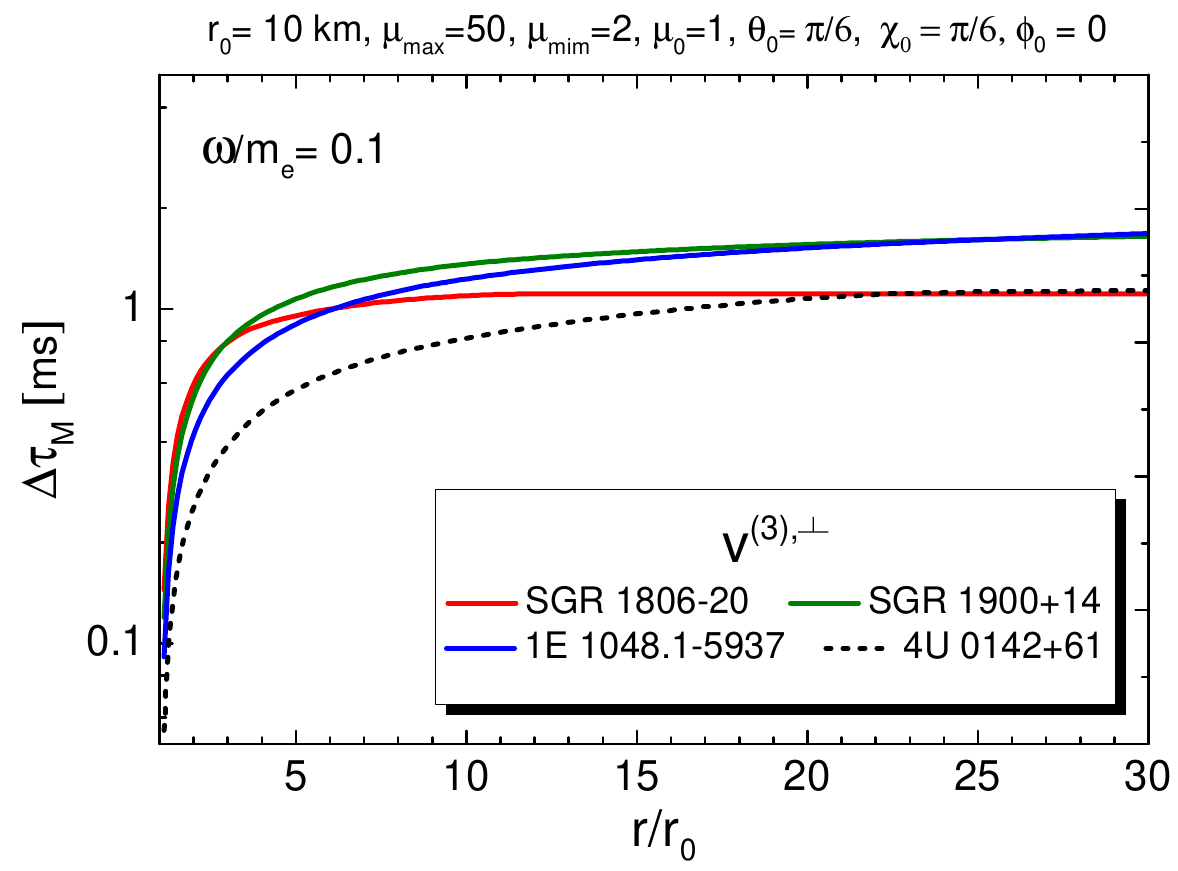}
		\caption{Photon time delay $ \Delta \tau_M $ for different magnetars as a function of the distance traveled by the photon from the surface of the star to its position, where $ r_0 $ is the radius of the star. 
			Perpendicular propagation to the magnetic field using mode $i=3$. SGR 1806-20 ($B_0/B_c$=45.35),
			SGR 1900+14 ($B_0/B_c$=15.47),
			1E 1048.1-5937 ($B_0/B_c$=8.84), and
			4U 0142+61 ($B_0/B_c$=2.94). The magnetic field used has a toroidal shape Eq.~(\ref{toroidal}) and the chemical potential variable (Eq. \ref{mudistance}). Here a surface magnetic field was used for the magnetars in the catalog \cite{Olausen2014}, in addition to $ \omega/m_e=0.1 $, $ M=2M_s $, $ r_0= 10 $ km, $\mu_0/m_e=1 $, $\mu_{min}/m_e=2 $, $\mu_{max}/m_e=50 $, $ \theta=\pi/6 $, $ \varphi =0 $ and $ \chi=\pi/6 $.
		}
		\label{timedelay_magnetars}
	\end{figure}

	Figure~\ref{timedelay_magnetars} illustrates the photon delay time induced by the magnetized plasma medium surrounding different magnetars, as photons propagate radially outward from the neutron star surface
	This panel, depicting perpendicular propagation  ($\vec{k} \perp \vec{B}$, mode $i=3$), reveals significant variations in delay time depending on the magnetar's surface magnetic field strength. Notably, the pulsar with the weakest magnetic field, \textit{4U 0142+61} ($B_0/B_c=2.94$), exhibits negligible delay effects, demonstrating minimal dispersion due to the relatively weak magnetization of the surrounding medium.
	
	Conversely, pulsars with fields above the critical magnetic field $B_c$ exhibit pronounced delays, reaching millisecond scales. Interestingly, although the magnetar \textit{SGR 1806-20} has the strongest surface magnetic field in the sample, \textit{1E 1048.1-5937 } presents a larger delay time. 
	This difference indicates that the photon delay depends not only on the
	magnetic field strength but also on the medium parameters, including the
	particle density and chemical potential profile.
	
	We observe from Figure\ref{TD_magnetars_Vacuum_om01} that the vacuum contribution to the photon delay time is significantly smaller than the contribution arising from the magnetized medium. Specifically, vacuum-induced delays are typically one to two orders of magnitude lower than medium-induced delays for gamma-ray photons, and this difference becomes even more pronounced, reaching up to four orders of magnitude, for photon energies in the X-ray regime. This underscores the dominant role of medium-induced effects in determining photon propagation delays around magnetars.

	\begin{center}
		\begin{tabular}{|c|c|c|}
			\hline 
			Pulsar &  $ B_s/B_c $  &    $ \Delta \tau_V^{\perp} (ns) $   \\
			\hline
			\hline
			J010043.1-721134 & 8.84 & 180 \\
			\hline
			4U 0142+61 & 2.94 & 33.5  \\
			\hline
			SGR 0418+5729 & 0.13 & 0.104  \\
			\hline
			SGR 0501+4516 & 4.30 & 61.6  \\
			\hline
			J164710.2-455216 &1.49 & 10.4 \\
			\hline
			SGR 1900+14 & 15.8 & 401 \\
			\hline
			1E 1547.0-5408  &  7.25 & 135 \\
			\hline
			J1808-2024  & 46.7 & 1548 \\
			\hline
		\end{tabular}
	\end{center}
	\textit{\small{Table 1: Photon time delay of different magnetars using the Vacuum contribution. Parameters: $r_0=10$ km, $\theta=\pi/6$, $\varphi=0$, and $\chi=\pi/6$.}
	}

	\begin{center}
		\small
		\begin{tabular}{|c|c|c|c|c|c|c|c|c|c|}
			\hline 
			\multicolumn{2}{|c|}{} &
			\multicolumn{4}{|c|}{$\Delta \tau_M^{\perp} \, (\mathrm{ms})$ for the specified energy $\omega$ } &
			\multicolumn{4}{|c|}{$\Delta \tau_M^{\parallel} \, (\mathrm{ms})$ for the specified energy $\omega$ } \\
			\hline
			\hline
			Pulsar & {\footnotesize $B_s/B_c$} &
			{\footnotesize 1 keV} &
			{\footnotesize 5 keV} &
			{\footnotesize 10 keV} &
			{\footnotesize 100 keV} &
			{\footnotesize 1 keV} &
			{\footnotesize 5 keV} &
			{\footnotesize 10 keV} &
			{\footnotesize 100 keV} \\
			\hline
			\hline
			{\footnotesize J010043.1-721134} & 8.84 &
			123.7 & 24.0 & 11.6 & 0.99 &
			168.0 & 32.9 & 16.0 & 1.84 \\
			\hline
			{\footnotesize 4U 0142+61} & 2.94 &
			99.3 & 19.2 & 9.26 & 0.37 &
			166.8 & 32.8 & 16.2 & 1.13 \\
			\hline
			{\footnotesize SGR 0418+5729} & 0.13 &
			24.8 & 3.00 & 0.86 & $\sim 0$ &
			49.1 & 10.1 & 3.04 & $\sim 0$ \\
			\hline
			{\footnotesize SGR 0501+4516} & 4.30 &
			112.5 & 21.8 & 10.5 & 0.59 &
			177.2 & 34.8 & 17.1 & 1.76 \\
			\hline
			{\footnotesize J164710.2-455216} & 1.49 &
			51.3 & 10.0 & 5.36 & 0.032 &
			133.1 & 26.3 & 13.6 & 0.46 \\
			\hline
			{\footnotesize SGR 1900+14} & 15.8 &
			119.6 & 23.2 & 11.2 & 0.78 &
			147.0 & 28.7 & 13.9 & 1.10 \\
			\hline
			{\footnotesize 1E 1547.0-5408} & 7.25 &
			121.8 & 23.6 & 11.4 & 0.95 &
			171.8 & 33.6 & 16.4 & 2.11 \\
			\hline
			{\footnotesize J1808-2024} & 46.7 &
			69.9 & 13.6 & 6.66 & 0.47 &
			72.8 & 14.2 & 6.94 & 0.50 \\
			\hline
		\end{tabular}
	\end{center}
	\textit{\small{
			Table 2: Photon time delay for parallel and perpendicular propagation
			for different magnetars, taking into account the medium contribution (mode $i=3$).
			We used photons with energies of 1, 5, 10, and 100 keV. Parameters: $\mu_0/m_e=1$, $\mu_{min}/m_e=2$, $\mu_{max}/m_e=50$, $\theta=\pi/6$, $\varphi=0$, and $\chi=\pi/6$}
	}

	Let us now compare the theoretical results of our model with the observational data.

	Recent X-ray polarimetric observations have strengthened the observational
	motivation for testing strong-field QED effects in magnetars. In particular,
	IXPE observations of 1E~1547.0-5408 have reported a high linear polarization
	degree, stimulating detailed discussions of whether the signal is dominated by
	source geometry, magnetospheric propagation, or vacuum-birefringence effects
	\cite{Taverna_2026,Stewart:2025cmt}. These studies show that magnetar polarimetry is
	a promising probe of QED propagation effects, although the interpretation
	requires detailed modeling of the emission geometry, plasma environment, and
	magnetospheric birefringence.
	
	\subsection{Comparison of photon time delay with observational data}
	
	According to \cite{Denisov17}, for the pulsar \textit{J1808-2024}, a delay time is estimated for the vacuum contribution between the two polarization modes that reach the detector of $ \Delta \tau(2, 3)_{Denisov}\sim 0.3 \mu$s. While, according to our theoretical estimates for the magnetized vacuum, we obtain a delay time of $ \Delta \tau(2,3)_{teo}= 1.40 \mu$s using Eq. (\ref{leyesdispersionvacio}); which, compared to the value of \cite{Denisov17}, is an order of magnitude higher, but this can be attributed to the model used and the parameters set (see table 1).

	According to \cite{Stollman}, the time delay of the quasiperiodic oscillation source of the Cygnus X-2 pulsar between photons with energy in the spectrum of soft $1 - 4$ ms. According to table 2.2, the values of the delay time $ \Delta\tau(\omega_1,\omega_2)=\Delta\tau(\omega_2)-\Delta\tau(\omega_1) $ for energies between $10-100 $ keV give a value of the order of $ 10^1-10^2 $ ms, which is within the expected range of values (see table 2), although depending on the energy values it may be an order of magnitude above the value of \cite{Stollman}.
	
	Other sources such as \cite{Uddipta16} and \cite{Kuzmin} measure for pulsars that emit in the radio wave spectrum, obtaining a delay time of the order of $10^2$ ms. These values fit our results better, but our model does not include that range of energies.
	
	Furthermore, it must be taken into account that current instruments allow us to obtain the arrival times of the pulses emitted by a pulsar with an uncertainty, generally between about 10 to 100 nanoseconds ($ \delta \tau =0.01-0.1 $ $\mu $s) \cite{Bailes2009,Lam2015} depending on the stability of the pulsar and the proximity to the ground detectors. Therefore, measuring the delay time associated with vacuum remains a challenge, because this time is around the value of its uncertainty.

	\section{Conclusions}\label{sec7}
	
	We have investigated photon propagation in strongly magnetized environments
	using the one-loop photon self-energy in QED for propagation parallel and
	perpendicular to the external magnetic field. The corresponding
	polarization-dependent dispersion relations were obtained in vacuum and in a
	magnetized electron-positron medium at finite temperature and chemical
	potential.
	
	The full one-loop calculation was analyzed together with its long-wavelength
	and low-energy limits. In vacuum, the perpendicular propagation modes exhibit
	birefringence, with different refractive indices and phase velocities, whereas
	for parallel propagation the physical mode remains on the light cone. In the
	degenerate magnetized medium ($T\ll\mu$), the numerical solutions show
	additional dispersive effects and oscillatory structures associated with
	Landau quantization.
	
	In the infrared limit, the one-loop photon self-energy agrees with the result
	obtained from the second-order expansion of the Euler-Heisenberg effective
	Lagrangian. The two descriptions therefore coincide in their common regime of
	validity.
	
	The calculated phase velocities were then used to estimate photon time delays
	in a simplified neutron-star environment with a toroidal magnetic-field
	profile, a radially dependent chemical potential, and Schwarzschild redshift
	effects. The delays depend on the polarization mode, propagation direction,
	photon energy, and medium parameters. Within this model, vacuum-induced delays
	are typically in the nanosecond-to-microsecond range for magnetar-scale fields,
	whereas the medium contribution can reach the millisecond scale in the X-ray
	regime.
	
	These delays were calculated using the phase velocity $v_p$. In strongly
	dispersive regions, particularly near resonances where the imaginary part of
	the photon self-energy becomes significant, the group velocity
	$v_g=d\omega/dk$ may differ substantially from $v_p$. A calculation based on
	the group velocity is therefore left for future work.
	
	These estimates should therefore be regarded as theoretical results obtained within a simplified astrophysical model. In particular, we have assumed a toroidal magnetic-field geometry and an electron plasma, used the phase velocity to estimate the delay, and neglected light bending. These idealizations limit a direct comparison with observations, and a more realistic assessment of the potential observability of the effect would require a more complete treatment of the magnetospheric structure, plasma composition, photon propagation, and spacetime geometry.
	
	The results connect the full one-loop QED calculation with the corresponding
	effective nonlinear electrodynamics description and quantify the contributions
	of vacuum birefringence and the magnetized medium to photon propagation. The
	same formalism can be extended to more realistic magnetospheric models,
	finite-temperature media, and nonlinear or axion-like extensions of QED
	\cite{Jorge:2024tok,Brevik2025,Chaichian2026}.

	\section*{Acknowledgments}
	A.R.J, A.P.M, and E.R.Q. were supported by the project of No. NA211LH500-002 from AENTA-CITMA, Cuba. A.R.J acknowledges the financial support from the Giersch Foundation.
	
	\section*{AI-use statement}
	
	AI-assisted tools were used only for language polishing and readability
	improvements. No scientific results, derivations, data analysis, or physical
	interpretations were generated by AI. All authors reviewed and approved the
	final version of the manuscript and take full responsibility for its content.
	
	\makeatletter
	\def\doi#1{}
	\def\url#1{}
	\makeatother
	%\bibliographystyle{unsrt}
	%\bibliographystyle{Definitions/mdpi}
	%\bibliography{references.bib}

	\appendix
	
	\section{Vacuum eigenvalues at infrared regime}
	\label{appen1}
	We can integrate Eq (\ref{appr2223})
	The integral expressions of eigenvalues of the vacuum photon self-energy  Eq. (\ref{vacuumpolarizationtensor2}), in the low-energy limit ($\omega<<2 m$), considering only perpendicular propagation to the direction of the magnetic field ($k_\parallel=0$) 
	\begin{eqnarray}
		&&	\kappa^{(1)}_V  =\dfrac{ \alpha k_\perp^2}{2\pi}\int_{0}^{\infty} \dfrac{d \tau}{ \tau}  e^{-\dfrac{\tau}{b}}   \left[ 
		\dfrac{2}{3} -
		\dfrac{\coth{\tau}}{ \tau }+
		\dfrac{1}{ \sinh^2{\tau} }
		\right] ,\label{appr3}\\
		&&\kappa^{(2)}_V  =\dfrac{ \alpha  k_\perp^2}{2\pi}\int_{0}^{\infty}\dfrac{d \tau}{ \tau}  e^{-\dfrac{\tau}{b}}  \left[ 
		\dfrac{2}{3} -
		\dfrac{\coth{\tau}}{\tau}+
		\dfrac{1}{ \sinh^2{\tau} }
		\right] ,\label{appr5}\\
		&&\kappa^{(3)}_V  =\dfrac{ \alpha k_\perp^2}{2\pi}\int_{0}^{\infty} \dfrac{d \tau}{ \tau}  e^{-\dfrac{\tau}{b}}   \left[ 
		\dfrac{2}{3} +\dfrac{2}{\sinh^2{\tau}}-\dfrac{2\tau\cosh{\tau}}{ \sinh^2{\tau}}  
		\right],\label{appr6}
	\end{eqnarray}
	These integrals can be solved analytically for arbitrary values of the magnetic field, as done in \cite{Aurora2023}, starting from the nonlinear Euler-Heisenberg Lagrangian. They also coincide with the corresponding coefficients in the vacuum expansion of the polarization operator.

	Now we can derive the phase velocity $v_{(i)}=\omega/k_\perp$ from the dispersion equation \ref{dispersionequationV},
	\begin{align}\label{leyesdispersionvacio}
		v_V^{(i)}=\left(1+\frac{\kappa^{(i)}_V(B)}{k_\perp^2}\right)^{-1/2}
		\approx 1-\frac{1}{2k_\perp^2} \kappa^{(i)}_V(B) ,
	\end{align}
	leading to,
	\begin{eqnarray}
		v^{(1)}_V  =\left( 1- \dfrac{ \alpha}{4\pi}\int_{0}^{\infty}\dfrac{d \tau}{ \tau}  e^{-\dfrac{\tau}{b}}   \left[ 
		\dfrac{2}{3} -
		\dfrac{\coth{\tau}}{ \tau }+ \right. \right. \nonumber \\
		\left. \left. 
		\dfrac{1}{ \sinh^2{\tau} }
		\right]  \right),%=1-\mathcal{L}_{\mathcal{F}},
		\label{appr71}
	\end{eqnarray}
	\begin{eqnarray}
		v^{(2)}_V  =\left( 1- \dfrac{ \alpha}{4\pi}\int_{0}^{\infty}\dfrac{d \tau}{ \tau}  e^{-\dfrac{\tau}{b}}  \left[ 
		\dfrac{2}{3} -
		\dfrac{\coth{\tau}}{\tau}+\right. \right. \nonumber \\
		\left. \left. 
		\dfrac{1}{ \sinh^2{\tau} }
		\right]  \right),%=1-\mathcal{L}_{\mathcal{F}},
		\label{appr7}
	\end{eqnarray}
	\begin{eqnarray}
		v^{(3)}_V  =\left( 1- \dfrac{ \alpha}{4\pi}\int_{0}^{\infty} \dfrac{d \tau}{ \tau} e^{-\dfrac{\tau}{b}}  \left[ 
		\dfrac{2}{3} +\dfrac{2}{\sinh^2{\tau}}-\right. \right. \nonumber \\
		\left. \left. \dfrac{2\tau\cosh{\tau}}{ \sinh^2{\tau}}  
		\right] \right).\label{appr81}
	\end{eqnarray}
	
	Now in order to solve the integrals in Eq. (\ref{appr71}-\ref{appr81}) we will use the small fields limit ($eB<<m^2$), due to this limit the functions inside the integrals in $v^{(2)}_V$ and $v^{(3)}_V$ are significantly different for zero only for $\tau<<1$ and we can expand the following expressions around $\tau=0$ and retain the first few terms,
	\begin{eqnarray}
		v^{(1)}_V  \approx\left( 1- \dfrac{ \alpha}{2\pi}\int_{0}^{\infty}d \tau e^{-\dfrac{\tau}{b}}  \left[ 
		\dfrac{\tau}{45}- \dfrac{\tau^3}{315}+...
		\right] \right),
	\end{eqnarray}
	\begin{eqnarray}
		v^{(2)}_V  \approx\left( 1- \dfrac{ \alpha}{2\pi}\int_{0}^{\infty}d \tau  e^{-\dfrac{\tau}{b}}  \left[ 
		\dfrac{\tau}{45}- \dfrac{\tau^3}{315}+...
		\right] \right),
	\end{eqnarray}
	\begin{eqnarray}
		v^{(3)}_V  \approx\left( 1- \dfrac{ \alpha}{2\pi}\int_{0}^{\infty}d \tau  e^{-\dfrac{\tau}{b}}  \left[ 
		\dfrac{\tau}{15} -\dfrac{\tau^3}{63}+...
		\right] \right),\label{appr11}
	\end{eqnarray}
	Neglecting the terms of higher order than $\tau$ we integrate,
	\begin{eqnarray}\nonumber
		v^{WF,(1)}_V =v^{WF,(2)}_V  &=& \left( 1- \dfrac{ \alpha}{2\pi}
		\dfrac{b^2}{45}
		\right),\\
		&=& \left( 1- \dfrac{ \alpha}{90\pi}
		\dfrac{B^2}{B_c^2}
		\right),\\ \nonumber
		v^{WF,(3)}_V  &=& \left( 1- \dfrac{ \alpha}{2\pi}
		\dfrac{b^2}{15}
		\right),\\
		&=& \left( 1- \dfrac{ \alpha}{30\pi}
		\dfrac{B^2}{B_c^2}
		\right).
	\end{eqnarray}
	
	A similar procedure for a strong magnetic field limit ($m^2<<eB$ ) using the asymptotic behavior of Eq. (\ref{appr71}-\ref{appr81})  gave us
	\begin{align}\label{sf_vel}
		v_p^{SF,(1)}=v_p^{SF,(2)}
		= \ \left (1-\frac{\alpha}{3\pi}\frac{B}{B_{c}}\right ),\\
		v_p^{SF,\,(3)}=
		\left (1-\frac{\alpha}{3\pi}\right),
	\end{align}
	where we define here the constant $\xi=7\alpha/45B_c$.
	\appendix
	%\section[\appendixname~\thesection]
	\section[\appendixname~\thesection]{Power Series Expansion of the Vacuum photon self-energy  Eigenvalues}\label{VacuumPowerExpansion}
	
	The eigenvalues of the tensor polarization in the vacuum are derived using the one-loop approximation \cite{Shabad75,Romero20} as in Eq. \ref{vacuumpolarizationtensor2}. 
	Let's develop the eigenvalues of the photon self-energy  in power series of $k^2$, in the limit of low frequencies and strong magnetic fields ($k^2<<eB$), which correspond to slight deviations of the light cone  $k^\mu k_\mu=k^2=\textbf{k}^2-\omega^2=0$. 
	\begin{eqnarray}
		\kappa^{(i)}_V=\sum_{l=0}^{\infty}\chi_{il}(k^2)^l,
	\end{eqnarray}
	the functions of the development: $ \chi_{il} $ 
	\begin{eqnarray} \label{chi}
		\chi_{il}=\dfrac{2 \alpha}{\pi}\int_{0}^{\infty}d\tau\int_{-1}^{1}d\eta \,  e^{-\dfrac{\tau}{b}} [ e^{\zeta_0} \left (\dfrac{\xi}{b m_{e}^2}\right )^{l} \nonumber \\  \dfrac{(-1)^{l}}{l!} \left( \rho_{0i}-l\left (\dfrac{\xi}{b m_{e}^2}\right )^{-1} \theta_{i} \right)+\delta_{1l}\dfrac{\bar{\eta}^2}{2\tau}   
		],
	\end{eqnarray}
	where $ \rho_{0i} $ and $ \theta_{i} $ are the new scalar functions for every mode of propagation bound to the previous ones by  $ \rho_{i}=\rho_{0i} + k^2\theta_{i} $
	\begin{equation}
		\zeta=\zeta_0-\dfrac{k^2\xi}{b m_{e}^2},
		\quad
		\zeta_{0}=\dfrac{(\omega^2-k^2_{\parallel})(\bar{\eta}^2 \tau-\xi)}{b m_{e}^2},
	\end{equation}
	\begin{equation}
		\rho_{01}=0,
		\quad
		\rho_{02}=\dfrac{\omega^2-k^2_{\parallel}}{2}(\bar{\eta}^2 \coth{\tau}-\varrho ),
	\end{equation}
	\begin{equation}
		\rho_{03}=\dfrac{\omega^2-k^2_{\parallel}}{2}(\varrho-\dfrac{\xi}{\sinh^2{\tau}} ),
	\end{equation}
	\begin{equation}
		\theta_{1}=\theta_{2}=-\dfrac{\varrho}{2},
		\quad \quad
		\theta_{3}=-\dfrac{\xi}{2 \sinh^2{\tau}}.
	\end{equation}
	%
	%This coefficients $\chi_{il}$ are calculated analytically using the asymptotic expressions in the limit when $\tau >> 1$ in the \ref{eq1} and numerically in the other case \ref{chi}
	%
	If we truncate the first four terms of the power series,
	\begin{equation}
		\kappa^{(i)}_V\approx\sum_{l=0}^{j}\chi_{il}(k^2)^l,
	\end{equation}
	for $j=3$, involve to a cubic equation in  $ k^2 $. This equation has been  solved in \cite{Rojas14} using Cardano formulas for polynomials of third degree. But, numerical calculations with quadratic and cubic roots are    numerically delicate, so we solve it with the help of the hypergeometric functions \cite{Sturmfels20} as
	\begin{equation}
		\kappa^{(i)}_V=-\sum_{j_2,j_3,...,j_n=0}^{\infty}\dfrac{(-1)^{j_1}j_1!}{(j_0+1)!j_2!j_3!}\dfrac{\chi_{0}^{j_0+1} \chi_{i2}^{j_2} \chi_{i3}^{j_3} }{(\chi_{i1}-1)^{j_1+1}}
	\end{equation}
	where $   j_0=j_2+2j_3+...+(n-1)j_n $ and $  j_1=2j_2+3j_3+...+nj_n $, assuming the perpendicular propagation  ($ k_{3}=0 $) we can compute the phase velocity of the photons $        v=\omega/k_{\perp} $ as
	\begin{equation}
		v^{(i)}_{V}=\left (1-\dfrac{1}{\omega^2}\sum_{j_2,j_3=0}^{\infty}\dfrac{(-1)^{j_1}j_1!}{(j_0+1)!j_2!j_3!}\dfrac{\chi_{i0}^{j_0+1} \chi_{i2}^{j_2} \chi_{i3}^{j_3}  }{(\chi_{i1}-1)^{j_1+1}}   \right )^{-1/2}.
	\end{equation}
	
	The case of propagation parallel to the magnetic field, $\mathbf{k}\parallel\mathbf{B}$, is not particularly interesting because the propagation speed of the electromagnetic radiation coincides with the speed of light, $c$. This can be verified by setting $k_\perp=0$ in the eigenvalues of the photon self-energy, Eq.~(\ref{vacuumpolarizationtensor2}). In this limit,
	$
	\rho_i=\theta_i k_\nu k^\nu ,
	$
	where
	$
	k_\nu k^\nu = k_\parallel^2-\omega^2 ,
	$
	with
	$
	\theta_1=\theta_3=-\frac{\varrho}{2},
	$
	and
	$
	\theta_2=\frac{\bar{\eta}^2}{2}\coth{\tau}.
	$
	Substituting these expressions for $\rho_i$ into Eq.~(\ref{vacuumpolarizationtensor2}) and factoring out the common term $k_\nu k^\nu$, we obtain
	%************************
	\begin{eqnarray}
		& k_\nu k^\nu \left (1-
		\dfrac{2 \alpha}{\pi}\int_{0}^{\infty}d\tau\int_{-1}^{1}d\eta  \,  e^{-\dfrac{\tau}{b}}   [  
		\theta_{i} e^{\zeta}+\dfrac{\bar{\eta}^2}{2\tau}  ]
		\right ) =0,
	\end{eqnarray}
	the solution inside the parentheses must equal one for it to hold for all values of B and $\omega$, which is not a physically possible solution. The only physically meaningful solution is $k^2=0$, corresponding to the light cone.
	\appendix
	\section[\appendixname~\thesection]{Eigenvalues  of  photon self-energy
		$ \Pi_{\mu\nu} $ at finite density and temperature.}\label{appEigenfinite}
	
	To solve  the Eq. dispersion, to do this, we must first diagonalize the photon self-energy
	$ \Pi_{\mu\nu} $.
	In \cite{Rojas79,Shabad84,Rojas87} the structure of the photon self-energy was obtained and widely discussed and it was shown that it can be expanded in terms of nine independent perpendicular tensors $ \Psi^{(i)} _{\mu\nu} $, six of them are symmetric tensors and three are antisymmetric.
	After doing the sum over $p_4$ 
	the scalars $q,t,r$ and $s$ are in the most general case ($k_\perp,k_3 \ne 0$) expressed as an integral over $p_3$ as 
	\begin{align}
		q(k,\omega,\mu,B) 
		=\left(1-\frac{k_{\perp}^2}{\omega^2}\right)t^+,
		\label{scalar_q}\\
		r(k,\omega,\mu,B) 
		=-i\dfrac{\alpha e B}{2\pi \omega} 
		\left(1-\frac{k_{\perp}^2}{\omega^2}\right)^{1/2}
		&\times 
		\sum^{\infty}_{n,n^{\prime}}F_{n,n^{\prime}}^{(3)}
		\int\limits_{-\infty}^{\infty}dp_3 \,
		\frac{ J^{+}_{n,n^{\prime}} }{ Q_{n,n^{\prime}}(p_3)}(n_{e}(p_3) - n_{p}(p_3)),
		\label{scalar_r}\\
		t^{\pm}(k,\omega,\mu,B) 
		= \dfrac{\alpha e B}{4\pi \omega^2}
		\sum^{\infty}_{n,n^{\prime}}
		\int\limits_{-\infty}^{\infty} 
		\dfrac{dp_3}{\varepsilon_{p_3,n}}
		&\times \Bigg[ 
		-F_{n,n^{\prime}}^{(2)}
		\left( 4\omega^2
		+\frac{J^{+}_{n,n^{\prime}}(2p_3k_{3} +J^{-}_{n,n^{\prime}})} { 2 Q_{n,n^{\prime}}(p_3)}\right) \nonumber\\
		&\pm 2  k_{\perp}^2
		N_{n,n^{\prime}}^{(1)}   
		\frac{ (2p_3k_{3}+J^{-}_{n,n^{\prime}} )}
		{Q_{n,n^{\prime}}(p_3)} \Bigg] (n_{e}(p_3) + n_{p}(p_3)),
		\label{scalar_t}\\[6pt]
		s(k,\omega,\mu,B) 
		=-\dfrac{2\alpha e B}{\pi } 
		\sum^{\infty}_{n,n^{\prime}} 
		\int\limits_{-\infty}^{\infty}\frac{dp_3}{\varepsilon_{p_3,n}}
		&\times\Bigg[
		\frac{eB(n-n^{\prime})}{k^2_{3}-\omega^2} F_{n,n^{\prime}}^{(1)}\nonumber\\
		&-\left(
		\frac{e^2B^2(n^{\prime}-n)^2}{k^2_{3}-\omega^2}
		+(2m_e^2+eB(n+n^{\prime})) \right)F_{n,n^{\prime}}^{(1)}\nonumber\\
		&- 2 e B (n n^{\prime})^{1/2} G_{n,n^{\prime}}^{(1)}  
		\Bigg]  
		\frac{2p_3k_3 +J^{+}_{n,n^{\prime}}}
		{ 4 \omega^2 Q_{n,n^{\prime}}(p_3)} 
		(n_{e}(p_3) + n_{p}(p_3)).
		\label{scalar_s}
	\end{align}
	In order to evaluate the corresponding integrals $I^{(i)}$ that appear in $q, t, r,$ and $s$, we must deal with the singularities arising from the denominator $Q_{n,n^{\prime}}(p_3)$ (see \cite{Rojas79}). 
	
	The phase velocities are obtained from the real part of these integrals, while the imaginary parts must be taken into account to describe the damping processes produced by the exponential factor $e^{-\mathrm{Im}[\omega]\, t}$. The amplitude of the photon wave is reduced by this term in the form
	\[
	A(y,t) = A_{0}\, e^{-\mathrm{Im}[\omega]\, t}\, e^{i \left(\mathrm{Re}[\omega]\, t - k y\right)}.
	\]

	The real and imaginary parts of $I^{(i)}$  are,

	\begin{eqnarray}
		Re[I^{(j)}] &= & \left\{ \begin{array}{lll}
			& \quad \quad  \quad \quad   \quad \quad \quad  \\
			I^{(1)}= -\dfrac{1}{\omega\sqrt{J_{nn^{\prime}}^2-4\omega^2m_n}} 
			ln{\left(\dfrac{ 2p_F\omega+\mu\sqrt{J_{nn^{\prime}}-4m_n^2\omega^2}}{2p_F\omega-\mu\sqrt{J_{nn^{\prime}}-4m_n^2\omega^2}}
				\right) },
			\\
			& \quad \quad  \quad \quad   \quad \quad \quad \\
			I^{(2)}= -\dfrac{1}{\sqrt{J_{nn^{\prime}}^2-4\omega^2m_n}} 
			ln{\left(\dfrac{ p_FJ_{nn^{\prime}}+\mu\sqrt{J_{nn^{\prime}}-4m_n^2\omega^2}}{p_FJ_{nn^{\prime}}-\mu\sqrt{J_{nn^{\prime}}-4m_n^2\omega^2}}
				\right) } ,
			& \quad \quad  \quad \quad  \quad \quad \quad,
			\vspace{0.3cm}\\
			I^{(3)} =\log{\left(\dfrac{p_F+\mu}{m_{n}}\right) ,} 
			& \quad \quad  \quad \quad  \quad \quad \quad,
		\end{array} \right.
	\end{eqnarray}
	
	\begin{eqnarray}  
		Im[I^{(j)}] &= & \left\{ \begin{array}{lll}
			I^{(1)}= \dfrac{\pi\omega}{\sqrt{J_{nn^{\prime}}^2-4m_n^2\omega^2}} 
			\left( \theta(\mu-2eB/\omega+\omega)+\theta(-\mu+2eB/\omega+\omega)\right),
			& \quad \quad \quad \\
			I^{(2)}= \dfrac{2 \omega^2\pi}{\sqrt{J_{nn^{\prime}}^2-4m_n^2\omega^2}J_{n,n^{\prime}}^{+}} 
			\left( \theta(\mu-2eB/\omega+\omega)+\theta(-\mu+2eB/\omega+\omega)\right),
			& \quad \quad     \quad,
			I^{(3)}= 0,
			& \quad \quad      \quad.
		\end{array} \right.
	\end{eqnarray}
	\appendix
	\section[\appendixname~\thesection]{Infrared regime of photon self-energy from Non linear electrodynamics}
	\label{NLEDIR}
	The second order expansion of non-linear Euler Heisenberg Lagrangian at finite density and temperature has the form 
	
	\begin{align}\nonumber
		\mathcal{L}^{\mu,T}=\frac{1}{2}{\mathcal{L}}^{\mu,T}_{\mathcal F} (F^{\mu\nu}f_{\mu\nu})&+\frac{1}{4}{\mathcal{L}}^{\mu,T}_{\mathcal F} (f^{\mu\nu}f_{\mu\nu}) +\mathcal{L}^{\mu,T}_{\mathcal{U}}(u^{\mu}f_{\mu\alpha}f^{\alpha\nu}u^{\nu} )+ \frac{1}{8}\mathcal{L}^{\mu,T}_{\mathcal{FF}} (f^{\mu\nu}F^B_{\mu\nu})^2
		+\\
		+&\frac{1}{8}\mathcal{L}^{\mu,T}_{\mathcal{GG}} (f^{\mu\nu}\tilde{F}^B_{\mu\nu})^2) + e\frac{\partial \mathcal{L}^{\mu,T}}{\partial \mu} a_{0}+ \frac{1}{2}e^2\frac{\partial^2 \mathcal{L}^{\mu,T}}{\partial \mu^2} (a_{0})^2+ \frac{1}{4}e\frac{\partial \mathcal{L}^{\mu,T}_{\mathcal F}}{\partial \mu} (F^{\mu\nu}f_{\mu\nu})a_{0}\label{NLEDT}.
	\end{align}
	The scalar $\mathcal{F}$ and the tensors $F$,  $\tilde{F}$ are assumed to be space- and time-independent and  ${u}_{\mu}$ is the four-velocity of the medium.  The last three terms arises due to the presence of the non-zero chemical potential, which is related to the Polyakov loop of the electromagnetic field (\cite{Hattori:2023egw}, \cite{Megias:2003}).
	The explicit form of derivatives $\mathcal{L}_{\mathcal{F},\mathcal{FF},\mathcal{GG},\mathcal{U}}^{(\mu,T)}$ is the following   
	
	\begin{align}
		\mathcal{L}^{T}_{\mathcal{F}}
		&=
		\frac{\partial\mathcal{L}^{T}}{\partial \mathcal{F}}
		\Big|_{f=0}
		\nonumber\\
		&=
		\frac{\alpha}{2\pi}
		\underbrace{
			\int_0^{\infty} \frac{dz}{z} e^{-z/b}
			\left(
			\frac{2 e^{2}}{3}
			-
			\frac{\coth z}{z}
			+
			\frac{1}{\sinh^{2} z}
			\right)
		}_{\kappa_V^{(1)}}
		\nonumber\\
		&\quad
		+
		\frac{\alpha}{2\pi}
		\int_0^{\infty}
		\frac{dz}{z}
		e^{-z/b}
		\left(
		-
		\frac{\coth z}{z}
		+
		\frac{1}{\sinh^{2} z}
		\right)
		\sum_{k=1}^{\infty}
		(-1)^{k}
		e^{-\frac{\beta^{2} k^{2}eB}{4z}}
		\cosh\!\left(e\mu\beta k\right),
	\end{align}
	
	\begin{eqnarray} \nonumber
		\mathcal{L}^{T}_{\mathcal{GG}}&=&\frac{\partial^2 \mathcal{L}^{T}}{\partial {\mathcal G}^2} \Big|_{f=0} 
		\\   
		&=&\frac{\alpha}{3\pi B^2} \underbrace{\int_0^{\infty} \frac{dz}{z} e^{-z/b}
			\left [ \left( z  - \frac{3}{2z}\right)\coth(z) - \frac{3}{2\sinh^2(z)} \right]  }_{\kappa^{(2)}_V}
		\\  \nonumber
		&+& \frac{\alpha}{3\pi B^2}\int_0^{\infty} \frac{dz}{z} e^{-z/b}
		\left [ \left( z  - \frac{3}{2z}\right)\coth(z) - \frac{3}{2\sinh^2(z)} \right ]  \sum^{\infty}_{k=1} (-1)^{k} e^{-\frac{\beta^{2} k^{2}eB}{4z}} \cosh \! \left(e \mu  \beta  k \right) \\  \nonumber
		&+& \frac{\alpha B^2}{3 \pi }\underbrace{\int_0^{\infty} \frac{dz}{z^2} e^{-z/b}
			\coth(z) \left (z  + \frac{3}{z}- 3\coth(z) \right ) \sum^{\infty}_{k=1} (-1)^{k} \frac{e\beta^{2} k^{2}}{4 B^3} e^{-\frac{\beta^{2} k^{2} eB}{4z}} \cosh \! \left(e \mu  \beta  k \right)}_{\rightarrow 0},
	\end{eqnarray}
	
	\begin{eqnarray}  \nonumber
		\mathcal{L}^{T}_{\mathcal{FF}} &=& \frac{\partial^2 \mathcal{L}^{T}}{\partial {\mathcal F}^2} \Big|_{f=0} \\
		&=&\frac{\alpha}{2\pi B^2} \underbrace{\int_0^{\infty} \frac{dz}{z} e^{-z/b}
			\left(
			z\coth(z) +\frac{1}{\sinh^2(z)} -\frac{2z\coth(z)}{\sinh^2(z)}
			\right)}_{\kappa^{(3)}_V}
		\\  \nonumber
		&+&\frac{\alpha}{2\pi B^2} \int_0^{\infty} \frac{dz}{z} e^{-z/b}
		\left(
		z\coth(z) +\frac{1}{\sinh^2(z)} -\frac{2z\coth(z)}{\sinh^2(z)}
		\right) \sum^{\infty}_{k=1} (-1)^{k} e^{-\frac{\beta^{2} k^{2}eB}{4z}} \cosh \! \left(e \mu  \beta  k \right),
	\end{eqnarray}
	
	\begin{eqnarray} \nonumber
		\mathcal{L}^{T}_{\mathcal{U}}&=&\frac{\partial \mathcal{L}^{T}}{\partial {\mathcal U}} \Big|_{f=0} 
		\\  \nonumber
		&=& \frac{\alpha B^2}{3 \pi }\underbrace{\int_0^{\infty} \frac{dz}{z^2} e^{-z/b}
			\coth(z) \left (-\frac{1}{z} +\coth(z) \right ) 2 \sum^{\infty}_{k=1} (-1)^{k} \frac{e\beta^{2} k^{2}}{4 B} e^{-\frac{\beta^{2} k^{2} eB}{4z}} \cosh \! \left(e \mu  \beta  k \right).}_{T\to 0,\rightarrow 0}
	\end{eqnarray}
	
	The expansion (\ref{NLEDT})  was previously used by us \cite{Jorge:2024tok} to study the thermodynamic properties of Dirac materials at finite density and zero temperature, assuming that the coefficients associated with the medium four-velocity and with the derivatives with respect to the chemical potential can be neglected. In what follows, we will assume this approximation as valid. Nevertheless, a more complete calculation  must consider all terms in the expression (\ref{NLEDT}) and will be addressed in a forthcoming article.
	
	In \cite{Shabad:2011hf}, the eigenvalues of the photon self-energy in the infrared limit were identified with the corresponding coefficients of the quadratic Euler-Heisenberg effective nonlinear Lagrangian. Given that, in the zero-temperature and finite-density limit, the coefficient associated with the medium four-velocity in the expansion of the effective Lagrangian vanishes, the expansion of the photon self-energy takes a form analogous to that obtained in Ref.~\cite{Shabad:2011hf}, with the coefficients depending not only on the magnetic field but also on the density. Its explicit form is given by
	$\Pi^{\rm IR}_{\mu\tau}(k,B,\mu,T)$
	\begin{equation}\label{effecIR}
		\begin{aligned}
			\Pi^{\rm IR}_{\mu\tau}(k,B,\mu,T) =
			&\;
			\frac{\partial \mathcal{L}^T(\mathcal{F},0,\mu,T) }{\partial \mathcal{F}}
			\left( \delta_{\mu\tau} k^2 - k_\mu k_\tau \right) \\
			&+ \frac{\partial^2 \mathcal{L}^T(\mathcal{F},0,\mu,T)}{\partial \mathcal{F}^2}
			(F_{\mu\alpha} k_\alpha)(F_{\tau\beta} k_\beta) \\
			&+ \left.
			\frac{\partial^2 \mathcal{L}^T(\mathcal{F},0,\mu,T)}{\partial \mathcal{G}^2}
			\right|_{\mathcal{G}=0}
			(\tilde{F}_{\mu\alpha} k_\alpha)(\tilde{F}_{\tau\beta} k_\beta).
		\end{aligned}
	\end{equation}
	
	\begin{equation}
		\mathcal{L}^T = \mathcal{L}^{V}(B) + \mathcal{L}^{M}(B,\mu,T) 
	\end{equation}
	The equivalence, within the one-loop approximation, between the effective Euler-Heisenberg (EH) Lagrangian in the infrared vacuum regime and the off-shell photon photon self-energy \cite{Shabad:2011hf} allows us to generalize to the medium the correspondence between the eigenvalues of the photon self-energy and the coefficients of the expansion, namely,

	\begin{align}
		\mathcal{L}^T_{\mathcal{F}} &= \kappa^{(1)}\big|_{k \to 0}, \\
		2\mathcal{F}\,\mathcal{L}^T_{\mathcal{FF}} &= \left(\kappa^{(1)}_T - \kappa^{(3)}_T\right)\big|_{k \to 0}, \\
		2\mathcal{F}\,\mathcal{L}^T_{\mathcal{GG}} &= \left(\kappa^{(1)}_T - \kappa^{(2)}_T\right)\big|_{k \to 0}.
	\end{align}
	
	\noindent and 
	\begin{equation}
		\kappa^i_T=\kappa^{1,2,3}_V(B) +\kappa_{M}^{1,2,3}(B,\mu,T)
	\end{equation}
	
	\begin{align}
		&\mathcal{L}^T_{\mathcal{F}}(B,\mu,T)=\mathcal{L}^{V}_{\mathcal{F}}(B)+\mathcal{L}^{M}_{\mathcal{F}}(B,\mu,T)\\
		&\mathcal{L}^T_{\mathcal{FF}}(B,\mu,T)=\mathcal{L}^{V}_{\mathcal{FF}}(B)+\mathcal{L}^{M}_{\mathcal{FF}}(B,\mu,T)\\
		&\mathcal{L}^T_{\mathcal{GG}}(B,\mu,T)=\mathcal{L}^{V}_{\mathcal{GG}}(B)+\mathcal{L}^{M}_{\mathcal{GG}}(B,\mu,T)
	\end{align}
	
	To verify the correspondence between $\mathcal{L}_{\mathcal{F}}$, $\mathcal{L}_{\mathcal{FF}}$, and $\mathcal{L}_{\mathcal{GG}}$ and the medium eigenvalues (see Section 4 (\ref{eqdispaprox2})-(\ref{eigeninfra3})), we exploit the equivalence between the Schwinger proper-time and imaginary-time formulations of the effective Lagrangian 
	\begin{align}
		&\mathcal{L}^{\mu,T}=\sum_{k=1}^{\infty}\frac{eB}{8\pi^2}\int\frac{ds}{s^{2}}e^{m^2s-\frac{\beta^2k^2}{4s}}\coth(eBs)\cosh(\beta\mu k)\\
		&=\frac{eB}{8\pi^2}\sum_{n} \alpha_n\int dp_3 \left (ln(1+e^{(\mu-E_n)\beta})ln(1+e^{(\mu+E_n)\beta})\right ),\label{Lagmu}
	\end{align}
	that at $T\to 0$ $\mu\ne 0$ (\ref{Lagmu}) becomes \cite{PhysRevD.51.2026}
	$$\mathcal{L}^{\mu,0}=\frac{\alpha b m^4}{8\pi^2}\sum_{n} \alpha_n\int dp_3 (E_n-\mu)\theta(\mu-E_n)=\frac{bm^4}{8\pi^2}\sum_{n} \alpha_n\left (\mu p_F-(1+2bn)\right)\ln\left (\frac{\mu+p_F}{\sqrt{(1+2bn)}}\right)$$
	Let us compute  the derivatives of Lagrangian, from the  expressions in imaginary time formalism which corresponds to the second expression of (\ref{Lagmu}). The rules to get the three derivatives are the following
	\[
	\mathcal{L}_{\mathcal{F}}^{\mu,0}=\left(
	X_0^{\mu,0}+h^2\frac{\partial X_0^{\mu,0}}{\partial h}\right)\]
	\[
	B^2\mathcal{L}^{\mu,0}_{\mathcal{FF}}=-\mathcal{L}_{\mathcal{F}}^{\mu,0}+h\frac{\partial^2 X_0^{\mu,0}}{\partial^2 h^2}\]
	\[B^2\mathcal{L}_{\mathcal{GG}}
	=
	\mathcal{L}_{\mathcal{F}}^{\mu,0}
	-
	\frac{2}{3h}
	\frac{\partial^2 X_0^{\mu,0}}{\partial m^2}\]
	
	where we have used  $\mathcal{L}^{\mu,0}=-bX_0=1/hX_0$

	Doing the derivatives, we get the integrals $\mathcal{L}^{(\mu)}_{\mathcal{F}}(\mu,B)$, $\mathcal{L}^{(\mu)}_{\mathcal{FF}}(\mu,B) $  and $\mathcal{L}^{(\mu)}_{\mathcal{GG}}(\mu,B)$ which corresponds to the eigenvalues $\kappa_{1}$,$\kappa_{2}$,$\kappa_{3}$ for perpendicular propagation.
	
	\begin{align}
		\mathcal{L}^{(\mu)}_{\mathcal{F}}(\mu,B) &\sim -\frac{\alpha}{2\pi B_c} \sum_{n=0}^{n_{\mu}} \alpha_n n
		\int \frac{dp_3\,\theta(\mu - E_{n})}{\sqrt{p_3^2 + m_n^2}} \nonumber \\
		&= -\frac{\alpha}{2\pi B_c} \sum_{n=0}^{n_{\mu}} \alpha_n n \ln\left(\frac{\mu + p_F}{m_n} \right)\to t/b\to 0 \\[1ex]
		B^2\mathcal{L}^{(\mu)}_{\mathcal{FF}}(\mu,B) &\sim \frac{\alpha eB}{2\pi B_c} \sum_{n=0}^{n_{\mu}} \alpha_n \int \frac{dp_3\,\theta(\mu - E_{n})}{2\sqrt{p_3^2 +m_n^2}} \nonumber \\
		&= \frac{\alpha eB}{2\pi B_c} \sum_{n=0}^{n_{\mu}} \frac{\alpha_n n}{2} \ln\left( \frac{\mu + p_F}{m_n} \right)\to t \label{LGGLFFM} \\[1ex]
		B^2\mathcal{L}^{(\mu)}_{\mathcal{GG}}(\mu,B)
		&\sim -\frac{\alpha eB}{4\pi B_c}
		\sum_{n=0}^{n_{\mu}} 4\alpha_n
		\int
		\frac{dp_3\,\theta(\mu-E_n)}{E_n^3}
		\nonumber \\
		&= -\frac{\alpha eB}{4\pi B_c} \sum_{n=0}^{n_{\mu}} 4  \alpha_n\frac{p_F}{\mu\,} \to s \label{LGGLFFM2}
	\end{align}
	
	In summary, the derivatives of the effective Lagrangian reproduce the infrared
	limits of the one-loop polarization eigenvalues for vacuum and perpendicular propagation of photon, providing an internal consistency check of the effective description used in the main text.

\end{document}